\documentclass[11pt,english]{article}
\usepackage[T1]{fontenc}
\usepackage[latin9]{inputenc}
\usepackage[letterpaper]{geometry}
\usepackage{color}
\usepackage{babel}
\usepackage{array}
\usepackage{verbatim}
\usepackage{multirow}
\usepackage{amsmath}
\usepackage{amssymb}
\usepackage{graphicx}
\usepackage{rotating}
\usepackage{setspace}
\usepackage[authoryear]{natbib}
\usepackage[unicode=true,pdfusetitle,
 bookmarks=true,bookmarksnumbered=false,bookmarksopen=false,
 breaklinks=false,pdfborder={0 0 0},pdfborderstyle={},backref=false,colorlinks=false]
 {hyperref}
\hypersetup{pdfborderstyle=,pdfborderstyle=}

\makeatletter

\providecommand{\tabularnewline}{\\}

\@ifundefined{date}{}{\date{}}

\usepackage{babel}
\usepackage{array}
\usepackage{multirow}
\usepackage{amsfonts}
\usepackage{xcolor}
\usepackage{array}
\usepackage{rotating}
\usepackage{booktabs}
\usepackage{url}
\usepackage{multirow}
\usepackage{colortbl}
\usepackage{subcaption}

\usepackage{authblk}

\providecommand{\tabularnewline}{\\}
\DeclareCaptionSubType*[alph]{figure}

\renewcommand{\hat}{\widehat}
\newtheorem{assumption}{Assumption}
\newenvironment{proof}[1][Proof]{\noindent \textbf{#1.} }{\  \rule{0.5em}{0.5em}}
\newtheorem{theorem}{Theorem}

\newtheorem{corollary}{Corollary}

\newtheorem{example}{Example}

\newtheorem{lemma}{Lemma}

\newtheorem{remark}{Remark}

\input{tcilatex}

\makeatother

\begin{document}
\title{On LASSO for Predictive Regression}
\author[a]{Ji Hyung Lee}
\author[b]{Zhentao Shi \thanks{ Corresponding author. 
Email addresses: \url{jihyung@illinois.edu} (J.~Lee),  \url{zhentao.shi@cuhk.edu.hk} (Z.~Shi), \url{zhangao@usc.edu} (Z.~Gao).
 }
 }
\author[c]{Zhan Gao}
\affil[a]{Department of Economics, University of Illinois. 1407 W. Gregory Dr., 214 David Kinley Hall, Urbana, IL 61801, USA.}
\affil[b]{Department of Economics, the Chinese University
of Hong Kong, Sha Tin, New Territories, Hong Kong SAR, China.}
\affil[c]{Department of Economics, University of Southern California. Kaprielian Hall, 3620 South Vermont Avenue, Los Angeles, CA 90089, USA.}

\renewcommand\Affilfont{\small \itshape}
\renewcommand\Authfont{\large}

\maketitle
\begin{abstract}
Explanatory variables in a predictive regression typically exhibit low signal strength and various degrees of persistence. Variable selection
in such a context is of great importance. In this paper, we explore
the pitfalls and possibilities of the LASSO methods in this predictive
regression framework. In the presence
of stationary, local unit root, and cointegrated predictors, we show
that the adaptive LASSO cannot asymptotically eliminate all cointegrating
variables with zero regression coefficients. This new finding motivates a novel post-selection adaptive LASSO, which we call the \emph{twin
adaptive LASSO} (TAlasso), to restore variable selection consistency.
Accommodating the system of heterogeneous
regressors, TAlasso achieves the well-known oracle property. In contrast,
conventional LASSO fails to attain coefficient estimation consistency
and variable screening in all components simultaneously. We 
apply these LASSO methods to evaluate the
short- and long-horizon predictability of S\&P 500 excess returns.
\end{abstract}



\thispagestyle{empty}

\newpage

\section{{Introduction\label{sec:Introduction}}}

Predictive regressions are used extensively in empirical
finance. A leading example is the stock return regression, for
which predictability has long been a primary goal. We focus on this type of predictive regression in this paper. The first central
econometric issue in these models is severe test size distortion in
the presence of highly persistent predictors coupled with regression
endogeneity. When persistence and endogeneity are high, the conventional
inferential apparatus designed for stationary data can be misleading.
Another major challenge in predictive regressions is the low signal-to-noise
ratio (SNR). One explanation is that the noise
in short-term predictive models may obscure the signal given by slow-moving (persistent) predictors. 

Even though certain predictors
have been suggested based on economic theory, there is no consensus regarding
the predictive ability of any predictor. A researcher may use \emph{ex
ante} a pool of candidate regressors, hoping to identify a few important
regressors. However, the more variables the researcher includes, the greater is
the need for a data-driven routine for variable selection as
many of these variables \emph{ex post} demonstrate little or no predictive ability
because of the competitive nature of the market. A low SNR naturally
calls for variable selection.

Advancements in machine learning techniques, driven by an unprecedented
abundance of data sources from many disciplines, offer valuable opportunities
for economic data analysis. In the era of big data, shrinkage methods
are becoming increasingly popular in econometric inference and prediction
because of their variable selection and regularization properties.
In particular, the \emph{least absolute shrinkage and selection operator}
\citep[LASSO;][]{tibshirani1996regression} has received intensive
study in the past two decades.

This paper investigates LASSO methods in predictive regressions.
The LASSO methods are attractive in predictive regressions because they enable researchers
to identify pertinent and exclude irrelevant predictors. However,
time series predictors in predictive regressions carry heterogeneous degrees
of persistence, which we call \textit{mixed roots}.
Some may exhibit short memory (e.g., Treasury bills),
whereas others may be highly persistent (e.g., most financial/macro predictors).
Moreover, a multitude of persistent predictors can be cointegrated.
For example, the dividend price (DP) ratio is essentially a
cointegrating residual between dividend and price, and the so-called
\emph{cay} data \citep{lettau2001consumption} is another cointegrating
residual between consumption, asset holdings, and labor income.

Given the difficulty of classifying time series predictors, we take an
agnostic view of the types of the regressors and examine whether the LASSO
methods can cope with heterogeneous regressors. We consider a predictive
regression with three categories of predictors: short memory (I(0))
regressors, cointegrated local-to-unity persistent regressors, and
non-cointegrated persistent regressors. Our focus is on variable selection.
It is well known that in cross-sectional regressions, the \emph{adaptive
LASSO} \citep[Alasso;][]{zou2006adaptive} enjoys \emph{variable selection
consistency}, which means that the estimated zeros exactly coincide with
the true zero coefficients when the sample size is sufficiently large.
The heterogeneous time series regressors challenge the conventional
wisdom.

The main contribution of this paper is its study and refinement of Alasso
in the time series context. As we consider a fixed number of regressors,
we use OLS as Alasso's initial estimator. Alasso, with a proper choice
of tuning parameter, consistently selects the non-cointegrating stationary and persistent
variables, but it may over-select inactive cointegrating variables,
where ``inactive'' means that these variables have zero regression coefficients, so they are excluded in the
true data-generating process (DGP). The convergence rate of the OLS
estimator is $\sqrt{n}$ for the cointegrated variables, and the resulting
Alasso penalty weight is too small to eliminate the true inactive
cointegrated groups because each cointegrating time series behaves as a
local unit root process. Nevertheless, as is formally
stated in Theorem \ref{thm:ada_3type}, with probability approaching
one (wpa1), variables forming an inactive cointegrating group cannot
all survive Alasso's selection.

This finding suggests a straightforward method of reclaiming the desirable oracle property: simply run a second-round
Alasso on the variables selected by the first-round Alasso. As
the first-round Alasso has broken the cointegration relationship between
inactive cointegrating variables, the over-selected variables become
non-cointegrated local-to-unity regressors in the second-round regression. In
the post-selection OLS, the speed of convergence of these variables
is boosted from the slow $\sqrt{n}$ rate to the fast $n$ rate.
Then, the second-round Alasso can successfully suppress all of the inactive
coefficients wpa1. We call this post-selection Alasso procedure \emph{twin
adaptive LASSO} and abbreviate it as TAlasso. 

TAlasso achieves the
oracle property \citep{fan2001variable}, which implies that the rate
of convergence is the same as the OLS when the relevant variables are
known in advance and that the variable selection is consistent in
the presence of a mixture of heterogeneous regressors. To the best of our knowledge,
this paper is the first to establish these desirable properties in
a nonstationary time series context. Note that the name TAlasso
distinguishes it from the \emph{post-selection double LASSO} \citep{belloni2014inference}.
Double LASSO is named for the double inclusion of selected variables
in cross-sectional data to correct for LASSO's shrinkage bias in order to achieve uniform
statistical inference. In contrast, our TAlasso features double exclusion
in predictive regressions with mixed roots. 

In developing asymptotic theory, we consider a linear process of
time series innovations that encompasses a general ARMA structure
arising in many practical applications, such as long-horizon return
prediction. To focus on the distinctive feature of nonstationary time
series, we adopt a simple asymptotic framework in which the number
of regressors $p$ is fixed and the number of time periods $n$ passes
to infinity. Our exploration in this paper is a stepping stone
toward automated variable selection in high-dimensional predictive
regressions.

We also explore the \emph{plain LASSO} \citep[Plasso henceforth;][]{tibshirani1996regression},
the \emph{standardized LASSO} (Slasso, in which the $l_{1}$-penalty
is multiplied by the sample standard deviation of each regressor)
in the predictive regression with heterogeneous regressors. We call
 the phenomenon \emph{variable screening effect} whereby an estimator
coerces some coefficients to exactly zero \citep{fanlv2008jrssb}.
We find that neither Plasso nor Slasso maintains variable screening
and consistent coefficient estimation in all of the components simultaneously,
as Plasso imposes equal penalty weight regardless of the nature
of the regressor, whereas Slasso's penalty only changes according to
the scale variation of individual regressors and ignores their connection
via cointegration.

In Monte Carlo simulations, we use various DGPs to examine the finite sample
performance of Alasso/TAlasso compared with Plasso/Slasso, assessing
their mean squared prediction errors and variable selection
success rates. TAlasso is much more capable in selecting the correct
model, thus improving prediction accuracy. These LASSO methods are
further evaluated in a real data application. We use the 12 predictors from \citet{welch2008comprehensive}
to predict the S\&P 500 returns. These 12
predictors exemplify the necessity of
considering the three types of heterogeneous regressors. Please see Section \ref{sec:empirical} for the details. Alasso and TAlasso attain better forecasting performance than their competitors and are robust in various estimation windows and prediction
horizons.

\paragraph{Literature Review}

Since the original LASSO paper by \citet{tibshirani1996regression} 
and the basis pursuit of \citet*{chen2001atomic}, a variety of important
extensions of LASSO have been proposed, such as Alasso \citep{zou2006adaptive}
and the elastic net \citep{zou2005regularization}. In econometrics, \citet{caner2009lasso},
\citet{caner2014adaptive}, \citet{shi2016estimation}, \citet*{su2016identifying},
and \citet{kock2019inference} use LASSO-type procedures in cross-sectional
and panel data models. \citet*{belloni2011L1}, \citet*{belloni2012sparse},
\citet*{belloni2014inference}, and \citet*{Belloni2015uniforminf} develop
methodologies and inferential theories in a variety of microeconometric
settings.

\citet{ng2013variable} surveys the variable selection methods in
predictive regressions. Unlike the extensive literature
on cross-sectional environments, the theoretical properties of shrinkage
methods have received relatively little attention in time series models despite great
empirical interest in macro and financial applications \citep{giannone2017economic,gu2018empirical,chinco2019sparse,rapach2019industry,freyberger2020dissecting}.
\citet{medeiros2016L1} study Alasso in high-dimensional stationary
time series, and \citet{kock2015oracle} discuss LASSO in a vector autoregression
(VAR) system. In time series forecasting, \citet{inoue2008useful}
apply various model selection and model averaging methods to forecast
U.S. consumer price inflation. \citet{sun2020time} study model averaging methods under structural changes, and \citet{hirano2017forecasting} develop
a local asymptotic framework with independently and identically distributed
(iid) orthonormalized predictors to study the risk properties of several
machine learning estimators. There are even fewer papers on LASSO with nonstationary data. \citet{caner2013alternative} discuss the bridge estimator,
a generalization of LASSO, for the augmented Dickey-Fuller test in
autoregressions, and \citet{kock2016consistent}
studies Alasso under the same setup. Earlier works on variable selection in the I(1) case
utilize the different stochastic limit of penalty functions with stochastically
trending regressors, such as \citet{phillips1994posterior} and \citet{ng2001lag}.
For example, the stochastic limit of penalty factors is different with integrated regressors compared with the stationary
case \citep[pp. 1529-1530]{ng2001lag}. 

In predictive regressions, \citet{kostakis2014robust}, \citet{lee2016predictive},
and \citet{phillips2013predictive,phillips2016robust} provide inferential
procedures in the presence of multiple predictors with various degrees
of persistence. \citet{xu2018testing} studies variable selection
and inference with possible cointegration between the I(1) predictors.
In a vector error correction model (VECM), \citet{liao2015automated}
use Alasso for cointegration rank selection. Under a similar setting
with high-dimensional I(0) and I(1) regressors and one cointegration
group, \citet{koo2016high} investigate Plasso's variable estimation
consistency and non-standard asymptotic distribution. We
differ from \citet{koo2016high} in two aspects. (i) In an asymptotic
framework that accommodates high-dimensional regressors, they focus
on the rate of convergence of Plasso. In contrast, our focus is the oracle
property in the mixed root model with three types of regressors. (ii)
The unit root variables enter their model via the cointegrating groups
only, so they do not deal with the balance of the predictive regression,
whereas in our analysis, we allow local-to-unity persistent variables
and local-to-zero small coefficients. Finally, \citet{smeekes2018macroeconomic}
demonstrate LASSO's numerical performance via simulations and empirical
examples, while we are the first to systematically explore the theory concerning
variable selection under mixed regressor persistence.

\paragraph{Notation}

We use standard notation.
 We define $\left\Vert \cdot\right\Vert _{1}$
and $\left\Vert \cdot\right\Vert $ as the usual vector $l_{1}$-
and $l_{2}$-norms, respectively. The arrows $\Longrightarrow$ and
$\overset{p}{\to}$ represent weak convergence and convergence in
probability, respectively. $\asymp$ indicates ``of the same asymptotic
order, '' and $\sim$ signifies ``being distributed as'' either exactly
or asymptotically, depending on the context. $\left\lfloor \cdot\right\rfloor $
is the floor function. $a \wedge b = \min\{a,b\}$.
The symbols $O\left(1\right)$ and $o(1)$
($O_{p}\left(1\right)$ and $o_{p}(1)$) denote (stochastically) asymptotically
bounded or negligible quantities. For a generic set $M$, let $\vert M\vert$
be its cardinality. For a generic vector $\theta=(\theta_{j})_{j=1}^{p}$
with $p\geq\vert M\vert$, let $\theta_{M}=(\theta_{j})_{j\in M}$
be the subvector of $\theta$ associated with the index set $M$.
$I_{p}$ is the $p\times p$ identity matrix, $\boldsymbol{1}_{p}$
is a column of $p$ ones, and $I\left(\cdot\right)$ is the indicator
function.

\bigskip{}
 The rest of this paper is organized as follows. Section \ref{sec:UR}
introduces unit root regressors into a simple LASSO framework to fix
ideas of Alasso. This model is substantially generalized in Section
\ref{sec:MRC} to include I(0), local-to-unity, and cointegrated
regressors, and the asymptotic properties of Alasso and TAlasso are established.
The mixed root model is carried over into Section \ref{sec:Conventional-LASSO-with}
to explore the behaviors of Plasso and Slasso. The theoretical results
are confirmed through a set of empirically motivated simulation designs
in Section \ref{sec:Monte-Carlo-Simulation}. Finally, we examine
stock return predictions via these LASSO methods in Section \ref{sec:empirical}.

\section{{Adaptive LASSO with Unit Roots\label{sec:UR}}}

In this section, we study LASSO with $p$ unit root regressors. To
fix ideas, we investigate the asymptotic behavior of Alasso under
a simple nonstationary regression model to better understand
the technical issues arising from nonstationary predictors. Section
\ref{sec:MRC} generalizes the model to include the I(0), (local)
unit roots, and cointegrated predictors.

\subsection{Model}

Assume that the dependent variable $y_{i}$ is generated from a linear
model 
\begin{equation}
y_{i}=\sum_{j=1}^{p}x_{ij}\beta_{jn}^{\ast}+u_{i}=x_{i\cdot}\beta_{n}^{\ast}+u_{i},\ i=1,\ldots,n,\label{eq:DGP1}
\end{equation}
where $n$ is the sample size. The $p\times1$ true coefficient is
$\beta_{n}^{\ast}=(\beta_{jn}^{\ast}=\beta_{j}^{0\ast}/n^{\delta_{j}})_{j=1}^{p}$,
where $\beta_{j}^{0\ast}\in\mathbb{R}$ is a fixed constant independent
of the sample size and $\delta_{j}\in\left(0,1\right)$ is the local-to-zero
rate, following \citet{phillips2013predictive} and \citet{timmermann2017monitoring}.
If $\beta_{j}^{0\ast}=0$, then $\beta_{jn}^{\ast}$ remains zero
regardless of the sample size; if $\beta_{j}^{0\ast}\neq0$, then
it varies with $n$. This type of local-to-zero coefficient is designed
to balance the I(0)--I(1) relation between the stock return and 
unit root predictors, and to model the weak SNR in predictive
regressions \citep{phillips2015halbert}.
The $1\times p$ regressor vector $x_{i\cdot}=(x_{i1},\ldots,x_{ip})$
follows a pure unit root process 
\begin{equation}
x_{i\cdot}=x_{(i-1)\cdot}+e_{i\cdot}=\sum_{k=1}^{i}e_{k\cdot},\label{DGP x}
\end{equation}
where $e_{k\cdot}=(e_{k1},...,e_{kp})$ is the innovation. For simplicity,
we assume the initial value $e_{0\cdot}=0$. We use the following
iid assumption on the innovations.

\begin{assumption} \label{ass:INNOV} The innovations $e_{i\cdot}$
and $u_{i}$ are generated from 
\[
\left(e_{i\cdot},u_{i}\right)^{\prime}\sim iid\text{ }\left(0,\Sigma\right),
\]
where $\Sigma=\left(\begin{array}{cc}
\Sigma_{ee} & \Sigma_{eu}\\
\Sigma_{eu}^{\prime} & \sigma_{u}^{2}
\end{array}\right)$ is positive-definite. \end{assumption}

The regression equation (\ref{eq:DGP1}) can be equivalently written
as 
\begin{equation}
y=\sum_{j=1}^{p}x_{j}\beta_{jn}^{\ast}+u=X\beta_{n}^{\ast}+u,\label{DGP y x}
\end{equation}
where $y=(y_{1},\ldots,y_{n})^{\prime}$ is the $n\times1$ response
vector, $u=(u_{1},\ldots,u_{n})^{\prime}$, $x_{j}=(x_{1j},\ldots,x_{nj})^{\prime}$,
and $X=(x_{1},\ldots,x_{p})$ is the $n\times p$ predictor matrix.
This pure I(1) regressor model in (\ref{DGP y x}) is a direct extension
of the common predictive regression application with a single unit
root predictor (e.g., DP ratio). The mixed roots case in Section \ref{sec:MRC} has multiple predictors and is more realistic in practice.

The literature focuses on the non-standard statistical inference
caused by persistent regressors and weak signals. The asymptotic theory
is usually confined to a small number of candidate predictors. Following the literature on predictive regressions, we consider
the asymptotic framework in which $p$ is fixed and the sample size
$n\rightarrow\infty$. This simple asymptotic framework allows us
to concentrate on the contrast between the standard iid setting and
the predictive regression involving nonstationary regressors.

In this model, the unknown coefficients $\beta_{n}^{\ast}$ can be obtained from the data by running OLS 
\[
\widehat{\beta}^{\mathrm{ols}}=\arg\min_{\beta}\Vert y-X\beta\Vert^{2},
\]
whose asymptotic behavior is well understood \citep{phillips1987time}.
Assumption \ref{ass:INNOV} implies the following functional central
limit theorem: 
\begin{equation}
\frac{1}{\sqrt{n}}\sum_{k=1}^{\left\lfloor nr\right\rfloor }\left(\begin{array}{c}
e_{k\cdot}^{\prime}\\
u_{k}
\end{array}\right)\Longrightarrow\left(\begin{array}{c}
B_{e}(r)\\
B_{u}(r)
\end{array}\right)\equiv BM\left(\Sigma\right).\label{FCLT}
\end{equation}
To represent the asymptotic distribution of the OLS estimator, define
$u_{i}^{+}=u_{i}-\Sigma_{eu}^{\prime}\Sigma_{ee}^{-1}e_{i\cdot}^{\prime}$, and then $n^{-1/2}\sum_{i=1}^{\left\lfloor nr\right\rfloor }u_{i}^{+}\Longrightarrow B_{u^{+}}(r)$.
By definition, $\mathrm{cov}\left(e_{ij},u_{i}^{+}\right)=0$ for
all $j$ so that 
\[
\frac{X^{\prime}u}{n}\Longrightarrow\zeta:=\int_{0}^{1}B_{e}(r)dB_{u^{+}}(r)+\int_{0}^{1}B_{e}(r)\Sigma_{eu}^{\prime}\Sigma_{ee}^{-1}dB_{e}(r)^{\prime},
\]
which is the sum of a (mixed) normal random vector and a non-standard
random vector. The OLS limit distribution is 
\begin{equation}
n\left(\hat{\beta}^{\mathrm{ols}}-\beta_{n}^{\ast}\right)=\left(\frac{X^{\prime}X}{n^{2}}\right)^{-1}\frac{X^{\prime}u}{n}\Longrightarrow\Omega^{-1}\zeta,\label{eq:ols_pure}
\end{equation}
where $\Omega:=\int_{0}^{1}B_{e}(r)B_{e}(r)^{\prime}dr$. This result
implies that when we inflate $\hat{\beta}_{j}^{\mathrm{ols}}$ by
the factor $n^{\delta_{j}}$ so that its magnitude is comparable to
the constant $\beta_{j}^{0\ast}$, we attain consistency in that 
\[
n^{\delta_{j}}\left(\hat{\beta_{j}}^{\mathrm{ols}}-\beta_{jn}^{\ast}\right)=n^{\delta_{j}}\hat{\beta_{j}}^{\mathrm{ols}}-\beta_{j}^{0\ast}=O_{p} (n^{\delta_{j}-1})=o_{p}\left(1\right)\ \mbox{for all }j\leq p.
\]

Even if $\beta_{j}^{0*}\neq0$, when  $\delta_{j}$ is close to 1,
the signal of $x_{j}$ is weak, and therefore
the rate of convergence is slow. However, some true coefficients
$\beta_{j}^{0\ast}$ in (\ref{DGP y x}) could be exactly zero, where
the associated predictors would be redundant (inactive) in the regression. Let
$M^{\ast}=\{j:\beta_{j}^{0\ast}\neq0\}$ be the index set of regressors
relevant to the regression, $p^{\ast}=\left\vert M^{\ast}\right\vert $,
and $M^{\ast c}=\left\{ 1,\ldots,p\right\} \backslash M^{\ast}$ be
the set of redundant regressors. For simplicity, we refer to $M^{*}$
as the \emph{active set, }meaning that it plays an active role in
the regression, and we call $M^{*c}$ the \emph{inactive set}. If
we had knowledge about $M^{\ast}$, ideally, we would estimate the
unknown parameters by OLS in the active set $M^{\ast}$ only. Define
$\widehat{\beta}^{\mathrm{ora}}=\left(\widehat{\beta}_{M^{*}}^{\mathrm{ora}\prime},\widehat{\beta}_{M^{*c}}^{\mathrm{ora}\prime}\right)^{\prime}$,
where
\[
\widehat{\beta}_{M^{*}}^{\mathrm{ora}}=\arg\min_{\beta}\Vert y-\sum_{j\in M^{\ast}}x_{j}\beta_{j}\Vert^{2}
\]
and $\widehat{\beta}_{M^{*c}}^{\mathrm{ora}}=\boldsymbol{0}$. We
call $\widehat{\beta}^{\mathrm{ora}}$ the \emph{oracle} estimator,
which is based on the infeasible ``oracle'' information of $M^{*}$.
Eq.(\ref{eq:ols_pure}) implies that its asymptotic distribution is 
\[
n\left(\hat{\beta}^{\mathrm{ora}}-\beta_{n}^{\ast}\right)_{M^{\ast}}\Longrightarrow\Omega_{M^{\ast}}^{-1}\zeta_{M^{\ast}},
\]
where $\Omega_{M^{\ast}}$ is the $p^{\ast}\times p^{\ast}$ submatrix
$\left(\Omega_{jj^{\prime}}\right)_{j,j^{\prime}\in M^{\ast}}$ and
$\zeta_{M^{\ast}}$ is the $p^{\ast}\times1$ subvector $\left(\zeta_{j}\right)_{j\in M^{\ast}}$.

\subsection{{Adaptive LASSO \label{sec:alasso_ur}}}

We study the asymptotic behavior of Alasso in predictive regressions
with these pure unit root regressors. Alasso for (\ref{eq:DGP1})
is defined as 
\begin{equation}
\hat{\beta}^{\mathrm{A}}=\arg\min_{\beta}\bigg\{\Vert y-X\beta\Vert^{2}+\lambda_{n}\sum_{j=1}^{p}\hat{\tau}_{j}|\beta_{j}|\bigg\},\label{eq:aLasso}
\end{equation}
where the weight $\hat{\tau}_{j}=|\hat{\beta}_{j}^{\mathrm{init}}|^{-\gamma}$
for some initial estimator $\widehat{\beta}_{j}^{\mathrm{init}}$,
and $\lambda_{n}$ and $\gamma$ are the two tuning parameters. In
practice, $\gamma$ is often fixed at either 1 or 2, and $\lambda_{n}$
is selected as the primary tuning parameter. Thus, we discuss the case of
a fixed $\gamma\geq1$ and the initial estimator $\widehat{\beta}^{\mathrm{init}}=\widehat{\beta}^{\mathrm{ols}}$
because of the fixed $p$ setting.

Alasso enjoys the oracle property in regressions with weakly dependent
regressors \citep{medeiros2016L1}. The following Theorem \ref{thm:ada_unit}
confirms that Alasso maintains the oracle property in regressions
with unit root regressors. Let $\widehat{M}^{\mathrm{A}}=\{j:\hat{\beta}_{j}^{\mathrm{A}}\neq0\}$
be Alasso's estimated active set, and $\bar{\delta}=\max\{(\delta_{j})_{j=1}^{p}\}$
be the fastest speed such that $\beta_{jn}^{*}$ shrinks to 0. 

\medskip{}

\begin{theorem} \label{thm:ada_unit} Suppose that the linear model (\ref{eq:DGP1})
satisfies Assumption \ref{ass:INNOV}. If the tuning parameters $\lambda_{n}$
and $\gamma$ are chosen such that 
\begin{equation}
\frac{\lambda_{n}}{n^{1-\bar{\delta}\gamma}}+\frac{1}{\lambda_{n}}\rightarrow0\label{eq:rate_pure_alasso}
\end{equation}
and $\bar{\delta}\gamma<1$, then
\begin{enumerate}
\item Variable selection consistency: $P(\widehat{M}^{\mathrm{A}}=M^{\ast})\rightarrow1.$
\item Asymptotic distribution: $n(\hat{\beta}^{\mathrm{A}}-\beta_{n}^{\ast})_{M^{\ast}}\Longrightarrow\Omega_{M^{\ast}}^{-1}\zeta_{M^{\ast}}.$ 
\end{enumerate}
\end{theorem}

\bigskip{}

As shown in Theorem \ref{thm:ada_unit} (a), the estimated active set
$\widehat{M}^{\mathrm{A}}$ coincides with the true active set $M^{*}$
wpa1; in other words, $\widehat{\beta}_{j}^{\mathrm{A}}\neq0$ if $j\in M^{*}$
and $\widehat{\beta}_{j}^{\mathrm{A}}=0$ if $j\in M^{*c}$. (b) indicates
that Alasso's asymptotic distribution in the true active set is as
if the oracle $M^{*}$ is known. In this nonstationary regression,
Alasso's adaptiveness is maintained through the proper choice of $\hat{\tau}_{j}=|\hat{\beta}_{j}^{\mathrm{ols}}|^{-\gamma}$.
When the true coefficient is nonzero, $\hat{\tau}_{j}$
delivers a penalty of a negligible order $\lambda_{n}/n^{1-\delta_{j}\gamma}\to0$,
recovering the OLS limit theory. When the true coefficient
is zero, $\hat{\tau}_{j}$ imposes a heavier penalty of the order
$\lambda_{n}/n^{1-\gamma}\rightarrow\infty$, thereby achieving consistent
variable selection. The intuition in \citet[Remark 2]{zou2006adaptive}
under deterministic design is generalized in our proof of the setting
with nonstationary regressors.

Both $\bar{\delta}$ and $\gamma$ appear in the rate condition (\ref{eq:rate_pure_alasso})
for generality. In practice, because $\bar{\delta}$ is an unknown feature
of the DGP and $\gamma$ can be controlled by
the user, we recommend setting $\gamma=1$, which simplifies (\ref{eq:rate_pure_alasso})
as $ \lambda_{n} / n^{1-\bar{\delta}}+ 1/ \lambda_{n} \rightarrow0.$
A conservative choice of $\lambda_{n}$ diverging more slowly than
polynomial orders of $n$, say, $\lambda_{n}\asymp\log\log n$\textbf{,
}satisfies the above condition for all $\bar{\delta}>0$. The practical
choice of the tuning parameter in the numerical simulations and the
empirical application will be  discussed in the corresponding sections.


\section{{Adaptive LASSO with Mixed Roots \label{sec:MRC} }}

In practice, we often encounter a multitude of candidate predictors
that exhibit various dynamic patterns. Some are stationary, whereas
others can be highly persistent and/or cointegrated. In this section,
we discuss the theoretical properties of Alasso under a mixed persistence
environment. We extend the model in Section \ref{sec:UR} to accommodate
I(0) and (local) unit root regressors, with possible cointegration
in the latter. The theory in this section provides general
guidance for multivariate predictive regressions.

\subsection{Model}

We introduce three types of predictors into the model. A $1\times p_{c}$
cointegrated system $x_{i\cdot}^{c}=(x_{i1}^{c},...,x_{ip_{c}}^{c})$
has cointegration rank $p_{1}$, so that $p_{2}=p_{c}-p_{1}$ is the
number of local unit roots in this cointegration system. Let $x_{i\cdot}^{c}=(x_{1i\cdot}^{c}, x_{2i\cdot}^{c})$ admit a triangular representation 
\begin{gather}
\underset{p_1 \times 1}{x_{1i\cdot}^{c\prime}} -\underset{p_{1}\times p_{2}}{A_{1}} \underset{p_2 \times 1}{x_{2i\cdot}^{c\prime}} =\underset{p_{1}\times1}{v_{1i\cdot}^{\prime}},\label{TRI CO}\\
(I_{p_{2}}-R_{2}L)x_{2i\cdot}^{c\prime}=v_{2i\cdot}^{\prime},\nonumber 
\end{gather}
where  $R_{2}=I_{p_{2}}+c_{2}/n$ with
$c_{2}=\mathrm{diag}\left( \check{c}_{1},...,\check{c}_{p_{2}}\right)$,
$L$ is the lag operator, and the vector $v_{1i\cdot}$ is the cointegrating
residual, whose initialization is an $O_{p}(1)$ stationary variable. Each local-to-unity parameter
 $\check{c}_{l}$ is finite, thus including both stationary and nonstationary local unit root regions. 
This local-to-unity specification includes the unit root process as
a special case when $\check{c}_{l}=0$.
The triangular representation \citep{Phillips1991cointinference,elliott1998}
is a convenient and general form of cointegrated systems. \citet{xu2018testing}
also adopts this structure in predictive regressions.

Assume that $y_{i}$ is generated from the linear model 
\begin{equation}
y_{i}=\sum_{l=1}^{p_{z}}z_{il}\alpha_{l}^{\ast}+\sum_{l=1}^{p_{1}}v_{1il}\phi_{1l}^{\ast}+\sum_{l=1}^{p_{x}}x_{il}\beta_{l}^{\ast}+u_{i}=z_{i\cdot}\alpha^{\ast}+v_{1i\cdot}\phi_{1}^{\ast}+x_{i\cdot}\beta^{\ast}+u_{i}.\label{eq:DGP_XYZ_infeasible}
\end{equation}
Each time series $z_{l}=(z_{1l},...,z_{nl})'$ is a stationary regressor.
Each $x_{l}=(x_{1l},...,x_{nl})'$ is a local-to-unity process (initialized
from $O_{p}(1)$ stationary process) such that $x_{i\cdot}=(x_{i1},...,x_{ip_{x}})$
satisfies
$(I_{p_{x}}-R_{x}L)x_{i\cdot}^{\prime}=e_{i\cdot}^{\prime},$
where $R_{x}=I_{p_{x}}+c_{x}/n$ with $c_{x}=\mathrm{diag}\left(\tilde{c}_{1},...,\tilde{c}_{p_{x}}\right)$
for finite $\tilde{c}_{l}$.

Eq.(\ref{eq:DGP_XYZ_infeasible}) is \emph{infeasible}  because
the cointegrating residual $v_{1i\cdot}$ is unobservable without
\emph{a priori} knowledge about the cointegration relationship. What
we observe is the vector $x_{i\cdot}^{c}$ that contains cointegration
groups. Substituting (\ref{TRI CO}) into (\ref{eq:DGP_XYZ_infeasible}),
we obtain a \emph{feasible} regression equation 
\begin{equation}
y_{i}=z_{i\cdot}\alpha^{\ast}+x_{1i\cdot}^{c}\phi_{1}^{\ast}+x_{2i\cdot}^{c}\phi_{2}^{\ast}+x_{i\cdot}\beta^{\ast}+u_{i}=z_{i\cdot}\alpha^{\ast}+x_{i\cdot}^{c}\phi^{\ast}+x_{i\cdot}\beta^{\ast}+u_{i},\label{DGP XYZ}
\end{equation}
where $\phi_{2}^{\ast}=-A_{1}^{\prime}\phi_{1}^{\ast}$ and $\phi^{\ast}=\left(\phi_{1}^{\ast\prime},\phi_{2}^{\ast\prime}\right)^{\prime}$.
Stacking the sample of $n$ observations, the infeasible and feasible
regressions can be written as 
\begin{align}
y & =Z\alpha^{\ast}+V_{1}\phi_{1}^{\ast}+X\beta^{\ast}+u\label{DGP XYZ 1}\\
 & =Z\alpha^{\ast}+X_{1}^{c}\phi_{1}^{\ast}+X_{2}^{c}\phi_{2}^{\ast}+X\beta^{\ast}+u\nonumber \\
 & =Z\alpha^{\ast}+X^{c}\phi^{\ast}+X\beta^{\ast}+u,\label{DGP XYZ 2}
\end{align}
where the variable $V_{1}$ is the $n\times p_{1}$ matrix that stacks
$\left(v_{1i\cdot}\right)_{i=1}^{n}$, and $X_{1}^{c}$, $X_{2}^{c}$,
$X^{c}$, $Z$, and $X$ are defined similarly. We explicitly define
$\alpha^{\ast}=\alpha^{0\ast}$ and $\phi^{\ast}=\phi^{0\ast}$ as
two coefficients independent of the sample size, which are associated
with $Z$ and $X^{c}$, respectively. As in Section \ref{sec:UR},
the coefficients associated with $X$ are specified as local-to-zero
sequences $\beta^{\ast}=\beta_{n}^{\ast}=(\beta_{l}^{0\ast}/n^{\delta_{j}})_{l=1}^{p_{x}}$,
where $\beta_{l}^{0\ast}$ is invariant to the sample size.

We assume a linear process for the innovation and cointegrating residual
vectors. In contrast to the simplistic iid assumption in Section \ref{sec:UR},
the linear process assumption is fairly general, including as special
cases many practical dependent processes such as the stationary autoregressive
and moving average processes. Let $v_{i\cdot}=\left(v_{1i\cdot},v_{2i\cdot}\right)$
and $p=p_{z}+p_{c}+p_{x}$.

\begin{assumption} \label{ass:INNOV coint} {[}Linear Process{]}
The vector of stacked innovation and stationary predictors follows the linear process: 
\[
\underset{(p+1)\times1}{\xi_{i}}:=(z_{i\cdot},v_{i\cdot},e_{i\cdot},u_{i})^{\prime}=F(L)\varepsilon_{i}=\sum_{k=0}^{\infty}F_{k}\varepsilon_{i-k},
\]
where $\underset{(p+1)\times1}{\varepsilon_{i}}=\begin{pmatrix}\varepsilon_{i\cdot}^{(z)\prime}\\
\varepsilon_{i\cdot}^{(v)\prime}\\
\varepsilon_{i\cdot}^{(e)\prime}\\
\varepsilon_{i}^{(u)\prime}
\end{pmatrix}\sim\mathrm{iid}\left(0,\Sigma_{\varepsilon}=\left(\begin{array}{cccc}
\Sigma_{zz} & \Sigma_{zv} & \Sigma_{ze} & 0\\
\Sigma_{zv}^{\prime} & \Sigma_{vv} & \Sigma_{ve} & 0\\
\Sigma_{ze}^{\prime} & \Sigma_{ve}^{\prime} & \Sigma_{ee} & \Sigma_{eu}\\
0 & 0 & \Sigma_{eu}^{\prime} & \Sigma_{uu}
\end{array}\right)\right)$, $F_{0}=I_{p+1},$ $\sum_{k=0}^{\infty}k\left\Vert F_{k}\right\Vert <\infty,$
$F(x)=\sum_{k=0}^{\infty}F_{k}x^{k}$ and $F(1)=\sum_{k=0}^{\infty}F_{k}>0.$
\end{assumption}

\begin{remark} Following the cointegration and predictive regression
literature, we allow the correlation between the innovation of the regression
error $\varepsilon_{i}^{\left(u\right)}$ and the innovation of nonstationary
predictors $\varepsilon_{i}^{\left(e\right)}$. However, to ensure identification, we rule out the correlation between $\varepsilon_{i}^{\left(u\right)}$
and the innovation of the stationary or cointegrated predictors. \end{remark}

\subsection{{OLS\label{subsec: MRC ols SEC}}}

As Alasso attaches a penalty term to the OLS criterion function,
we first study the asymptotic distribution of the OLS estimator 
\[
\widehat{\theta}^{\mathrm{ols}}=(W^{\prime}W)^{-1}W^{\prime}y
\]
under the mixed roots, where $W = \left( Z, X^c, X \right)$  
is the observed predictor matrix. To state the result, we define the
true coefficients $\theta_{n}^{*}=\left(\alpha^{0\ast\prime},\phi_{1}^{0\ast\prime},\phi_{2}^{0\ast\prime},\beta_{n}^{\ast\prime}\right)^{\prime}$,
where $\phi_{2}^{0*}=-A_{1}^{\prime}\phi_{1}^{0*}$, a diagonal normalizing
matrix $R_{n}=\mathrm{diag}\left(\left(\sqrt{n}\boldsymbol{1}_{p_{z}+p_{1}}^{\prime},n\boldsymbol{1}_{p_{2}+p_{x}}^{\prime}\right)\right)$,
and a rotation matrix $Q=\begin{pmatrix}I_{p_{z}} & 0 & 0 & 0\\
0 & I_{p_{1}} & 0 & 0\\
0 & A_{1}^{\prime} & I_{p_{2}} & 0\\
0 & 0 & 0 & I_{p_{x}}
\end{pmatrix}$. 

\begin{theorem} \label{thm:OLS} If the linear model (\ref{DGP XYZ})
satisfies Assumption \ref{ass:INNOV coint}, then 
\begin{align}
R_{n}Q\left(\widehat{\theta}^{\mathrm{ols}}-\theta_{n}^{\ast}\right)= & \begin{pmatrix}\sqrt{n}(\widehat{\alpha}^{\mathrm{ols}}-\alpha^{0*})\\
\sqrt{n}(\widehat{\phi}_{1}^{\mathrm{ols}}-\phi_{1}^{0*})\\
n(A_{1}^{\prime}\widehat{\phi}_{1}^{\mathrm{ols}}+\widehat{\phi}_{2}^{\mathrm{ols}})\\
n(\widehat{\beta}^{\mathrm{ols}}-\beta_{n}^{*})
\end{pmatrix}\Longrightarrow\left(\Omega^{+}\right)^{-1}\zeta^{+}.\label{eq:OLS_Q_rate}
\end{align}
where $\Omega^{+}$ is the weak limit of $R_{n}^{-1}Q^{\prime-1}W^{\prime}WQ^{-1}R_{n}^{-1}$
and $\zeta^{+}$ is the weak limit of $R_{n}^{-1}Q^{\prime-1}W^{\prime}u$,
whose explicit expressions are spelled out in (\ref{eq:ols_term_quad})
and (\ref{eq:ols_term_cross}) in Appendix Section \ref{subsec:Proofs-in-Section3},
respectively. \end{theorem}

\begin{remark} To see the effect of rotation matrix $Q$, note
that the definitions of $\Omega^{+}$ and $\zeta^{+}$ include
the transformation $Q^{\prime-1}W=\left(Z^{+},X^{+}\right)$, where
$Z^{+}=\left(Z,V_{1}\right)$ is the ``extended'' stationary regressors
and $X^{+}=\left(X_{2}^{c},X\right)$ is the ``extended'' nonstationary
regressors. The feasible regressors in $W$ are mapped by $Q^{\prime-1}$
into the infeasible regressors $\left(Z^{+},X^{+}\right)$ in which
the stationary and nonstationary components are separated. Then, the
scaling factor $R_{n}^{-1}$ sends the building blocks of the regression
analysis to their respective asymptotic weak limits as $n\to\infty$, in which
$\Omega^{+}$ is the weak limit of the Gram matrix and
$\zeta^{+}$ is the weak limit of the empirical process. \end{remark}

\begin{remark} \label{rmk:OLS} Because we keep an agnostic view about
the identities of the stationary, local unit root, and cointegrated
regressors, Theorem \ref{thm:OLS} is not useful for statistical inference,
as we do not know which coefficients converge at the $\sqrt{n}$-rate
and which at the $n$-rate. (\ref{eq:OLS_Q_rate}) shows that with the
help of the rotation $Q$, the estimators $\widehat{\phi}_{1}^{\mathrm{ols}}$
and $\widehat{\phi}_{2}^{\mathrm{ols}}$ are tightly connected in
the sense that $A_{1}^{\prime}\widehat{\phi}_{1}^{\mathrm{ols}}+\widehat{\phi}_{2}^{\mathrm{ols}}=O_{p}\left(n^{-1}\right)$.
Without the rotation that is unknown in practice, the OLS components
associated with the stationary and cointegration systems converge
at the $\sqrt{n}$ rate, whereas only those associated with the non-cointegration
local unit roots converge at the $n$ rate. 
Let $R_{n}^{\mathrm{f}}=\mathrm{diag}((\sqrt{n}\boldsymbol{1}_{p-p_{x}}^{\prime},n\boldsymbol{1}_{p_{x}}^{\prime}))$
be the normalizing matrix for the feasible components; then, 
\[
R_{n}^{\mathrm{f}}\left(\widehat{\theta}^{\mathrm{ols}}-\theta_{n}^{\ast}\right)\Longrightarrow\left(\lim_{n\to\infty}R_{n}^{\mathrm{f}}Q^{-1}R_{n}^{-1}\right)\left(\Omega^{+}\right)^{-1}\zeta^{+},
\]
or, more explicitly, 
\begin{equation}
\begin{pmatrix}\sqrt{n}(\widehat{\alpha}^{\mathrm{ols}}-\alpha^{0\ast})\\
\sqrt{n}(\widehat{\phi}_{1}^{\mathrm{ols}}-\phi_{1}^{0\ast})\\
\sqrt{n}(\widehat{\phi}_{2}^{\mathrm{ols}}-\phi_{2}^{0\ast})\\
n(\widehat{\beta}^{\mathrm{ols}}-\beta_{n}^{\ast})
\end{pmatrix}\Longrightarrow\begin{pmatrix}I_{p_{z}} & 0 & 0 & 0\\
0 & I_{p_{1}} & 0 & 0\\
0 & -A_{1}^{\prime} & 0 & 0\\
0 & 0 & 0 & I_{p_{x}}
\end{pmatrix}\left(\Omega^{+}\right)^{-1}\zeta^{+}.\label{eq:OLS_marginal}
\end{equation}
Although each variable in the cointegration system appears as a local-to-unity
process, as a group they are extremely comoving, reducing their rate
of convergence from $n$ to $\sqrt{n}$. This effect is analogous
to the deterioration of the convergence rate of the OLS estimator for
nearly perfectly collinear regressors. \end{remark}

\subsection{Adaptive LASSO}

Similar to Section \ref{sec:alasso_ur}, Alasso for model (\ref{DGP XYZ 2})
is estimated as 
\begin{equation}
\hat{\theta}^{\mathrm{A}}=\arg\min_{\theta}\bigg\{\Vert y-W\theta\Vert^{2}+\lambda_{n}\sum_{j=1}^{p}\hat{\tau}_{j}|\theta_{j}|\bigg\},\label{Alasso mr}
\end{equation}
where $\hat{\tau}_{j}=|\hat{\theta}_{j}^{\mathrm{ols}}|^{-\gamma}$.
The literature on Alasso has established the oracle property in many
models. \citet{caner2013alternative} and \citet{kock2016consistent}
study Alasso's rate adaptiveness in a pure autoregressive setting
with iid error processes. In their cases, the potential nonstationary
regressor is the first-order lagged dependent variable, and the other
regressors are stationary. Therefore, the components of different
convergence rates are known in advance. We complement this line of
nonstationary LASSO literature by allowing a general regression framework
with mixed degrees of persistence. We also generalize the error processes
to the commonly used dependent processes, which is important in practice.
For example, the long-horizon return regression in Section \ref{sec:empirical}
requires this type of dependence in its error structure because
of the overlapping return construction.

Surprisingly, Theorem \ref{thm:ada_3type} shows that in the mixed
root model, Alasso's oracle property holds partially, but not for
all regressors. To discuss variable selection in this context, we
introduce the following notations. We partition the index set of all
regressors $\mathcal{M}=\left\{ 1,\ldots,p\right\} $ into four components:
$\mathcal{I}_{0}$ (I(0) variables associated with $Z$), $\mathcal{C}_{1}$
(associated with $X_{1}^{c}$), $\mathcal{C}_{2}$ (associated with
$X_{2}^{c}$), and $\mathcal{I}_{1}$ (nonstationary variables associated
with $X$). Let $\mathcal{C}=\mathcal{C}_{1}\cup\mathcal{C}_{2}$
and $\mathcal{I}=\mathcal{I}_{0}\cup\mathcal{I}_{1}$. Let $M^{\ast}=\left\{ j:\theta_{j}^{0\ast}\neq0\right\} $
be the true active set for the feasible representation, and let $\widehat{M}^{\mathrm{A}}=\{j:\hat{\theta}_{j}^{\mathrm{A}}\neq0\}$
be Alasso's estimated active set. Next, let $\mathcal{M}_{Q}=\mathcal{I}\cup\mathcal{C}_{1}$
be the set of coordinates that are invariant to the rotation in $Q$.
Similarly, let $M_{Q}^{\ast}=M^{\ast}\cap\mathcal{M}_{Q}$ be the
active set in the infeasible regression equation (\ref{eq:DGP_XYZ_infeasible}),
and let $M_{Q}^{\ast c}=\mathcal{M}\backslash M_{Q}^{\ast}$ be the corresponding
inactive set. The DGP (\ref{DGP XYZ 1}) obviously implies $\mathcal{C}_{2}\cap M_Q^{\ast}=\emptyset$
and $\mathcal{C}_{2}\subseteq M_{Q}^{\ast c}$. For a generic index set
$M\subseteq\mathcal{M}$, let $\mathrm{CoRk}\left(M\right)$ be the
cointegration rank of the variables in $M$.

\begin{theorem} \label{thm:ada_3type} Suppose that the linear model
    (\ref{DGP XYZ}) satisfies Assumption \ref{ass:INNOV coint} and that
    for all of the coefficients in the set $\mathcal{C}_{2}$, we have 
    \begin{equation}
    \phi_{2l}^{0*}\neq0\ \text{ if }\sum_{s=1}^{p_{1}}\left|A_{1sj}\phi_{1s}^{0*}\right|\neq0.\label{eq:no_pathology}
    \end{equation}
    If the tuning parameters $\lambda_{n}$ and $\gamma$ are chosen such
    that 
    \begin{equation}
    \frac{\lambda_{n}}{n^{\left(1-\bar{\delta}\gamma\right)\wedge0.5}}+\frac{1}{\lambda_{n}}\rightarrow0\label{eq:lambda_rate}
    \end{equation}
    and $\bar{\delta}\gamma<1$, then we have the following results:
    \begin{enumerate}
    \item Consistency and asymptotic distribution: 
    \begin{align}
    (R_{n}Q(\hat{\theta}^{\mathrm{A}}-\theta_{n}^{\ast}))_{M_{Q}^{*}} & \Longrightarrow(\Omega_{M_{Q}^{*}}^{+})^{-1}\zeta_{M_{Q}^{*}}^{+}\label{eq:alasso_active_dist}\\
    (R_{n}Q(\hat{\theta}^{\mathrm{A}}-\theta_{n}^{\ast}))_{M_{Q}^{*c}} & \overset{p}{\to}0.\label{eq:alasso_inactive_dist}
    \end{align}
    \item Partial variable selection consistency: 
    \begin{align}
    P\left(M^{*}\cap\mathcal{\mathcal{I}}=\widehat{M}^{\mathrm{A}}\cap\mathcal{\mathcal{I}}\right) & \rightarrow1,\label{eq:sel_I01}\\
    P\left(\left(M^{*}\cap\mathcal{C}\right)\subseteq(\widehat{M}^{\mathrm{A}}\cap\mathcal{C})\right) & \rightarrow1,\label{eq:sel_C}\\
    P\left(\mathrm{CoRk}(M^{*})=\mathrm{CoRk}(\widehat{M}^{\mathrm{A}})\right) & \to1.\label{eq:sel_CC}
    \end{align}
    \end{enumerate}
\end{theorem}

\begin{remark} Condition (\ref{eq:no_pathology}) is an extra assumption
    that rules out the pathological case that some nonzero elements in
    $A_{1s}\phi_{1s}^{0*}$, $s=1,\ldots,p_{1}$, happen to exactly cancel
    out one another and render $\phi_{2l}^{0*}$ inactive. In other
    words, it ensures that if an $x_{l}^{c}$ is involved in more than
    one active cointegration group, it must be active in (\ref{DGP XYZ}).
    This condition holds in general, as it is violated only under very
    specific configurations of $\phi_{1}^{0*}$ and $A_{1}$. For instance,
    in the demonstrative example in Table \ref{tab:Demonstration-of-Cointegration},
    the condition breaks down if $x_{g}^{c}$'s true coefficient $\spadesuit_{ga}\times\bigstar_{a}+\spadesuit_{gb}\times\bigstar_{b}=0$.

\begin{table}[h]
    \caption{Diagram of a cointegrating system in predictors}
    \label{tab:Demonstration-of-Cointegration}\smallskip{}
    
    \begin{centering}
    \begin{tabular}{c|ccccc|cccc|cc}
    \hline 
    $\left(\phi_{1}^{0*},\phi_{2}^{0*}\right)$  & $\bigstar_{a}$  & $\bigstar_{b}$  & 0  & 0  & 0  & $\clubsuit_{f}$  & $\clubsuit_{g}$  & 0  & 0  & coint.  & \tabularnewline
    $\left(\mathcal{C}_{1},\mathcal{C}_{2}\right)$  & $x_{a}^{c}$  & $x_{b}^{c}$  & $x_{c}^{c}$  & $x_{d}^{c}$  & $x_{e}^{c}$  & $x_{f}^{c}$  & $x_{g}^{c}$  & $x_{h}^{c}$  & $x_{q}^{c}$  & resid.  & $\phi_{1}^{0*}$\tabularnewline
    \hline 
    \multirow{5}{*}{$\left(I_{p_{1}},-A_{1}'\right)$} & 1  & 0  & 0  & 0  & 0  & $\spadesuit_{fa}$  & $\spadesuit_{ga}$  & 0  & 0  & $v_{a}$  & $\bigstar_{a}$\tabularnewline
     & 0  & 1  & 0  & 0  & 0  & 0  & $\spadesuit_{gb}$  & 0  & 0  & $v_{b}$  & $\bigstar_{b}$\tabularnewline
     & 0  & 0  & 1  & 0  & 0  & \cellcolor{lightgray}0  & \cellcolor{lightgray}0  & $\spadesuit_{hc}$  & 0  & $v_{c}$  & 0\tabularnewline
     & 0  & 0  & 0  & 1  & 0  & \cellcolor{lightgray}$\spadesuit_{fd}$  & \cellcolor{lightgray}0  & $\spadesuit_{hd}$  & $\spadesuit_{qd}$  & $v_{d}$  & 0\tabularnewline
     & 0  & 0  & 0  & 0  & 1  & \cellcolor{lightgray}0  & \cellcolor{lightgray}$\spadesuit_{ge}$  & 0  & 0  & $v_{e}$  & 0\tabularnewline
    \hline 
    \end{tabular}
    \par\end{centering}
    \medskip{}
    {\footnotesize{}Note: The diagram represents a cointegration system
    of 9 variables $x_{a}^{c},x_{b}^{c},\ldots$ of cointegrating rank
    5. The last column represents the coefficients in $\phi_{1}^{0*}$,
    with $\bigstar$ as a nonzero entry. In the matrix $-A_{1}^{\prime}$,
    $\spadesuit$ is nonzero, and the coefficients in gray cells are irrelevant
    to the value $\phi_{2}^{0*}=-A_{1}^{\prime}\phi_{1}^{0*}$ no matter
    zero or nonzero. The first row displays the coefficients $\phi_{1}^{0*}$
    (the same as in the last column, with $\bigstar$ for non-zeros) and
    $\phi_{2}^{0*}$ ($\clubsuit$ for non-zeros). In this example, $\left(x_{a}^{c},x_{b}^{c},x_{f}^{c},x_{g}^{c}\right)$
    are in the active set $M^{*}$. There are two active cointegration
    groups, $\left(x_{a}^{c},x_{f}^{c},x_{g}^{c}\right)$ and $\left(x_{b}^{c},x_{g}^{c}\right)$,
    and three inactive cointegration groups, $\left(x_{c}^{c},x_{h}^{c}\right)$,
    $\left(x_{d}^{c},x_{f}^{c},x_{g}^{c},x_{q}^{c}\right)$, and $\left(x_{e}^{c},x_{g}^{c}\right)$.
    } 
\end{table}

\end{remark}

Were we informed of the oracle about the true active variables in
$M^{*}$\emph{\ }and the cointegration matrix $A_{1}$, we would
transform the cointegrated variables into cointegrating residuals,
discard the inactive variables, and then run OLS. Ideally, we would
conduct an estimation with variables in $M_{Q}^{*}$ only. Such an oracle
OLS shares the same asymptotic distribution as its Alasso counterpart
in (\ref{eq:alasso_active_dist}). Outside of the active set $M_{Q}^{*}$,
(\ref{eq:alasso_inactive_dist}) shows that all of the other (transformed)
variables in $M_{Q}^{*c}$ consistently converge to zero.

The variable selection results in Theorem \ref{thm:ada_3type}
are novel and interesting. (\ref{eq:sel_I01}) shows that variable selection
is consistent for the pure I(0) and local unit root variables, which
is in line with the well-known oracle property of Alasso. However,
instead of confirming the oracle property, (\ref{eq:sel_C}) indicates
that in the cointegration set $\mathcal{C}$, the selected $\widehat{M}^{\mathrm{A}}$
asymptotically contains the true active variables in $M^{\ast}$ but Alasso
may over-select inactive variables. The inconsistency stems from the mismatch
between the convergence rate of the initial estimator and the marginal
behavior of a cointegrated variable viewed in isolation. Consider
a pair of inactive cointegrating variables, such as $\left(x_{c}^{c},x_{h}^{c}\right)$
in Table \ref{tab:Demonstration-of-Cointegration}. The unknown cointegration
relationship precludes transforming this pair into the cointegrating
residual $v_{c}$. Without the rotation, OLS associated with the pair
can only achieve the $\sqrt{n}$ rate according to (\ref{eq:OLS_marginal}).
The resulting penalty weights are insufficient to remove these variables
that \emph{individually} appear as nonstationary. In consequence,
Alasso fails to eliminate \emph{both} $x_{c}^{c}$ and $x_{h}^{c}$
wpa1. To the best of our knowledge, this is the first case of Alasso's
variable selection inconsistency in an important empirical model.

Under condition (\ref{eq:no_pathology}) all variables
    in the active cointegration groups have nonzero coefficients, and Alasso
    selects them asymptotically according to (\ref{eq:sel_C}). Despite
    potential variable over-selection in $\mathcal{C}$, (\ref{eq:sel_CC})
    brings relief: in the limit, the cointegration rank in Alasso's selected
    set, $\mathrm{CoRk}(\widehat{M}^{\mathrm{A}})$, must equal the active
    cointegration rank $\mathrm{CoRk}(M^{*})$. Note that under our
    agnostic perspective, we do not need to know or use testing procedures
    to determine the value of $\mathrm{CoRk}(M^{*})$.
Let $C^{*c}=M^{*c}\cap\mathcal{C}$ be the index set of the inactive variables
in $\mathcal{C}$. (\ref{eq:sel_C}) implies that the variables in $C^{*c}\cap\widehat{M}^{\mathrm{A}}$---Alasso's mistakenly selected inactive variables---cannot form
cointegration groups. In mathematical expression, 
$P\left(\mathrm{CoRk}\left(C^{*c}\cap\widehat{M}^{\mathrm{A}}\right)=0\right)\to1.$

\begin{example}
Let us again take $\left(x_{c}^{c},x_{h}^{c}\right)$ in Table \ref{tab:Demonstration-of-Cointegration}
as an example. (\ref{eq:sel_CC}) indicates that Alasso is at least
partially effective in that it prevents the inactive $x_{c}^{c}$
and $x_{h}^{c}$ from entering $\widehat{M}^{\mathrm{A}}$ simultaneously, as it kills at least one variable in the pair to break the cointegration
relationship. The intuition is as follows. Suppose that $\left(x_{c}^{c},x_{h}^{c}\right)$
are both selected. Because of coefficient estimation consistency, in
the predictive regression these two variables together behave like
the cointegrating residual $v_{c}$, which is an I(0). 
In the limit Alasso will
not tolerate this inactive $v_{c}$, because these coefficients 
associated with the underlying $x_{c}$ and $x_{h}$
are subject to a penalty
weight $\widehat{\tau}_{j}=1/O_{p}\left(n^{-1/2}\right)$, which is
of the same order as the pure I(0) variables. Recall that Alasso removes
all inactive pure I(0) variables when $n\to\infty$ under the same
level of penalty weight.
\end{example}

\begin{example}
Similar reasoning applies in another inactive cointegration group
$\left(x_{e,}^{c},x_{g}^{c}\right)$. In the regression, $x_{e}^{c}$
is inactive but $x_{g}^{c}$ is active, as it is involved in the active
cointegration groups $(x_{a}^{c},x_{f}^{c},x_{g}^{c})$
and $\left(x_{b}^{c},x_{g}^{c}\right)$. $x_{g}^{c}$ is selected
wpa1, but let us suppose that $x_{e}^{c}$ is also selected. If so, $x_{e}^{c}$'s
contribution to the predictive regression would be equivalent to the
I(0) cointegrating residual $v_{e}$. Its corresponding penalty is
of the order $1/O_{p}\left(n^{-1/2}\right)$, which is sufficient
to remove the inactive $v_{e}$. Therefore, $x_{e}$ cannot survive
Alasso's variable selection.
\end{example}

The intuition gleaned in these examples can be generalized to cointegration
relationships involving more than two variables and multiple cointegrated
groups. 
The proof of Theorem \ref{thm:ada_3type} formalizes this argument by inspecting a linear combination of the corresponding Karush-Kuhn-Tucker (KKT) condition for the selected variables.

In the literature, Alasso achieved the oracle property in a single
implementation, so there was no need to run it twice. In our predictive
regression with mixed roots, it is possible that a single run of Alasso
over-selects inactive variables in the cointegration system. Given
that the cointegrating ties are all shattered wpa1 in (\ref{eq:sel_CC}),
further action can help fulfill the oracle property.

\subsection{Twin Adaptive LASSO}

When the sample size is sufficiently large, with high probability,
the mistakenly selected inactive variables have no cointegration
relationship, so they behave as isolated local-to-unity processes
in the post-selection regression equation of $y$ on $\left(w_{j}\right)_{j\in\widehat{M}^{\mathrm{A}}}$,
where $w_{j}=(w_{1j},\ldots,w_{nj})^{\prime}$ is the $j$-th time series regressors.
This observation suggests the need to run a post-selection
Alasso. We first obtain the post-Alasso initial OLS estimator 
\[
\widehat{\theta}^{\mathrm{po}}=\left(W_{\widehat{M}^{\mathrm{A}}}^{\prime}W_{\widehat{M}^{\mathrm{A}}}\right)^{-1}W_{\widehat{M}^{\mathrm{A}}}^{\prime}y.
\]
The post-selection OLS estimator $\widehat{\theta}_{j}^{\mathrm{po}}=O_{p}\left(n^{-1}\right)$
for the over-selected inactive cointegrated variables $j\in C^{*c}\cap\widehat{M}^{\mathrm{A}}$,
rather than $O_{p}\left(n^{-1/2}\right)$, as for the first-around initial
$\widehat{\theta}_{j}^{\mathrm{ols}}$. The resulting post-selection penalty level
$\hat{\tau}_{j}^{\mathrm{po}}=|\widehat{\theta}_{j}^{\mathrm{po}}|^{-\gamma}$
is sufficiently heavy to wipe out these redundant variables in another round
of Alasso.
We call this procedure TAlasso, for which 
$\hat{\theta}^{\mathrm{TA}}=\left(\hat{\theta}_{\widehat{M}^{A}}^{\mathrm{TA\prime}},\hat{\theta}_{\mathcal{M}\backslash\widehat{M}^{A}}^{\mathrm{TA\prime}}\right)^{\prime}$,
where $\hat{\theta}_{\mathcal{M}\backslash\widehat{M}^{A}}^{\mathrm{TA}}=\boldsymbol{0}$
and 
\begin{equation}
\hat{\theta}_{\widehat{M}^{A}}^{\mathrm{TA}}=\arg\min_{\theta}\bigg\{\Vert y-\sum_{j\in\widehat{M}^{\mathrm{A}}}w_{j}\theta_{j}\Vert^{2}+\lambda_{n}\sum_{j\in\widehat{M}^{\mathrm{A}}}\hat{\tau}_{j}^{\mathrm{po}}|\theta_{j}|\bigg\}.
\label{eq:alasso_twin}
\end{equation}
TAlasso asymptotically reclaims variable selection
consistency for all types of variables. Let $\widehat{M}^{\mathrm{TA}}$
be the active set for TAlasso.

\begin{theorem} \label{thm:alasso_twin} Under the same assumptions
and the same rate for $\lambda_{n}$ as in Theorem \ref{thm:ada_3type},
the estimator $\hat{\theta}^{\mathrm{TA}}$ satisfies
\begin{enumerate}
\item Asymptotic distribution: $(R_{n}Q(\hat{\theta}^{\mathrm{TA}}-\theta_{n}^{\ast}))_{M_Q^{*}}\Longrightarrow(\Omega_{M_Q^{*}}^{+})^{-1}\zeta_{M_Q^{*}}^{+}$;
\item Variable selection consistency: $P(\widehat{M}^{\mathrm{TA}}=M^{*})\rightarrow1.$ 
\end{enumerate}
\end{theorem}

Faced with a variety of potential predictors with unknown orders of
integration, we may not be able to sort them into different persistence
categories in predictive regressions without potential testing errors
\citep{smeekes2020unit}. Our research provides a valuable practical
algorithm to address this problem. TAlasso is the first estimator that achieves the desirable
oracle property without requiring prior knowledge on the persistence
of multivariate regressors. Despite the intensive study of LASSO 
in recent years, Theorems \ref{thm:ada_3type} and \ref{thm:alasso_twin}
are novel and important. With the cointegration system in the predictors,
the former shows that Alasso does not automatically adapt to the behavior
of a \emph{system} of regressors. Nevertheless, it at least breaks
all redundant cointegration groups so that its defects can be easily addressed
by another round of Alasso. This solution echoes the repeated implementation
of a machine learning procedure as in \citet{Phillips2019boosting}.

\section{Conventional LASSO with Mixed Roots\label{sec:Conventional-LASSO-with}}

Originally, LASSO was proposed as a plain $l_{1}$-penalized regression
without a sophisticated weighting scheme, motivated by the optimization
problem's variable screening effect: some coefficients are estimated
as exactly zeros in finite samples under a wide range of the tuning
parameter \citep{tibshirani1996regression}. As mentioned in the introduction,
in this paper, we call this original estimator
\begin{equation}
\hat{\theta}^{\mathrm{P}}=\arg\min_{\theta}\left\{ \Vert y-W\theta\Vert^{2}+\lambda_{n}\left\Vert \theta\right\Vert _{1}\right\} \label{lasso mr}
\end{equation}
Plasso. 
However, Plasso is scale-variant in the sense that if we change the
unit of $w_{j}$ by multiplying it by a nonzero constant $c$, such
a change is not reflected in the penalty term in (\ref{lasso mr}),
so the Plasso estimator does not change proportionally to $\widehat{\theta}_{j}^{\mathrm{P}}/c$.
To keep the estimation scale-invariant to the choice of the 
unit of $x_{j}$, researchers often scale-standardize LASSO as 
\begin{equation}
\hat{\theta}^{\mathrm{S}}=\arg\min_{\theta}\bigg\{\Vert y-W\theta\Vert^{2}+\lambda_{n}\sum_{j=1}^{p}\widehat{\sigma}_{j}\left\vert \theta_{j}\right\vert \bigg\}.\label{Slasso mr}
\end{equation}
where $\widehat{\sigma}_{j}=\sqrt{n^{-1}\sum_{i=1}^{n}\left(w_{ij}-\bar{w}_{j}\right)^{2}}$
is the sample standard deviation of $w_{j}$ and $\bar{w}_{j}$ is
the sample average. In this paper, we call (\ref{Slasso mr}) Slasso.
Such standardization is the default option for LASSO in many statistical
packages, such as the \texttt{R} package \texttt{glmnet}. Note
that Alasso is also scale-invariant with $\widehat{\theta}^{\mathrm{init}}=\widehat{\theta}^{\mathrm{ols}}$
and $\gamma=1$.

Both Plasso and Slasso are special cases of $\min_{\theta}\left\{ \Vert y-X\theta\Vert_{2}^{2}+\lambda_{n}\sum_{j=1}^{p}\hat{\tau}_{j}|\theta_{j}|\right\} $,
where the former uses $\hat{\tau}_{j}=1$ and the latter specifies
$\hat{\tau}_{j}=\hat{\sigma}_{j}.$ When Slasso's scale standardization
is carried out with stationary and weakly dependent regressors, each
$\widehat{\sigma}_{j}^{2}$ converges in probability to its finite
population variance. In the literature, the difference between the
asymptotic properties of Plasso and Slasso are not of interest because
Slasso uses another set of constant configurations for the
penalty levels.

In sharp contrast, for those $w_{j}$ associated with $\mathcal{C}\cup\mathcal{I}_{1}$
that are individually nonstationary, by (\ref{FCLT}), we have 
\begin{equation}
\frac{\widehat{\sigma}_{j}}{\sqrt{n}}=\sqrt{\frac{1}{n^{2}}\sum_{i=1}^{n}\left(w_{ij}-\bar{w}_{j}\right)^{2}}\Longrightarrow d_{j},\label{eq:dj}
\end{equation}
where $d_{j}$ is a non-degenerate random variable (whose expression
can be found in the proof of Corollary \ref{thm:std_lasso_3type}),
so that the sample standard deviation $\widehat{\sigma}_{j}=O_{p}\left(\sqrt{n}\right)$
diverges in the limit. As a result, it imposes a much heavier penalty
on the associated coefficients than the stationary time series does. 

Adopting a standard argument for LASSO as in \citet{knight2000asymptotics}, 
we derive the following results characterizing
Plasso's asymptotic behavior with various choices of $\lambda_{n}$.
For exposition, we define a function $D:\mathbb{R}^{3}\mapsto\mathbb{R}$
as $D\left(s,v,b\right)=s\left[v\cdot\mathrm{sgn}(b)I(b\neq0)+|v|I(b=0)\right]$,
where $\mathrm{sgn}(b)=I(b>0) -I(b<0) $
is the sign function.

\begin{corollary} \label{thm:lasso-mixture} Suppose that the linear model
(\ref{DGP XYZ}) satisfies Assumption \ref{ass:INNOV coint}.
\begin{enumerate}
\item If $\lambda_{n}\rightarrow\infty$ and $\lambda_{n}/\sqrt{n}\rightarrow0$,
then $R_{n}Q(\hat{\theta}^{\mathrm{P}}-\theta_{n}^{\ast})\Longrightarrow\left(\Omega^{+}\right)^{-1}\zeta^{+}.$
\item If $\lambda_{n}/\sqrt{n}\rightarrow c_{\lambda}\in(0,\infty)$, then
\begin{eqnarray*}
R_{n}Q(\hat{\theta}^{\mathrm{P}}-\theta_{n}^{\ast}) & \Longrightarrow & \arg\min_{v}\bigg\{ v^{\prime}\Omega^{+}v-2v^{\prime}\zeta^{+}+c_{\lambda}\sum_{j\in\mathcal{I}_{0}\cup\mathcal{C}}D\left(1,v_{j},\theta_{j}^{0*}\right)\bigg\}.
\end{eqnarray*}
\item If $\lambda_{n}/\sqrt{n}\rightarrow\infty$ and $\lambda_{n}/n\rightarrow0$,
then 
\begin{eqnarray*}
\frac{1}{\lambda_{n}}R_{n}Q(\hat{\theta}^{\mathrm{P}}-\theta_{n}^{\ast}) & \Longrightarrow & \arg\min_{v}\bigg\{ v^{\prime}\Omega^{+}v+\sum_{j\in\mathcal{I}_{0}\cup\mathcal{C}}D\left(1,v_{j},\theta_{j}^{0*}\right)\bigg\}.
\end{eqnarray*}
\end{enumerate}
\end{corollary}

In Corollary \ref{thm:lasso-mixture}(a), the tuning parameter is
 small and the limit distribution of Plasso is equivalent to that
of OLS, so there is no variable screening effect. Similar to Plasso in
the cross-sectional setting, there is no guarantee of consistent
variable selection. 
The screening effect occurs when the tuning parameter $\lambda_{n}$ becomes larger. In view of the OLS
rate of convergence in (\ref{eq:OLS_marginal}), we can call those
$\theta_{j}$ associated with $\mathcal{I}_{0}\cup\mathcal{C}$ the
\emph{slow coefficients} (at rate $\sqrt{n}$) and those associated
with $\mathcal{I}_{1}$ the \emph{fast coefficients} (at rate $n$).
When $\lambda_{n}$ is raised to the magnitude in (b), the term $D(1,v_{j},\theta_{j}^{0*})$
causes variable screening among the slow coefficients but not the
fast coefficients. If we further increase $\lambda_{n}$ to the level
of (c), then the convergence rate of the slow coefficients is dragged
down by the large penalty, but there is still no variable screening
effect for the fast coefficients. The local-to-zero coefficient $\delta_{l}$
for those in $\mathcal{I}_{1}$ plays no role in these asymptotic
results because $\widehat{\beta}_{l}$ converges to $\beta_{ln}^{*}$
at the fast rate $n$ regardless of $\delta_{l}$.

Corollary \ref{thm:lasso-mixture} reveals a major drawback of Plasso
in the mixed root model. Because it has one uniform penalty level for
all variables, it is not adaptive to these various types of predictors.
To induce variable screening in $\hat{\theta}_{\mathcal{I}_{1}}^{\mathrm{P}}$,
the tuning parameter must be ballooned to $\lambda/n\to c_{\lambda}\in(0,\infty]$,
but the consistency of the slow coefficients would collapse under
such a disproportionately heavy $\lambda_{n}$. 

\begin{remark} When nonstationary regressors are present, the results
in \citet[Section 2]{zou2006adaptive} are no longer applicable. Consider
the simple case $W=X$ such that all of the regressors are non-cointegrating local
unit roots. Recall that the random matrix $\Omega$, defined in the
line following (\ref{eq:ols_pure}), is the weak limit of $n^{-2}X'X$.
Corollary \ref{thm:lasso-mixture}(c) implies that in this case, the
limiting distribution of $\arg\min_{v}\left\{ v^{\prime}\Omega v+\sum_{j=1}^{p}D(1,v_{j},\theta_{j}^{0\ast})\right\} $
is non-degenerate because of the randomness of $\Omega$. This differs
from Lemma 3 of \citet{zou2006adaptive}, in which the Plasso estimator
degenerates to a constant. The distinction arises because in
our context $\Omega$ is random, whereas in Zou's (2006) iid setting
the counterpart of $\Omega$ degenerates to a non-random matrix. \end{remark}

We now turn to Slasso. Unlike the behavior of the slow coefficients
in Plasso, the penalty scheme of Slasso makes the cointegration components
in $\hat{\theta}^{\mathrm{S}}$ susceptible to variable screening
at much smaller scales of the tuning parameter. 

\begin{corollary} \label{thm:std_lasso_3type} Suppose that the linear
model (\ref{DGP XYZ}) satisfies Assumption \ref{ass:INNOV coint}.
\begin{enumerate}
\item If $\lambda_{n}\rightarrow0$, then $R_{n}Q(\hat{\theta}^{\mathrm{S}}-\theta_{n}^{\ast})\Longrightarrow\left(\Omega^{+}\right)^{-1}\zeta^{+}.$
\item If $\lambda_{n}\to c_{\lambda}\in(0,\infty)$, then 
\begin{eqnarray*}
R_{n}Q(\hat{\theta}^{\mathrm{S}}-\theta_{n}^{\ast}) & \Longrightarrow & \arg\min_{v}\bigg\{ v^{\prime}\Omega^{+}v-2v^{\prime}\zeta^{+}+c_{\lambda}\sum_{j\in\mathcal{C}}D\left(d_{j},v_{j},\theta_{j}^{0*}\right)\bigg\}.
\end{eqnarray*}
\item When $\lambda_{n}\rightarrow\infty$ and $\lambda_{n}/\sqrt{n}\rightarrow0$,
then 
\begin{eqnarray*}
\frac{1}{\lambda_{n}}R_{n}Q(\hat{\theta}^{\mathrm{S}}-\theta_{n}^{\ast}) & \Longrightarrow & \arg\min_{v}\bigg\{ v^{\prime}\Omega^{+}v+\sum_{j\in\mathcal{C}}D\left(d_{j},v_{j},\theta_{j}^{0*}\right)\bigg\}.
\end{eqnarray*}
\end{enumerate}
\end{corollary}

\begin{remark} The tuning parameter $\lambda_{n}$ in Corollary \ref{thm:std_lasso_3type}
is an $\sqrt{n}$ order smaller than that in Corollary \ref{thm:lasso-mixture}.
Part(a) produces the same asymptotic distribution as OLS. The distinction
of Plasso and Slasso arises from the coefficients in the set $\mathcal{C}$.
Their corresponding penalty terms have the multipliers $\widehat{\sigma}_{j}=O_{p}\left(\sqrt{n}\right)$
rather than the desirable $O_{p}\left(1\right)$ that is suitable for
their slow convergence rate under OLS. In other words, the penalty
level is overly heavy for these parameters. The overwhelming penalty
level causes a variable screening effect in (b) as soon as $\lambda_{n}\rightarrow c_{\lambda}\in\left(0,\infty\right)$.
While the first argument of $D(\cdot,v_{j},\theta_{j}^{0\ast})$ in
Plasso is $1$, it is replaced in Slasso by the random variable $d_{j}$,
which introduces an extra source of uncertainty in variable screening.
Moreover, (c) implies that for the consistency of $\widehat{\phi}^{\mathrm{S}}$,
the tuning parameter $\lambda_{n}$ must be small enough in the sense
$\lambda_{n}/\sqrt{n}\rightarrow0$; otherwise, they will be inconsistent.
In both (b) and (c) the penalty term $D(d_{j},v_{j},\theta_{j}^{0*})$
screens those variables in $\mathcal{C}$ only. Again, the local-to-zero
coefficient $\delta_{l}$ for those in $\mathcal{I}_{1}$ are irrelevant
asymptotically. \end{remark}

To summarize this section, neither Plasso nor Slasso induces consistent
parameter estimation and variable screening effect simultaneously
for all components of the mixed regressors. The weight $\hat{\tau}_{j}$
is a constant for Plasso, while it exploits only the marginal variation
of $w_{j}$ for Slasso. Plasso uses a single tuning parameter and does
not adapt to the different orders of magnitude of the slow and fast
coefficients. Slasso suffers from overwhelming penalties for 
coefficients associated with the cointegration groups. Because the cointegrated
regressors are individually local unit root processes, they can only form a linear combination to produce a stationary time series when classified
into a system. In contrast,
the penalty of Alasso/TAlasso reflects the cointegration system
via the initial/post-selection OLS.

\section{Simulations\label{sec:Monte-Carlo-Simulation}}

In this section, we examine via simulations the performance of the
LASSO methods in forecasting and variable screening. We consider
different sample sizes to demonstrate the quality of the asymptotic approximation 
in finite samples. Comparison is based on the one-period-ahead
out-of-sample forecast.

\subsection{Simulation Design}

Following the settings in Sections \ref{sec:UR}, \ref{sec:MRC}, and
\ref{sec:Conventional-LASSO-with}, we consider four DGPs. Each DGP
starts after a burn-in of $1000$ periods for the innovations to mimic stationary $O_{p}\left(1\right)$
realizations.

\bigskip{}

\textbf{DGP 1 (Pure unit roots)}. This DGP corresponds to the pure
unit root model in Section 2. Consider a linear model with 9 unit
root predictors, the same number as the persistent regressions in
Section \ref{sec:empirical}. Each regressor $x_{ij}$
is drawn from the random walk $x_{ij}=x_{i-1,j}+e_{ij}$, and the dependent
variable $y_{i}$ is generated from $y_{i}=\gamma^{\ast}+x_{i\cdot}\beta_{n}^{\ast}+u_{i}$,
where the innovations $\left(e_{i\cdot},u_{i}\right)\sim iid\;N\left(0,\Sigma\right)$.
The intercept $\gamma^{\ast}=0.25$, the slope coefficient $\beta_{n}^{\ast}=\left(1,1,1,0,0,0,0,0,0\right)^{\prime}/\sqrt{n}$,
and the covariance matrix $\Sigma$ is estimated from the data of \citet{welch2008comprehensive}, as detailed in Supplement S1.

\smallskip{}

\textbf{DGP 2 (Mixed roots and cointegration)}. This DGP is designed
for the mixed root model in Section 3, which emulates the kitchen-sink
approach in Section \ref{sec:empirical}. The dependent variable
\begin{equation}
y_{i}=\gamma^{\ast}+\sum_{l=1}^{3} \alpha_{l}^{\ast} z_{il}
+  \sum_{l=1}^{4} \phi_{l}^{\ast} x_{il}^{c}
+  \sum_{l=1}^{5} \beta_{ln}^{\ast} x_{il}
+u_{i},\label{eq:DGP2}
\end{equation}
 where $\gamma^{\ast}=0.3$, $\alpha^{\ast}=\left(0.4,0,0\right)$,
$\phi^{\ast}=\left(0.5,-0.5,0,0\right)$, and $\beta_{n}^{\ast}=\left(1/\sqrt{n},1/\sqrt{n},0,0,0\right)$.
The vector $\xi_{i}=( z_{i\cdot}, v_{i\cdot}, e_{i\cdot},u_{i} )$
follows a VAR(1) process $\xi_{i}=\Phi\xi_{i-1}+\varepsilon_{i}$,
where $\varepsilon_{i}\sim iid\;N\left(0,\Sigma_{\varepsilon}\right)$
in which $\Phi$ and $\Sigma_{\varepsilon}$ are empirically estimated
using the data of \citet{welch2008comprehensive}, as detailed in Supplement
S1. $x_{i\cdot}^{c}\in\mathbb{R}^{4}$
is a vector I(1) process with cointegration rank $2$ based on the
VECM, $\Delta x_{i\cdot}^{c}=\Gamma^{\prime}\Lambda x_{i-1,\cdot}^{c}+v_{i\cdot},$
where $\Lambda=\begin{pmatrix}1 & -1 & 0 & 0\\
0 & 0 & 1 & -1
\end{pmatrix}$ and $\Gamma=\begin{pmatrix}0 & 1 & 0 & 0\\
0 & 0 & 0 & 1
\end{pmatrix}$ are the cointegrating matrix and the loading matrix, respectively.
$\left(x_{il}\right)_{l=1}^{5}$ are non-cointegrating random walks
generated by $x_{il}=x_{i-1,l}+e_{il}$.

\smallskip{}

\textbf{DGP 3 (Mixed roots with local-to-unity predictors)} The coefficients and the innovation processes are the same as those
in DGP 2, but we replace the I(1) processes in DGP 2 with local-to-unity
processes. $x_{i\cdot}^{c}$ is a vector of a local-to-unity process
with cointegration rank $2$ based on the triangular representation
\begin{align*}
\begin{pmatrix}1 & -1 & 0 & 0\\
0 & 0 & 1 & -1
\end{pmatrix}x_{i\cdot}^{c} & =\begin{pmatrix}v_{i2}\\
v_{i4}
\end{pmatrix}\\
\left(I_{2}-\text{diag}\left(\left(1-\frac{4.8386}{n},1\right)\right)L\right)\begin{pmatrix}x_{i1}^{c}\\
x_{i3}^{c}
\end{pmatrix} & =\begin{pmatrix}v_{i1}\\
v_{i3}
\end{pmatrix}.
\end{align*}
$\left(x_{il}\right)_{l=1}^{5}$ are non-cointegrating local-to-unity
processes generated by $x_{il}=\left(1- \check{c}_{l}/n\right)x_{i-1,l}+e_{il}$,
$l=1,2,\cdots,5$, and $( \check{c}_{1},\check{c}_{2},\cdots,\check{c}_{5})=\left(7.4934,\,5.9152,\,8.9277,\,0,\,0\right)$,
where these local-unity coefficients are also estimated following
\citet{xu2018testing}. With some local-to-unity parameters being
zero, DGP 3 incorporates a mixture of I(0), I(1), and local-to-unity
processes.

\smallskip{}

\textbf{DGP 4 (Stationary autoregression)}. The autoregressive distributed
lag (ARDL) model is a classical specification for time series regressions.
In addition to including lags of $y_{i}$, it is common to accommodate
lags of predictors in predictive regressions, such as in \citet{medeiros2016L1}.
The stationary dependent variable in the following equation is generated
from the ARDL model
\[
y_{i}=\gamma^{\ast}+\rho^{\ast}y{}_{i-1}+\sum_{l=1}^{4}\left(\phi_{l1}^{\ast}x_{il}^{c}+\phi_{l2}^{\ast}x_{i-1,l}^{c}\right)+\sum_{l=1}^{5}\left(\beta_{l1n}^{\ast}x_{il}+\beta_{l2n}^{\ast}x_{i-1,l}\right)+\sum_{l=1}^{3}\left(\alpha_{l1}^{\ast}z_{il}+\alpha_{l2}^{\ast}z_{i-1,l}\right)+u_{i},
\]
where $\gamma^{\ast}=0.3$, $\rho^{\ast}=0.4$, $\phi_{1}^{\ast}=\left(0.75,-0.75,0,0\right)$,
$\phi_{2}^{\ast}=\left(0,0,0,0\right)$, $\beta_{1n}^{\ast}=\left(1.5/\sqrt{n},1,0,1/\sqrt{n},0\right)$,
$\beta_{2n}^{\ast}=\left(0,-1,0,0,0\right)$, $\alpha_{1}^{\ast}=\left(0.6,0.4\right)$,
$\alpha_{2}^{\ast}=\left(0.8,0\right)$, and $\alpha_{3}^{\ast}=\left(0,0\right)$.
The predictors $x_{i}^{c}$, $x_{i}$, and $z_{i}$ are generated in the same way as in DGP 3. 

\begin{remark}

Our theory allows the local-to-zero coefficients for $\mathcal{I}_{1}$
to be $\beta_{ln}^{*}=\beta_{l}^{0*}/n^{\delta_{l}}$ for any $\delta_{l}\in\left(0,1\right)$.
In the simulations, we fix $\delta_{l}=0.5$ for all $l\in\mathcal{I}_{1}$,
which is the knife-edge rate for the balance of the predictive regression.
When $\beta_{l}^{0*}\neq0$, an active pure local
unit root variable contributes to the dependent variable with 
\[
\beta_{ln}^{*} x_{\left\lfloor nr\right\rfloor ,l}=\beta_{l}^{0*}\left(x_{\left\lfloor nr\right\rfloor ,l}/\sqrt{n}\right)\Longrightarrow\beta_{l}^{0*}J_{c_{xl}}\left(r\right)=O_{p}\left(1\right),
\]
for $r\in\left[0,1\right]$, where $J_{c_{xl}}\left(r\right)$ is
an Ornstein--Uhlenbeck process. The limit process is of the same stochastic
order of the active stationary variables in $Z^{+}$, and so is $y$.
In contrast, when 
$\delta_{l}\in\left(0.5,1\right)$, 
we have $\beta_{ln}^{*} x_{\left\lfloor nr\right\rfloor ,l} =\left(\beta_{l}^{0*}/n^{\delta_{l}-0.5}\right)\left(x_{\left\lfloor nr\right\rfloor ,l}/\sqrt{n}\right)=o_{p}\left(1\right)$,
and it is asymptotically dominated by the active stationary variables
and the innovation $u_{i}$. Alasso/TAlasso would require a
very large sample size to detect such weak signal in reality. When
$\delta_{l}\in\left(0,0.5\right)$, the active stationary variables
in $Z^{+}$ are dominated by $\beta_{ln}^{*}x_{\left\lfloor nr\right\rfloor ,l}=O_{p}\left(n^{0.5-\delta_{l}}\right)$.
In this case, the stochastic order of $y$ is determined by the active
variable in $\mathcal{I}_{1}$ of the largest order, and therefore
$y_{n}=O_{p}\left(n^{0.5-\text{\ensuremath{\underbar{\ensuremath{\delta}}}}}\right)\to\infty$,
where $\underbar{\ensuremath{\delta}}=\min\{(\delta_{l})_{l\in\mathcal{I}_{1}}\}$.
This violates the consensus that stock returns in most predictive regressions
are stochastically bounded. Thus, we only consider $\delta_{l}=0.5$
in the simulations.

\end{remark}

\smallskip{}

As we develop our theory with regressors of fixed dimension, OLS is
a natural benchmark. Another benchmark is the oracle OLS estimator
under infeasible information. The sample sizes in our exercise are
$n=80$, $120,$ $200$, $400$, and $800$. We run $10,000$ replications
for each sample size and each DGP.

Each shrinkage estimator relies on its tuning parameter $\lambda_{n}$,
which is the appropriate rate multiplied by a constant $c_{\lambda}$.
We use fivefold cross validation (CV), where the sample is temporally
ordered and then partitioned into five consecutive blocks, to guide
the choice of $c_{\lambda}$. To ensure that the tuning parameter
changes according to the rate specified in the asymptotic theory,
we set $n=100$ and run an exploratory simulation 100 times for
each method that requires a tuning parameter. In each replication,
we use the fivefold CV to obtain $c_{\lambda}^{\left(1\right)},\ldots,c_{\lambda}^{\left(100\right)}$.
We then fix $c_{\lambda}=\mathrm{median}(c_{\lambda}^{\left(1\right)},\ldots,c_{\lambda}^{\left(100\right)})$
in the full-scale 10,000 replications. To specify the tuning parameters for other sample sizes,
we multiply the constant $c_{\lambda}$ that we have calibrated from
$n=100$ by the rates suggested by our asymptotic theory. We multiply $c_{\lambda}$ by $\sqrt{n}$ for Plasso and Slasso, and
by $\sqrt{n}/(\log \log n)$ for Alasso and TAlasso.

\subsection{Performance Comparison}

\begin{table}
    \caption{\label{tab:DGP1234: MPSE}MPSE and variable screening in simulations}
    \small %
    \noindent\begin{minipage}[t]{1\columnwidth}%
    \begin{center}
    \subcaption{MPSE (Relative to OLS)}%
    \begin{tabular}{>{\centering}p{1.5cm}>{\centering}p{1.35cm}>{\centering}p{1.35cm}>{\centering}p{1.35cm}>{\centering}p{1.35cm}>{\centering}p{1.35cm}>{\centering}p{1.35cm}>{\centering}p{1.35cm}}
    \hline 
        & $n$ & Oracle & OLS & Alas. & TAlas. & Plas. & Slas.\tabularnewline
    \hline 
    DGP 1 & $80$ & 0.8614 & 1.0000 & 0.9256 & 0.9450 & \textbf{0.8945} & 0.8953\tabularnewline
        & $120$ & 0.9090 & 1.0000 & 0.9622 & 0.9657 & \textbf{0.9347} & 0.9390\tabularnewline
        & $200$ & 0.9425 & 1.0000 & 0.9900 & 0.9896 & \textbf{0.9604} & 0.9652\tabularnewline
        & $400$ & 0.9699 & 1.0000 & 0.9976 & 0.9905 & \textbf{0.9861} & 0.9862\tabularnewline
        & $800$ & 0.9841 & 1.0000 & 0.9977 & \textbf{0.9912} & 0.9935 & 0.9931\tabularnewline
    \hline 
    DGP 2 & $80$ & 0.8537 & 1.0000 & 0.9351 & \textbf{0.9295} & 0.9467 & 0.9651\tabularnewline
        & $120$ & 0.8955 & 1.0000 & 0.9514 & \textbf{0.9447} & 0.9663 & 0.9831\tabularnewline
        & $200$ & 0.9344 & 1.0000 & 0.9649 & \textbf{0.9615} & 0.9775 & 0.9985\tabularnewline
        & $400$ & 0.9717 & 1.0000 & 0.9900 & \textbf{0.9854} & 0.9952 & 1.0145\tabularnewline
        & $800$ & 0.9875 & 1.0000 & 0.9932 & \textbf{0.9906} & 0.9986 & 1.0238\tabularnewline
    \hline 
    DGP 3 & $80$ & 0.8452 & 1.0000 & 0.9247 & \textbf{0.9183} & 0.9457 & 0.9519\tabularnewline
        & $120$ & 0.9046 & 1.0000 & 0.9562 & \textbf{0.9521} & 0.9746 & 0.9820\tabularnewline
        & $200$ & 0.9373 & 1.0000 & 0.9705 & \textbf{0.9678} & 0.9826 & 0.9951\tabularnewline
        & $400$ & 0.9701 & 1.0000 & 0.9887 & \textbf{0.9831} & 0.9954 & 1.0097\tabularnewline
        & $800$ & 0.9874 & 1.0000 & 0.9927 & \textbf{0.9904} & 0.9990 & 1.0205\tabularnewline
    \hline 
    DGP 4 & $80$ & 0.7188 & 1.0000 & 0.8896 & \textbf{0.8766} & 0.9237 & 0.9229\tabularnewline
        & $120$ & 0.8223 & 1.0000 & 0.9244 & \textbf{0.9129} & 0.9590 & 0.9765\tabularnewline
        & $200$ & 0.8914 & 1.0000 & 0.9497 & \textbf{0.9423} & 0.9818 & 1.0225\tabularnewline
        & $400$ & 0.9513 & 1.0000 & 0.9727 & \textbf{0.9686} & 0.9921 & 1.0431\tabularnewline
        & $800$ & 0.9796 & 1.0000 & 0.9897 & \textbf{0.9866} & 0.9997 & 1.0593\tabularnewline
    \hline 
    \end{tabular}
    \par\end{center}%
    \end{minipage}
    
    \medskip{}
    
    \centering{}%
    \noindent\begin{minipage}[t]{1\columnwidth}%
    \begin{center}
    \subcaption{Variable screening success rates}%
    \begin{tabular*}{1\textwidth}{@{\extracolsep{\fill}}@{\extracolsep{\fill}}cc|rrrr|rrrr|rrrr}
    \hline 
    \noalign{\vskip0.2cm}
    \multicolumn{2}{l|}{} & \multicolumn{4}{c|}{$SR$} & \multicolumn{4}{c|}{$SR_{1}$} & \multicolumn{4}{c}{$SR_{2}$}\tabularnewline
        & $n$ & Alas. & TAlas. & Plas. & Slas. & Alas. & TAlas. & Plas. & Slas. & Alas. & TAlas. & Plas. & Slas.\tabularnewline
    \hline 
    \multirow{5}{*}{\begin{turn}{90}
    DGP 1
    \end{turn}} & $80$ & 0.729 & \textbf{0.733} & 0.697 & 0.714 & 0.613 & 0.577 & 0.790 & \textbf{0.808} & 0.787 & \textbf{0.812} & 0.651 & 0.667\tabularnewline
        & $120$ & 0.764 & \textbf{0.769} & 0.688 & 0.731 & 0.674 & 0.644 & 0.862 & \textbf{0.863} & 0.809 & \textbf{0.831} & 0.602 & 0.664\tabularnewline
        & $200$ & 0.815 & \textbf{0.822} & 0.667 & 0.747 & 0.762 & 0.741 & \textbf{0.928} & 0.926 & 0.841 & \textbf{0.862} & 0.536 & 0.658\tabularnewline
        & $400$ & 0.879 & \textbf{0.887} & 0.623 & 0.763 & 0.865 & 0.856 & \textbf{0.976} & 0.974 & 0.886 & \textbf{0.903} & 0.446 & 0.658\tabularnewline
        & $800$ & 0.936 & \textbf{0.945} & 0.566 & 0.770 & 0.944 & 0.942 & \textbf{0.995} & 0.994 & 0.932 & \textbf{0.947} & 0.352 & 0.658\tabularnewline
    \hline 
    \multirow{5}{*}{\begin{turn}{90}
    DGP 2
    \end{turn}} & $80$ & 0.779 & \textbf{0.804} & 0.643 & 0.572 & 0.853 & 0.840 & 0.934 & \textbf{0.943} & 0.726 & \textbf{0.779} & 0.435 & 0.308\tabularnewline
        & $120$ & 0.820 & \textbf{0.846} & 0.634 & 0.584 & 0.897 & 0.889 & 0.967 & \textbf{0.968} & 0.765 & \textbf{0.816} & 0.397 & 0.309\tabularnewline
        & $200$ & 0.861 & \textbf{0.890} & 0.617 & 0.593 & 0.939 & 0.934 & \textbf{0.988} & 0.986 & 0.805 & \textbf{0.858} & 0.351 & 0.313\tabularnewline
        & $400$ & 0.905 & \textbf{0.936} & 0.593 & 0.601 & 0.977 & 0.976 & \textbf{0.998} & 0.997 & 0.853 & \textbf{0.908} & 0.303 & 0.318\tabularnewline
        & $800$ & 0.937 & \textbf{0.970} & 0.576 & 0.606 & 0.996 & 0.995 & \textbf{1.000} & \textbf{1.000} & 0.896 & \textbf{0.952} & 0.273 & 0.324\tabularnewline
    \hline 
    \multirow{5}{*}{\begin{turn}{90}
    DGP 3
    \end{turn}} & $80$ & 0.773 & \textbf{0.796} & 0.642 & 0.601 & 0.827 & 0.815 & 0.903 & \textbf{0.923} & 0.735 & \textbf{0.783} & 0.455 & 0.371\tabularnewline
        & $120$ & 0.812 & \textbf{0.836} & 0.637 & 0.614 & 0.872 & 0.864 & 0.947 & \textbf{0.952} & 0.768 & \textbf{0.816} & 0.416 & 0.373\tabularnewline
        & $200$ & 0.854 & \textbf{0.880} & 0.624 & 0.622 & 0.918 & 0.913 & \textbf{0.978} & 0.974 & 0.808 & \textbf{0.856} & 0.372 & 0.370\tabularnewline
        & $400$ & 0.901 & \textbf{0.931} & 0.605 & 0.633 & 0.967 & 0.965 & \textbf{0.996} & 0.994 & 0.854 & \textbf{0.907} & 0.326 & 0.375\tabularnewline
        & $800$ & 0.935 & \textbf{0.967} & 0.588 & 0.632 & 0.992 & 0.992 & \textbf{1.000} & 0.999 & 0.894 & \textbf{0.949} & 0.294 & 0.369\tabularnewline
    \hline 
    \multirow{5}{*}{\begin{turn}{90}
    DGP 4
    \end{turn}} & $80$ & 0.746 & \textbf{0.803} & 0.587 & 0.599 & 0.867 & 0.845 & \textbf{0.925} & 0.907 & 0.666 & \textbf{0.775} & 0.362 & 0.395\tabularnewline
        & $120$ & 0.782 & \textbf{0.845} & 0.587 & 0.616 & 0.903 & 0.890 & \textbf{0.950} & 0.924 & 0.701 & \textbf{0.815} & 0.344 & 0.411\tabularnewline
        & $200$ & 0.811 & \textbf{0.878} & 0.582 & 0.633 & 0.921 & 0.913 & \textbf{0.960} & 0.932 & 0.738 & \textbf{0.855} & 0.330 & 0.433\tabularnewline
        & $400$ & 0.836 & \textbf{0.908} & 0.576 & 0.648 & 0.933 & 0.930 & \textbf{0.966} & 0.927 & 0.772 & \textbf{0.892} & 0.316 & 0.463\tabularnewline
        & $800$ & 0.854 & \textbf{0.927} & 0.565 & 0.655 & 0.937 & 0.936 & \textbf{0.969} & 0.912 & 0.799 & \textbf{0.922} & 0.296 & 0.484\tabularnewline
    \hline 
    \end{tabular*}
    \par\end{center}
    {\footnotesize Note: Panel (a) compares
    the out-of-sample prediction accuracy in terms of MPSE $E\big[(y_{n}-\hat{y}_{n})^{2}\big]$,
    where that of OLS is normalized to be 1. Panel (b) compares the variable
    screening performance in terms of the success rates $SR$, $SR_{1}$, and
    $SR_{2}$ defined in \eqref{eq:SR1_SR2} and \eqref{eq:SR}.
    Bold numbers indicate the best performance in each category of measurement.
    }{\footnotesize\par}%
    \end{minipage}
\end{table}

Table \ref{tab:DGP1234: MPSE}(a) reports the out-of-sample prediction
accuracy in terms of the mean prediction squared error (MPSE) $E\big[(y_{n}-\hat{y}_{n})^{2}\big]$.
By the simulation design, unpredictable variation arises from
the variance of the idiosyncratic error $u_{i}$, which is 1. 

Plasso and Slasso achieve variable screening and consistent estimation,
as the predictors are unit root processes in DGP 1. They are slightly
better in forecasting than Alasso/TAlasso when the sample size is
small; however, as the sample size increases to $n=800$, TAlasso surpasses
the conventional LASSO estimators, which suggests that variable selection
is conducive to forecasting in a large sample. In DGPs 2--4
of mixed roots and cointegrated regressors, the settings are more
difficult for the conventional LASSO to navigate. TAlasso is the
best performer, followed by Alasso, which also outperforms the conventional
LASSO methods by a non-trivial margin.

Table \ref{tab:DGP1234: MPSE}(b) summarizes the variable screening
performance. Recall that $M^{\ast}=\{j:\theta_{j}^{\ast}\neq0\}$ is
the set of relevant regressors and $\widehat{M}=\{j:\hat{\theta}_{j}\neq0\}$
is the estimated active set. We define two \emph{success rates} for variable
screening: 
\begin{equation}
    SR_{\text{1}}=\left|M^{*}\right|^{-1}
    E\left[\left|   M^{\ast}\cap\widehat{M} \right|\right],\quad 
    SR_{\text{2}}=  \left|M^{*c}\right|^{-1} E\left[\left|  M^{\ast c}\cap\widehat{M}^{c} \right|\right].\label{eq:SR1_SR2}
\end{equation}
Here, $SR_{1}$ is the percentage of correct selections from the active
set, and $SR_{2}$ is the percentage of correct removals from the inactive set.
 We also report the overall success rate of classification
into zero or nonzero coefficients, defined as
\begin{equation}
    SR=p^{-1}E\left[\left|\{j:I(\theta_{j}^{*}=0)=I(\hat{\theta}_{j}=0)\right|\right].\label{eq:SR}
\end{equation}
 These expectations in $SR$, $SR_{1}$, and $SR_{2}$ are computed by
the average across the simulation replications.

In terms of the overall selection measure $SR$, TAlasso is the most
effective and Alasso takes the second place. As the sample size increases,
TAlasso's success rates approach 100\% in all DGPs, which supports
variable selection consistency. While the difference in $SR_{1}$
among these methods becomes negligible when the sample size is large,
the gain of TAlasso and Alasso stems largely from $SR_{2}$. The asymptotic
theory suggests that $\lambda_{n}\asymp\sqrt{n}$ is too small for
Plasso to eliminate $0$ coefficients corresponding to the nonstationary regressors.
Plasso and Slasso achieve high $SR_{1}$ at the cost of low $SR_{2}$.
In view of the results in Table \ref{tab:DGP1234: MPSE}, the advantage
of TAlasso's variable screening capability helps with forecast accuracy
in the context of predictive regressions where many included regressors
actually exhibit no predictive power.\footnote{In Table \ref{tab:DGP1234: MPSE}(b) Slasso has lower $SR_{2}$ than
Plasso in a few instances. However, due to the presence
of $\hat{\tau}_{j}=\hat{\sigma}_{j}=O_{p}\left(\sqrt{n}\right)$ in
the penalty term, in asymptotics it imposes a heavier penalty on coefficients
of I(1) regressors than Plasso does. The reason is that in our simulations
we fix $c_{\lambda}^{\mathrm{P}}$ (for Plasso) and $c_{\lambda}^{\mathrm{S}}$
(for Slasso) by CV separately. CV selects the tuning parameter $c_{\lambda}$
favoring lower MPSE and adjusts $c_{\lambda}$ in finite sample. For
example, in DGP 2, $c_{\lambda}^{\mathrm{P}}=0.6576$ is much larger than  $c_{\lambda}^{\mathrm{S}}=0.1676$.
If we fixed
$c_{\lambda}^{\mathrm{P}}$ by CV and set $c_{\lambda}^{\mathrm{S}}=c_{\lambda}^{\mathrm{P}}$,
Slasso would have a much higher $SR_{2}$.}

\begin{table}[h]
    \caption{\label{tab:coint_coef_analysis} Variable screening in DGPs 2--4
    for the inactive cointegrated group $\phi_{3}^{0*}=\phi_{4}^{0*}=0$}
    
    \small\medskip{}
    
    \begin{centering}
    \begin{tabular*}{1\textwidth}{@{\extracolsep{\fill}}ll|rrrr|rrrr|rrrr}
        \hline 
        \noalign{\vskip0.2cm}
        \multicolumn{2}{l|}{} & \multicolumn{4}{c|}{Both $\hat{\phi}_{3},\hat{\phi}_{4}=0$} & \multicolumn{4}{c|}{One and only one of $\hat{\phi}_{3},\hat{\phi}_{4}=0$} & \multicolumn{4}{c}{Neither $\hat{\phi}_{3},\hat{\phi}_{4}$$=0$}\tabularnewline[0.2cm]
            & $n$ & Alas. & TAlas. & Plas. & Slas. & Alas. & TAlas. & Plas. & Slas. & Alas. & TAlas. & Plas. & Slas.\tabularnewline
        \hline 
        \multirow{5}{*}{\begin{turn}{90}
        DGP 2
        \end{turn}} & $80$ & 0.443 & \textbf{0.597} & 0.182 & 0.181 & 0.480 & 0.344 & 0.572 & 0.541 & 0.077 & 0.059 & 0.246 & 0.277\tabularnewline
            & $120$ & 0.485 & \textbf{0.665} & 0.150 & 0.194 & 0.459 & 0.292 & 0.578 & 0.566 & 0.055 & 0.043 & 0.272 & 0.240\tabularnewline
            & $200$ & 0.529 & \textbf{0.738} & 0.112 & 0.214 & 0.436 & 0.234 & 0.566 & 0.581 & 0.036 & 0.028 & 0.322 & 0.205\tabularnewline
            & $400$ & 0.557 & \textbf{0.827} & 0.070 & 0.232 & 0.425 & 0.160 & 0.561 & 0.636 & 0.018 & 0.014 & 0.369 & 0.132\tabularnewline
            & $800$ & 0.603 & \textbf{0.907} & 0.050 & 0.273 & 0.392 & 0.090 & 0.551 & 0.645 & 0.006 & 0.004 & 0.399 & 0.082\tabularnewline
        \hline 
        \multirow{5}{*}{\begin{turn}{90}
        DGP 3
        \end{turn}} & $80$ & 0.424 & \textbf{0.596} & 0.144 & 0.241 & 0.491 & 0.338 & 0.624 & 0.604 & 0.085 & 0.066 & 0.232 & 0.156\tabularnewline
            & $120$ & 0.461 & \textbf{0.663} & 0.120 & 0.258 & 0.483 & 0.294 & 0.622 & 0.619 & 0.057 & 0.043 & 0.258 & 0.123\tabularnewline
            & $200$ & 0.499 & \textbf{0.734} & 0.089 & 0.262 & 0.467 & 0.239 & 0.602 & 0.637 & 0.034 & 0.028 & 0.309 & 0.102\tabularnewline
            & $400$ & 0.532 & \textbf{0.823} & 0.059 & 0.287 & 0.451 & 0.163 & 0.606 & 0.649 & 0.017 & 0.014 & 0.335 & 0.064\tabularnewline
            & $800$ & 0.568 & \textbf{0.907} & 0.042 & 0.314 & 0.427 & 0.089 & 0.591 & 0.634 & 0.005 & 0.004 & 0.368 & 0.052\tabularnewline
        \hline 
        \multirow{5}{*}{\begin{turn}{90}
        DGP 4
        \end{turn}} & $80$ & 0.326 & \textbf{0.561} & 0.156 & 0.343 & 0.526 & 0.366 & 0.510 & 0.510 & 0.149 & 0.073 & 0.334 & 0.148\tabularnewline
            & $120$ & 0.381 & \textbf{0.658} & 0.151 & 0.390 & 0.514 & 0.305 & 0.503 & 0.495 & 0.105 & 0.037 & 0.347 & 0.115\tabularnewline
            & $200$ & 0.418 & \textbf{0.740} & 0.134 & 0.409 & 0.506 & 0.240 & 0.496 & 0.508 & 0.076 & 0.020 & 0.371 & 0.083\tabularnewline
            & $400$ & 0.465 & \textbf{0.838} & 0.117 & 0.475 & 0.490 & 0.155 & 0.483 & 0.473 & 0.045 & 0.007 & 0.400 & 0.051\tabularnewline
            & $800$ & 0.499 & \textbf{0.903} & 0.099 & 0.509 & 0.476 & 0.096 & 0.457 & 0.459 & 0.025 & 0.002 & 0.444 & 0.032\tabularnewline
        \hline 
    \end{tabular*}
    
    \par\end{centering}
    \medskip{}
    
    \footnotesize Note: The numbers report the proportion of occurrences, over the
    replications,  of eliminating both inactive variables (the first column,
    desirable outcome), eliminating only one variable (the second
    column), and eliminating neither variable (the third column).
\end{table}
    
Finally, as shown in Table \ref{tab:coint_coef_analysis}, we check the variable screening of the inactive cointegration group $\left(x_{i3}^{c},x_{i4}^{c}\right)$
in DGPs 2--4, where the true coefficients $\phi_{3}^{0*}=\phi_{4}^{0*}=0$.
Alasso is effective in preventing both redundant variables from remaining
in the regression according to the third column, while in the second
column the chance of omitting one variable is around 40-50\%, indicating that it breaks the cointegration pair but cannot
screen out both inactive regressors simultaneously. In contrast, TAlasso successfully
identifies both redundant variables with significantly higher probabilities,
as in the first column. Plasso and Slasso break down in variable
selection consistency, as our theory suggests.

\section{Empirical Application\label{sec:empirical}}

\subsection{Data\label{subsec:short-horizon-application}}

\begin{figure}
\begin{centering}
\includegraphics[scale=0.75]{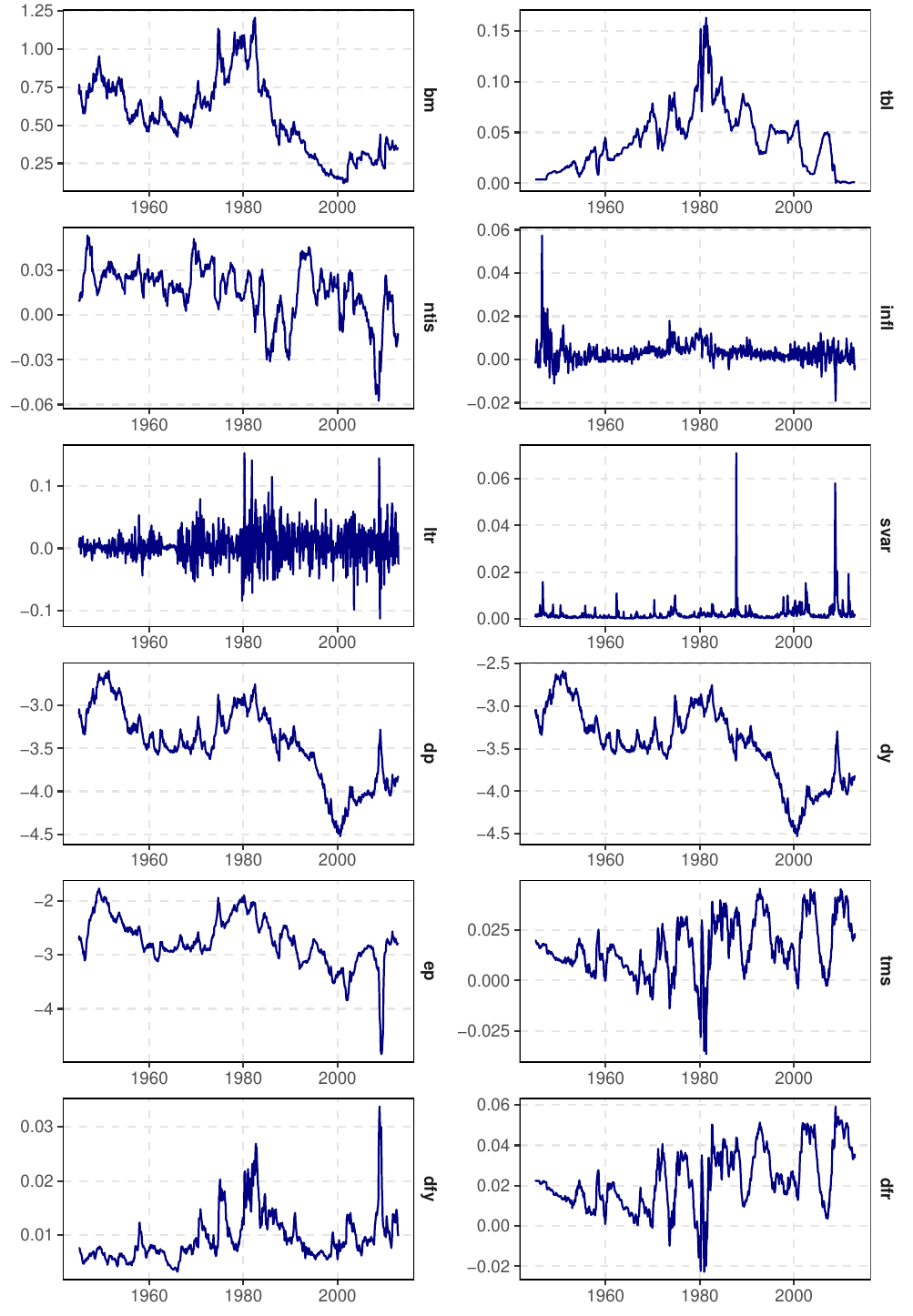}
\par\end{centering}

{\footnotesize Note: The persistent regressors
include the dividend price ratio (\texttt{dp}), dividend yield (\texttt{dy}),
earning price ratio (\texttt{ep}), term spread (\texttt{tms}), default
yield spread (\texttt{dfy}), default return spread (\texttt{dfr}),
book-to-market ratio (\texttt{bm}), net equity expansion (\texttt{ntis}),
and treasury bill rates (\texttt{tbl}), whose estimated AR(1) coefficients are greater than $0.95$. The stationary predictors are the long-term return
of government bonds (\texttt{ltr}), stock variance (\texttt{svar}),
and inflation (\texttt{infl}).}

\caption{\label{fig:Welch-Goyal-data} Time plots of the 12 predictors from the dataset of Welch and Goyal (2008)}
\end{figure}

We apply the LASSO methods to \citet{welch2008comprehensive}'s dataset
to predict stock returns. We focus on the improvement in terms of
prediction error and variable screening.

The dataset of \citet{welch2008comprehensive} is one of the most widely
used in predictive regressions. \citet*{koo2016high} update this
monthly data from January 1945 to December 2012, and we use the same
time span. The dependent variable is \emph{excess return} (ExReturn),
defined as the difference between the continuously compounded return
on the S\&P 500 index and the three-month Treasury bill rate. The
estimated AR(1) coefficient of the excess return is $0.149$, indicating
weak persistence. The 12 financial and macroeconomic predictors are
introduced and depicted in Figure \ref{fig:Welch-Goyal-data}. Three variables,
namely \texttt{ltr}, \texttt{infl}, and \texttt{svar}, oscillate around
the mean, and nine variables are highly persistent with estimated AR(1) coefficients
greater than $0.95$. The pairs (\texttt{tms}, \texttt{dfr}) and
(\texttt{dp}, \texttt{dy}) are visibly moving in a synchronized pattern
that suggests potential cointegration, and it is difficult to determine
whether cointegration holds among (\texttt{dp}, \texttt{dy}, \texttt{ep}).
\texttt{ep} fluctuates with \texttt{dy} before 2000, but the link dissolves
after 2000, and the two series diverge in opposite directions
during the Great Recession. The presence of stationary 
and persistent predictors fits the mixed roots environment studied in this
paper, and our agnostic approach avoids decision errors resulting from hypothesis
testing.

As recognized in the literature, the signal of persistent predictors
may become stronger in long-horizon return predictions \citep{cochrane2009asset}.
In addition to the one-month-ahead short-horizon forecast, we construct
the long-horizon excess return as the sum of the continuous compounded
monthly excess return on the S\&P 500 index
\[
\text{LongReturn}_{i}=\sum_{k=i}^{i+12h-1}\text{ExReturn}_{k},
\]
where $h$ is the length of the forecasting horizon, and $h=1$ indicates one year. We use $h=1/12$ (one month), 1/4 (three months),
1/2 (half a year), 1, 2, and 3 in our empirical exercises.

\subsection{Performance}

\begin{figure}[ph]
\begin{centering}
\includegraphics[scale=0.85]{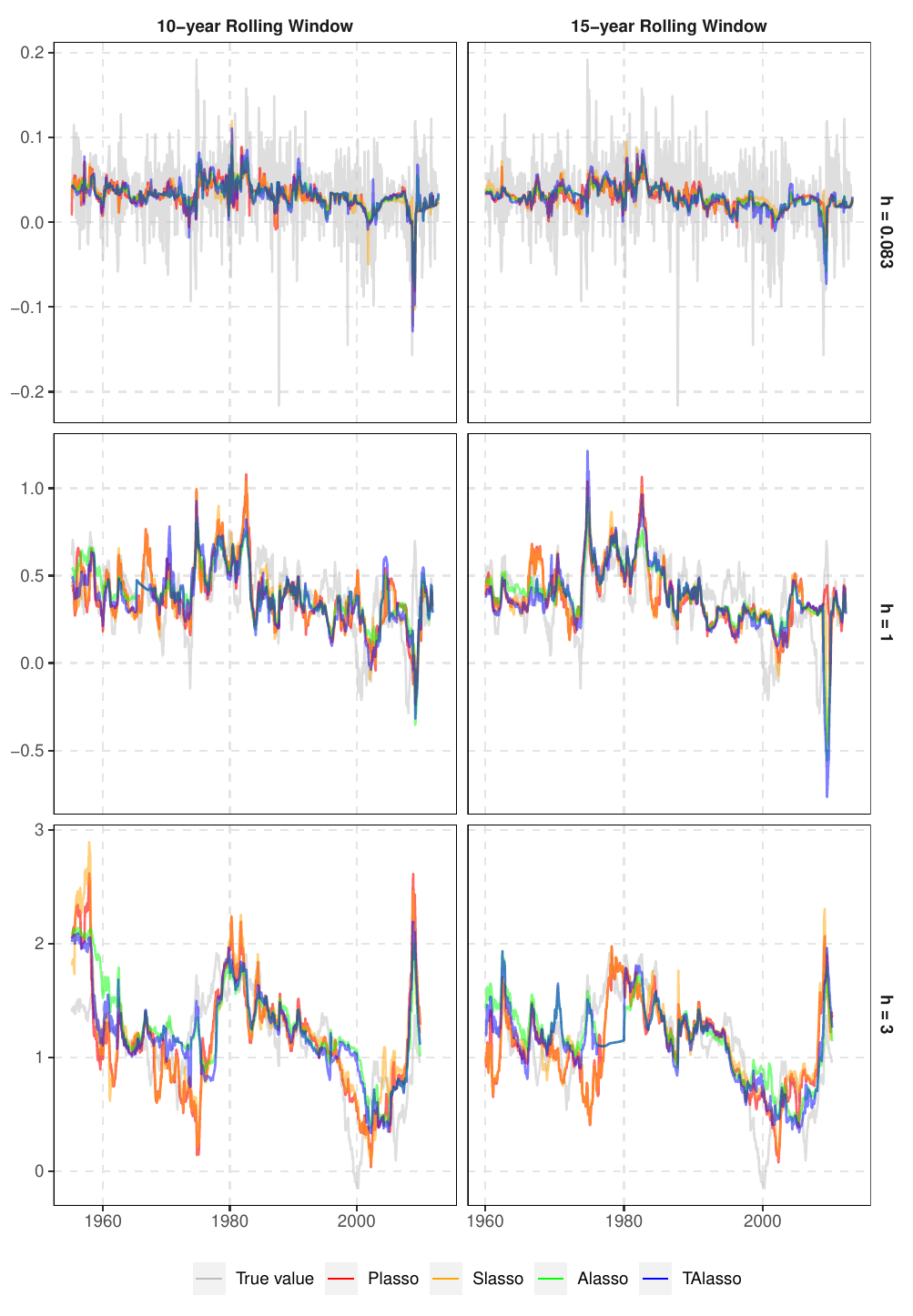}
\par\end{centering}
\medskip{}

\caption{Realized return versus predicted returns. $h$ is selected as $1/12$,
$1$, and 3 to present the short, medium, and long horizons,
respectively.}

\label{fig:koo_trend}
\end{figure}

We forecast short- and long-horizon stock returns recursively
with either a $10$-year or $15$-year rolling window. All 12 variables
are made available in the predictive regression, which \citet{welch2008comprehensive}
call the \emph{kitchen sink model}. The tuning parameters for
the shrinkage estimators are determined by a 10-fold CV on MPSE with
consecutive partitions in each estimation window. The CV method is
a preferred choice for prediction purposes. As a robustness check,
we also use the Bayesian information criterion (BIC) to decide the
tuning parameters for Alasso and TAlasso.\footnote{BIC is a popular choice geared toward variable selection, but in our context, it is incompatible
with the Plasso/Slasso, which cannot cope with variable
screening and consistent estimation simultaneously.}

The forecast returns of the LASSO methods are shown in Figure \ref{fig:koo_trend}
along with the true realized return in gray for $h=1/12=0.083$
(short horizon), $1$ (median horizon), and $3$ (long horizon). When
$h=0.083$, the realized excess return resembles white noise that
is extremely difficult to forecast. When the horizon is extended to
$h=3$, the dynamics of the long-run aggregated return become clearer.
For most of the time, Alasso and TAlasso track the realized return closely.

\begin{table}
    \caption{\label{tab:welch-goyal-mpae-rmpse} RMPSE and MPAE for predicting
    S\&P 500 excess return}
    
    \begin{centering}
    \medskip{}
    \begin{tabular}{ccrrrrrrrr}
        \hline 
            &  & \multicolumn{1}{c}{OLS} & \multicolumn{1}{c}{RWwD} & \multicolumn{1}{c}{Plas.} & \multicolumn{1}{c}{Slas.} & \multicolumn{1}{c}{Alas.} & \multicolumn{1}{c}{TAlas.} & \multicolumn{1}{c}{Alas.} & \multicolumn{1}{c}{TAlas.}\tabularnewline
        \multicolumn{2}{c}{tuning para.} & \multicolumn{1}{c}{NA} & \multicolumn{1}{c}{NA} & \multicolumn{1}{c}{CV} & \multicolumn{1}{c}{CV} & \multicolumn{1}{c}{CV} & \multicolumn{1}{c}{CV} & \multicolumn{1}{c}{BIC} & \multicolumn{1}{c}{BIC}\tabularnewline
        \hline 
            & $h$ & \multicolumn{8}{c}{RMPSE$\times100$}\tabularnewline
        \begin{turn}{90}
        \end{turn} & {\small{}1/12} & \textit{\small{}1.000} & {\small{}0.950} & {\small{}0.944} & {\small{}0.947} & \textbf{\small{}0.932} & {\small{}0.952} & {\small{}0.949} & {\small{}0.950}\tabularnewline
        10-year & {\small{}1/4} & \textit{\small{}1.000} & {\small{}0.842} & {\small{}0.945} & {\small{}0.900} & \textbf{\small{}0.806} & {\small{}0.821} & {\small{}0.835} & {\small{}0.839}\tabularnewline
        rolling & {\small{}1/2} & \textit{\small{}1.000} & {\small{}0.946} & {\small{}0.937} & {\small{}0.914} & \textbf{\small{}0.862} & {\small{}0.883} & {\small{}0.919} & {\small{}0.916}\tabularnewline
        window & {\small{}$1$} & {\small{}1.000} & \textit{\small{}1.123} & {\small{}0.953} & {\small{}0.933} & {\small{}0.958} & {\small{}0.947} & {\small{}0.943} & \textbf{\small{}0.912}\tabularnewline
            & {\small{}$2$} & {\small{}1.000} & \textit{\small{}1.297} & {\small{}0.962} & {\small{}0.904} & {\small{}0.898} & {\small{}0.905} & {\small{}0.893} & \textbf{\small{}0.865}\tabularnewline
            & {\small{}$3$} & {\small{}1.000} & \textit{\small{}1.173} & {\small{}0.892} & {\small{}0.941} & {\small{}0.854} & {\small{}0.806} & {\small{}0.855} & \textbf{\small{}0.774}\tabularnewline
            &  &  &  &  &  &  &  &  & \tabularnewline
        \multicolumn{1}{c}{\begin{turn}{90}
        \end{turn}} & {\small{}1/12} & \textit{\small{}1.000} & {\small{}0.982} & {\small{}0.956} & {\small{}0.960} & \textbf{\small{}0.952} & {\small{}0.963} & {\small{}0.980} & {\small{}0.984}\tabularnewline
        15-year & {\small{}1/4} & \textit{\small{}1.000} & {\small{}0.915} & {\small{}0.891} & {\small{}0.889} & \textbf{\small{}0.871} & {\small{}0.874} & {\small{}0.921} & {\small{}0.924}\tabularnewline
        rolling & {\small{}1/2} & \textit{\small{}1.000} & {\small{}0.924} & {\small{}0.958} & {\small{}0.919} & \textbf{\small{}0.882} & {\small{}0.917} & {\small{}0.903} & {\small{}0.918}\tabularnewline
        window & {\small{}$1$} & {\small{}1.000} & \textit{\small{}1.055} & \textbf{\small{}0.859} & {\small{}0.961} & {\small{}0.917} & {\small{}1.029} & {\small{}0.934} & {\small{}0.962}\tabularnewline
            & {\small{}$2$} & {\small{}1.000} & \textit{\small{}1.518} & {\small{}0.940} & {\small{}0.952} & {\small{}0.860} & \textbf{\small{}0.818} & {\small{}0.865} & {\small{}0.820}\tabularnewline
            & {\small{}$3$} & {\small{}1.000} & \textit{\small{}1.617} & {\small{}0.974} & {\small{}1.002} & {\small{}0.953} & {\small{}0.890} & {\small{}0.843} & \textbf{\small{}0.824}\tabularnewline
        \hline 
            & $h$ & \multicolumn{8}{c}{MPAE$\times100$}\tabularnewline
        \begin{turn}{90}
        \end{turn} & {\small{}1/12} & \textit{\small{}1.000} & {\small{}0.944} & {\small{}0.948} & {\small{}0.938} & \textbf{\small{}0.928} & {\small{}0.951} & {\small{}0.942} & {\small{}0.947}\tabularnewline
        10-year & {\small{}1/4} & \textit{\small{}1.000} & {\small{}0.925} & {\small{}0.933} & {\small{}0.907} & \textbf{\small{}0.862} & {\small{}0.876} & {\small{}0.903} & {\small{}0.920}\tabularnewline
        rolling & {\small{}1/2} & \textit{\small{}1.000} & {\small{}0.955} & {\small{}0.922} & {\small{}0.909} & \textbf{\small{}0.851} & {\small{}0.863} & {\small{}0.915} & {\small{}0.917}\tabularnewline
        window & {\small{}$1$} & {\small{}1.000} & \textit{\small{}1.140} & {\small{}0.957} & {\small{}0.944} & {\small{}0.929} & {\small{}0.939} & {\small{}0.925} & \textbf{\small{}0.895}\tabularnewline
            & {\small{}$2$} & {\small{}1.000} & \textit{\small{}1.356} & {\small{}0.903} & {\small{}0.886} & {\small{}0.824} & {\small{}0.834} & {\small{}0.831} & \textbf{\small{}0.807}\tabularnewline
            & {\small{}$3$} & {\small{}1.000} & \textit{\small{}1.391} & {\small{}0.940} & {\small{}0.960} & {\small{}0.912} & {\small{}0.836} & {\small{}0.896} & \textbf{\small{}0.814}\tabularnewline
            &  &  &  &  &  &  &  &  & \tabularnewline
        \begin{turn}{90}
        \end{turn} & {\small{}1/12} & \textit{\small{}1.000} & {\small{}0.979} & {\small{}0.971} & \textbf{\small{}0.969} & {\small{}0.970} & {\small{}0.984} & {\small{}0.979} & {\small{}0.989}\tabularnewline
        15-year & {\small{}1/4} & \textit{\small{}1.000} & {\small{}0.932} & {\small{}0.905} & {\small{}0.906} & \textbf{\small{}0.875} & {\small{}0.884} & {\small{}0.926} & {\small{}0.925}\tabularnewline
        rolling & {\small{}1/2} & \textit{\small{}1.000} & {\small{}0.983} & {\small{}0.962} & {\small{}0.942} & \textbf{\small{}0.893} & {\small{}0.934} & {\small{}0.903} & {\small{}0.928}\tabularnewline
        window & {\small{}$1$} & {\small{}1.000} & \textit{\small{}1.107} & {\small{}0.921} & {\small{}0.970} & \textbf{\small{}0.853} & {\small{}0.933} & {\small{}0.879} & {\small{}0.878}\tabularnewline
            & {\small{}$2$} & {\small{}1.000} & \textit{\small{}1.705} & {\small{}0.916} & {\small{}0.893} & {\small{}0.812} & {\small{}0.805} & {\small{}0.804} & \textbf{\small{}0.777}\tabularnewline
            & {\small{}$3$} & {\small{}1.000} & \textit{\small{}1.907} & {\small{}1.010} & {\small{}1.015} & {\small{}1.040} & {\small{}0.966} & {\small{}0.897} & \textbf{\small{}0.882}\tabularnewline
        \hline 
    \end{tabular}
    \par\end{centering}
    \begin{centering}
    \bigskip{}
    \par\end{centering}
    \footnotesize Note: The upper panel compares the
    root mean prediction squared error (RMPSE), defined as $\sqrt{E\left[\left(y_{n}-\hat{y}_{n}\right)^{2}\right]}$,
    and the lower panel compares the mean prediction absolute error $E\left[\left|y_{n}-\hat{y}_{n}\right|\right]$.
    The RMPSE and MPAE of OLS are normalized to be 1. Bold numbers
    indicate the best performance in each row, and italic numbers
    indicate the worst.
\end{table}

Table \ref{tab:welch-goyal-mpae-rmpse} quantifies the forecast error
in terms of the out-of-sample RMPSE (root MPSE) and mean predicted
absolute error (MPAE) $E\left[\left|y_{n}-\hat{y}_{n}\right|\right]$.
In addition to OLS, which includes all of the variables without any screening,
we include random walk with drift (RWwD), i.e., the historical average
of the excess returns, $\hat{y}_{n}=(n-1)^{-1}\sum_{i=1}^{n-1}y_{i}$,
as another benchmark that utilizes no information from regressors. The results show that OLS loses in the short horizon, whereas
RWwD suffers in the long horizon, indicating the ineffectiveness of 
the all-in and all-out approaches. Variable screening is essential for
balanced performance in this empirical example. 
Alasso/TAlasso forecasts are more precise than Plasso/Slasso in general.
When the horizon is $h=2$ or $h=3$, TAlasso can achieve
the smallest RMPSE, and Alasso is stronger than Plasso and
Slasso by a substantial margin. The results are robust when the tuning
parameters are chosen by either CV or BIC.

There is an exceptional case of $h=1$ with a 15-year rolling window.
In this particular instance, Alasso fails to foresee the recovery trend after the financial crisis
in 2008, and another round of Alasso further worsens TAlasso. As shown
in Figure \ref{fig:koo_trend}, when $h=1$, Plasso's forecast coincides with the movement of the realization during the recovery
period after 2008, whereas Alasso/TAlasso moves in the opposite direction.
Whereas large deviation exacerbates RMPSE, under MPAE, the gap between Plasso and Alasso/TAlasso in this case is narrowed or even reversed.
Under the 10-year rolling window, all of the methods encounter difficulty
around the financial crisis, and the difference between TAlasso and
Slasso is negligible. Thus, we view the unsatisfactory RMPSE of Alasso/TAlasso
here as an adverse case under the specific rolling window.

\begin{figure}[htbp]
\centering{}\medskip{}
\includegraphics[width=1\textwidth,height=0.8\textheight]{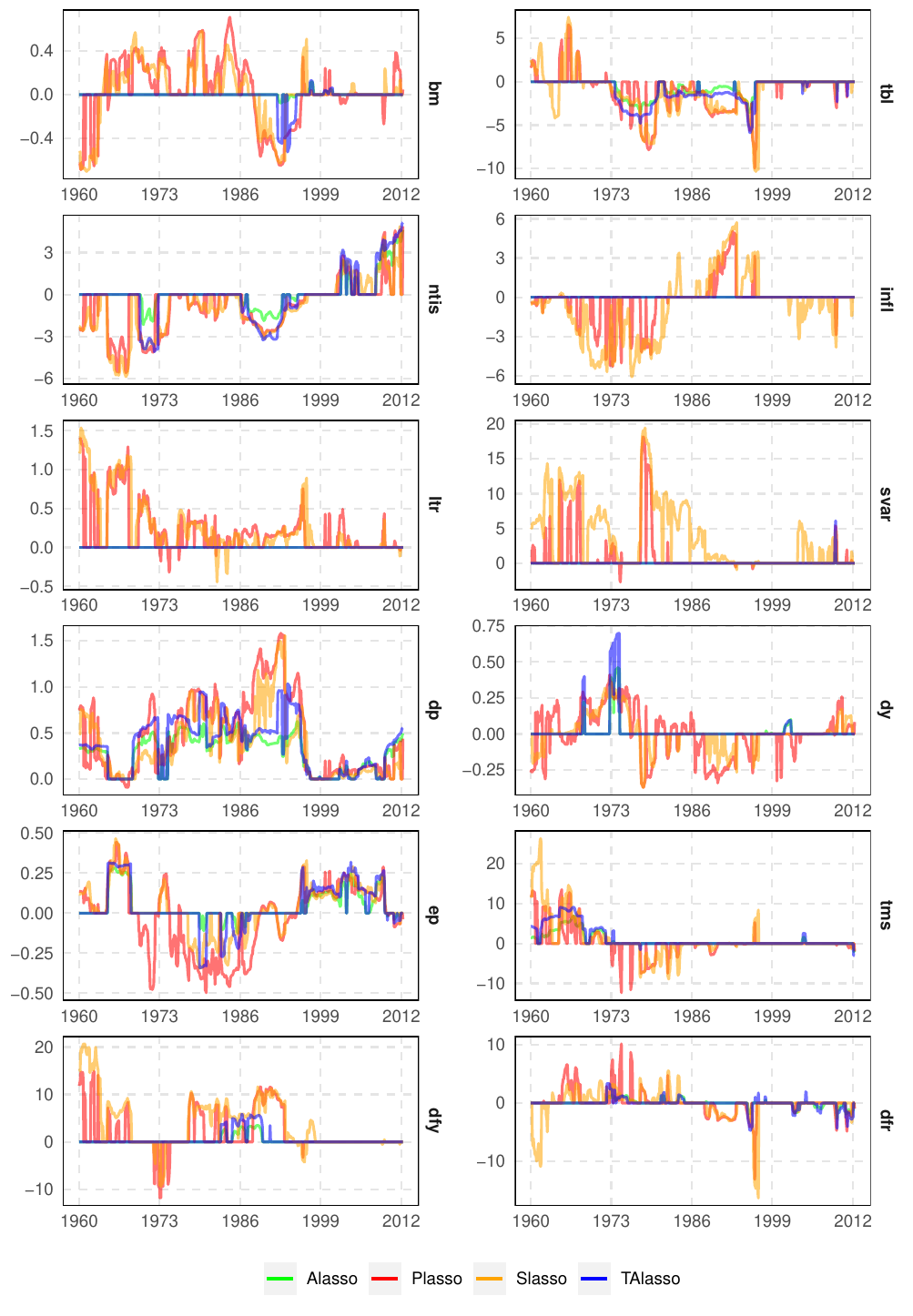}
\caption{\label{fig:welch-Goyal_CV_coef} Estimated coefficients for all 12
variables (15-year rolling window, $h=1/2$)}
\end{figure}

\begin{figure}[htbp]
    \begin{centering}
    \begin{tabular}{>{\raggedright}p{1.8cm}>{\centering}p{1.8cm}>{\centering}p{1.8cm}>{\centering}p{1.8cm}>{\centering}p{1.8cm}}
    \multicolumn{5}{c}{Fraction of active Alasso estimates eliminated by TAlasso}\tabularnewline
    \hline 
     & \multicolumn{1}{c}{\texttt{dp}} & \multicolumn{1}{c}{\texttt{dy}} & \multicolumn{1}{c}{\texttt{dfr}} & \multicolumn{1}{c}{\texttt{tms}}\tabularnewline
    \hline 
    \multirow{1}{1.8cm}{CV} & 0.018 & 0.250 & 0.263 & 0.014\tabularnewline
    \multirow{1}{1.8cm}{BIC} & 0.016 & 0.460 & 0.236 & 0.308\tabularnewline
    \hline 
    \end{tabular}
    \par\end{centering}
    \bigskip{}
    
    \begin{centering}
    \begin{minipage}[t]{0.5\textwidth}%
    \begin{center}
    \includegraphics[width=1\textwidth]{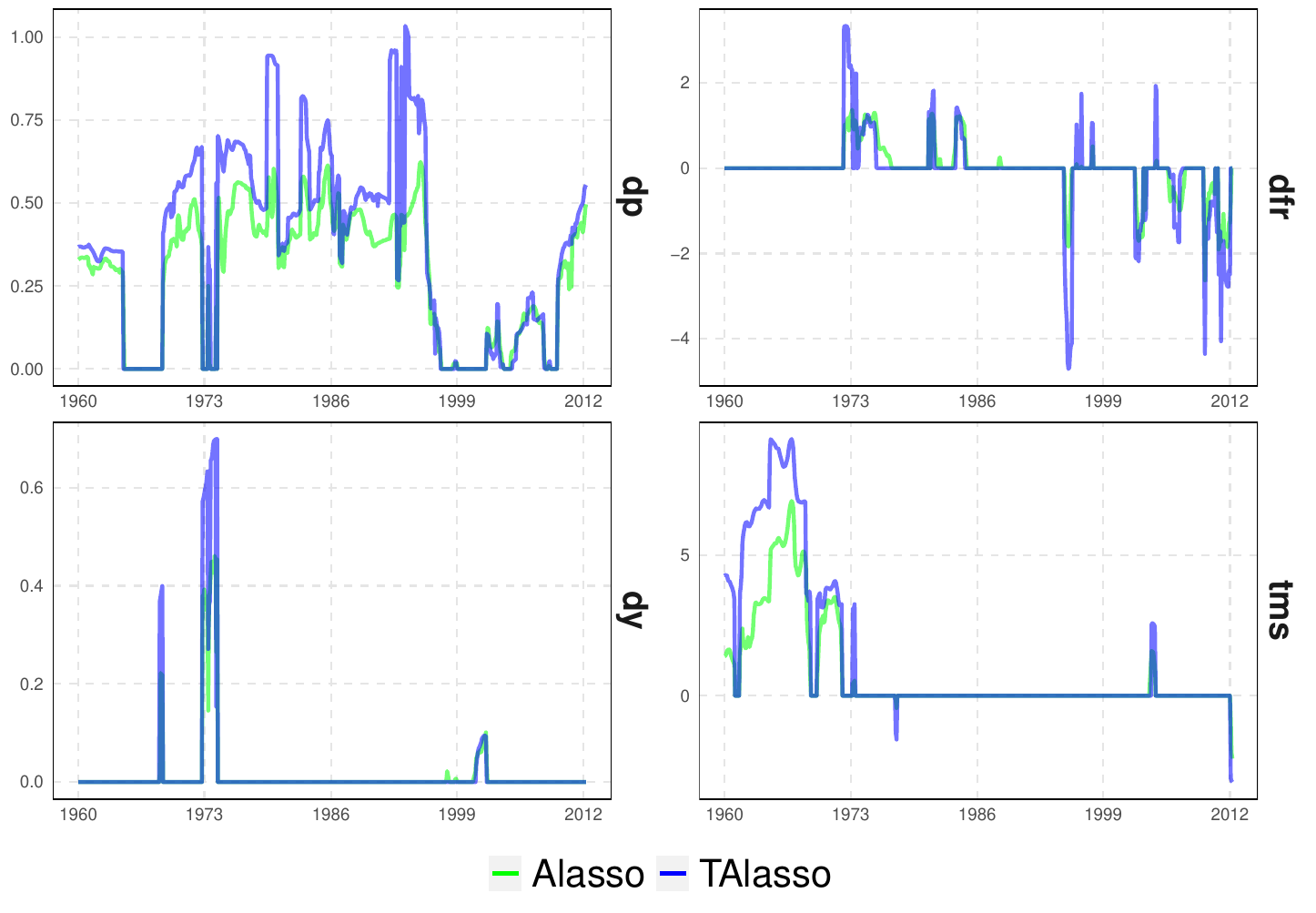}
    \par\end{center}
    \begin{center}
    \subcaption{CV}
    \par\end{center}%
    \end{minipage}\hfill{}%
    \begin{minipage}[t]{0.5\textwidth}%
    \begin{center}
    \includegraphics[width=1\textwidth]{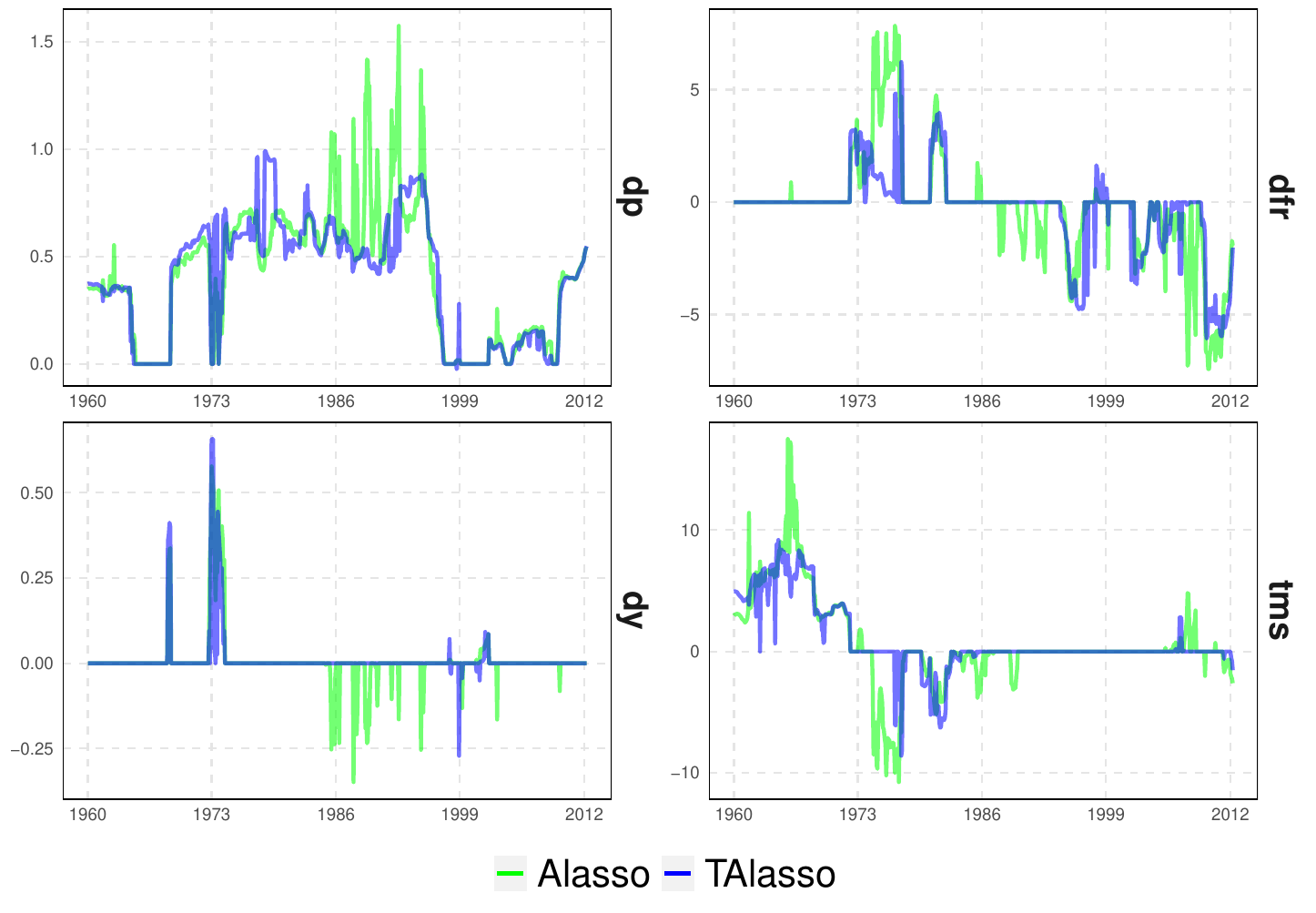}
    \par\end{center}
    \begin{center}
    \subcaption{BIC}
    \par\end{center}%
    \end{minipage}
    \par\end{centering}
    \caption{\label{fig:welch-goyal-coint-coef} Estimated coefficient of potentially cointegrated variables (15-year rolling window, $h=1/2$)}
\end{figure}

In terms of prediction performance, it is known that eliminating irrelevant
predictors is more important than including relevant predictors; see, for example, \citet{ploberger2003empirical}. The inclusion of the irrelevant predictors could be detrimental
in forecasting contexts, and including irrelevant nonstationary predictors
in predictive regression can be harmful because stock returns
are stationary. In this sense, Alasso and TAlasso provide more conservative
variable selection in predictive regressions. In Figure \ref{fig:welch-Goyal_CV_coef},
the instance with $h=1/2$ and the 15-year rolling window is used to illustrate
the estimated coefficients under CV. The shrinkage methods select
different variables over the estimation windows, indicating the evolution
of the predictive models across time. Alasso and TAlasso throw out
more variables than Plasso or Slasso and hence deliver more parsimonious
models. For example, they completely eliminate the variables \texttt{ltr} and \texttt{infl}. 

To highlight the potentially cointegrated variables, Figure
\ref{fig:welch-goyal-coint-coef}(a) presents the Alasso and TAlasso estimates
of (\texttt{dp}, \texttt{dy}) and (\texttt{dfr}, \texttt{tms}) by CV, and in Figure \ref{fig:welch-goyal-coint-coef}(b), BIC decides
the tuning parameters and then produces the estimates. The table ahead
of the subfigures lists the fraction of
the active Alasso estimates annihilated by TAlasso over the rolling windows. In more than a
quarter of the cases when Alasso selects \texttt{dy} or \texttt{tms}, TAlasso
further penalizes the corresponding coefficient to be $0$ in the
second round. In general, BIC tends to freeze out more variables
than CV, especially for \texttt{dy} and \texttt{tms}. Although the
numbers vary, similar patterns are found in other combinations of
forecast horizons and rolling window lengths.

\section{Conclusion}

We explore LASSO procedures in the presence of stationary, nonstationary,
and cointegrated predictors. Although it does not enjoy the well-known
oracle property, Alasso breaks the links between inactive cointegration
groups, and its repeated implementation TAlasso recovers the
oracle property because of the differentiated penalty on the zero and
nonzero coefficients. TAlasso is adaptive to a system of multiple
predictors with various degrees of persistence, unlike Plasso's uniform
penalty or Slasso's penalty based only on the marginal variation of
each predictor. Moreover, TAlasso saves the effort of sorting the
predictors according to their degrees of persistence, so we can be
agnostic regarding the time series properties of the predictors. The automatic
penalty adjustment of TAlasso guarantees consistent model selection
and the optimal rate of convergence. Such desirable properties may
improve the out-of-sample prediction in practice under complex predictive
environments with a mixture of regressors.

To focus on the mixed root setting, we adopt the simplest asymptotic
framework with a fixed $p$ and $n\to\infty$ to demonstrate the clear
contrast between OLS, Alasso, TAlasso, Plasso, and Slasso. This asymptotic
framework is in line with the state of the art of the predictive regression
studies in financial econometrics \citep*{kostakis2014robust,phillips2016robust,xu2018testing}.
However, the large number of potential regressors available
in the era of big data indicate the need for a theoretical extension to allow
for an infinite number of regressors in the limit. As the restricted
eigenvalue condition \citep*{bickel2009simultaneous} is unsuitable
in our context, in which the nonstationary part of the Gram matrix does
not degenerate, a few papers \citep{onatski2018alternative,onatski2020spurious,zhang2019identifying,bykhovskaya2020cointegration}
point to the spectral analysis of a large random matrix \citep{bai2010spectral}
as the foundation of asymptotic analysis. In future research, we will apply the new technical apparatus to deal with
the minimal eigenvalue of the Gram matrix to generalize the
insight gleaned from low-dimensional asymptotics to high-dimensional asymptotics. 

Another line of related literature concerns uniformly valid inference
and forecasting after LASSO model selection; see \citet*{belloni2018valid,belloni2018uniformly}
or \citet*{hirano2017forecasting}, for example. These papers allow
for model selection error by LASSO and provide valid inference or
prediction by introducing local limit theory with only small departures
from the true models. Combining these recent developments with our
current LASSO theory is anther exciting direction for future research.

\section*{Acknowledgements}
We would like to thank the Editor, Serena Ng, the Associate Editor, and two anonymous referees
for their thoughtful comments, which substantially improved this paper.
We thank Mehmet Caner, Zongwu Cai, Yoosoon Chang, Changjin
Kim, Bonsoo Koo, Zhipeng Liao, Tassos Magdalinos, Joon Park, Hashem Pesaran, Peter
Phillips, Kevin Song, Jing Tao, Keli Xu, Jun Yu, and the seminar participants
at Kansas, Indiana, Purdue, UBC, UW, Duke, KAEA, IPDC, AMES, and IAAE
conferences for helpful comments. We also thank Bonsoo Koo for sharing
the data for the empirical application. Shi acknowledges the financial support of the Hong Kong Research
Grants Council No. 24614817 and No. 14500118. All remaining errors are ours.

\bigskip \bigskip \bigskip

\appendix

\setcounter{footnote}{0}
\setcounter{table}{0} 
\setcounter{figure}{0} 
\setcounter{equation}{0} 

\renewcommand{\thefootnote}{A.\arabic{footnote}} 
\renewcommand{\theequation}{A.\arabic{equation}} 
\renewcommand{\thefigure}{A.\arabic{figure}} 
\renewcommand{\thetable}{A.\arabic{table}}

\section{Technical Appendix}

\subsection{Proofs in Section \ref{sec:UR}}

\begin{proof} {[}Proof of Theorem \ref{thm:ada_unit}{]} Since only
Alasso is considered in this proof, we denote $\widehat{\beta}=\widehat{\beta}^{\mathrm{A}}$
and $\widehat{M}=\widehat{M}^{\mathrm{A}}$ for conciseness. Let $\beta_{n}=\beta_{n}^{\ast}+n^{-1}v$
be a perturbation around the true parameter $\beta_{n}^{\ast}$, and
let 
\[
\Psi_{n}(v)=\Vert Y-\sum_{j=1}^{p}x_{j}(\beta_{jn}^{\ast}+\frac{v_{j}}{n})\Vert^{2}+\lambda_{n}\sum_{j=1}^{p}\hat{\tau}_{j}|\beta_{jn}^{\ast}+\frac{v_{j}}{n}|.
\]
Define $\hat{v}^{(n)}=n(\hat{\beta}-\beta_{n}^{\ast})$.
The fact that $\hat{\beta}$ minimizes (\ref{eq:aLasso})
implies $\hat{v}^{(n)}=\arg\min_{v}\Psi_{n}(v).$ Let 
\begin{eqnarray}
V_{n}(v) & = & \Psi_{n}(v)-\Psi_{n}(0)=\Vert u-\frac{X^{\prime}v}{n}\Vert^{2}-\Vert u\Vert^{2}+\lambda_{n}\bigg(\sum_{j=1}^{p}\hat{\tau}_{j}|\beta_{jn}^{\ast}+\frac{v_{j}}{n}|-\sum_{j=1}^{p}\hat{\tau}_{j}|\beta_{jn}^{\ast}|\bigg)\nonumber \\
 & = & v^{\prime}(\frac{X^{^{\prime}}X}{n^{2}})v-2\frac{u^{^{\prime}}X}{n}v+\lambda_{n}\sum_{j=1}^{p}\hat{\tau}_{j}(|\beta_{jn}^{\ast}+\frac{v_{j}}{n}|-|\beta_{jn}^{\ast}|).\label{eq:V_4}
\end{eqnarray}
The first and the second terms in the right-hand side of (\ref{eq:V_4})
converge in distribution, as $n^{-2} X^{\prime}X \Longrightarrow \Omega$
and $ n^{-1} X^{\prime}u = n^{-1} \sum_{i=1}^{n}x_{i\cdot}^{\prime}u_{i}\Longrightarrow\zeta$,
by the functional central limit theorem (FCLT) and the continuous
mapping theorem. The third term involves the weight $\hat{\tau}_{j}=|\widehat{\beta}_{j}^{\mathrm{ols}}|^{-\gamma}$
for each $j$. Since the OLS estimator $n\left(\widehat{\beta}^{\mathrm{ols}}-\beta_{n}^{\ast}\right)\Longrightarrow\Omega^{-1}\zeta=O_{p}(1),$
we have 
\begin{equation}
\hat{\tau}_{j}=\left\vert \beta_{jn}^{\ast}+O_{p}\left(n^{-1}\right)\right\vert ^{-\gamma}=|\beta_{j}^{0\ast}/n^{\delta_{j}}+O_{p}\left(n^{-1}\right)|^{-\gamma}.\label{eq:pure_ala1}
\end{equation}

If $\beta_{j}^{0\ast}\neq0$, then $\beta_{jn}^{\ast}$ dominates
$n^{-1}v_{j}$ for a large $n$ and 
\begin{equation}
|\beta_{jn}^{\ast}+\frac{v_{j}}{n}|-|\beta_{jn}^{\ast}|=n^{-1}v_{j}\text{sgn}(\beta_{jn}^{\ast})=n^{-1}v_{j}\text{sgn}(\beta_{j}^{0\ast}).\label{eq:pure_ala2}
\end{equation}
Now (\ref{eq:pure_ala1}) and (\ref{eq:pure_ala2}) imply 
\begin{align}
\lambda_{n}\hat{\tau}_{j}\cdot(|\beta_{jn}^{\ast}+\frac{v_{j}}{n}|-|\beta_{jn}^{\ast}|) & =\frac{\lambda_{n}}{n|\beta_{j}^{0\ast}/n^{\delta_{j}}+O_{p}\left(n^{-1}\right)|^{\gamma}}v_{j}\text{sgn}(\beta_{j}^{0\ast})=\frac{\lambda_{n}n^{\delta_{j}\gamma-1}}{|\beta_{j}^{0\ast}+o_{p}\left(1\right)|^{\gamma}}v_{j}\text{sgn}(\beta_{j}^{0\ast})\nonumber \\
 & =O_{p}\left(\lambda_{n}n^{\delta_{j}\gamma-1}\right)=o_{p}\left(1\right)\label{eq:pure_ala3}
\end{align}
by the given rate of $\lambda_{n}$. On the other hand, if $\beta_{j}^{0\ast}=0$,
then $(|\beta_{jn}^{\ast}+n^{-1}v_{j}|-|\beta_{jn}^{\ast}|)=n^{-1}|v_{j}|.$
For any fixed $v_{j}\neq0,$ 
\begin{equation}
\lambda_{n}\hat{\tau}_{j}\cdot(|\beta_{j}^{\ast}+\frac{v_{j}}{n}|-|\beta_{j}^{\ast}|)=\frac{\lambda_{n}}{n|\widehat{\beta}_{j}^{\mathrm{ols}}|^{\gamma}}\left\vert v_{j}\right\vert =\frac{\lambda_{n}n^{\gamma-1}}{|n\widehat{\beta}_{j}^{\mathrm{ols}}|^{\gamma}}\left\vert v_{j}\right\vert =\frac{\lambda_{n}n^{\gamma-1}}{O_{p}\left(1\right)}\left\vert v_{j}\right\vert \rightarrow\infty\label{eq:pure_ala4}
\end{equation}
since $\lambda_{n}\rightarrow\infty$, $\gamma\geq1$ and the OLS
estimator is asymptotically non-degenerate. Thus we have $V_{n}(v)\Longrightarrow V(v)$
for every fixed $v$, where 
\[
V(v)=\begin{cases}
v^{\prime}\Omega v-2v^{\prime}\zeta, & \mbox{if }v_{M^{\ast c}}=\mathbf{0}_{|M^{*c}|}\\
\infty, & \mbox{ otherwise}.
\end{cases}
\]
Both $V_{n}\left(v\right)$ and $V\left(v\right)$ are strictly convex
in $v$, and $V\left(v\right)$ is uniquely minimized at 
\[
\begin{pmatrix}v_{M^{\ast}}\\
v_{M^{\ast c}}
\end{pmatrix}=\begin{pmatrix}\Omega_{M^{\ast}}^{-1}\zeta_{M^{\ast}}\\
0
\end{pmatrix}.
\]
Applying the Convexity Lemma \citep{pollard1991asymptotics}, we have
\begin{equation}
\hat{v}_{M^{\ast}}^{(n)}=n(\hat{\beta}_{M^{\ast}}-\beta_{M^{\ast}}^{\ast})\Longrightarrow\Omega_{M^{\ast}}^{-1}\zeta_{M^{\ast}}\mbox{ \ \ \ \ and \ \ \ \ }\hat{v}_{M^{\ast c}}^{(n)}\Longrightarrow0.\label{eq:dev}
\end{equation}
The first part of the above result establishes Theorem \ref{thm:ada_unit}(b)
about the asymptotic distribution for the coefficients in $M^{*}$.

\medskip{}
 Next, we show variable selection consistency. The result $P(M^{\ast}\subseteq\widehat{M})\rightarrow1$
immediately follows by the first part of (\ref{eq:dev}), since $\hat{v}_{M^{\ast}}^{(n)}$
converges in distribution to a non-degenerate continuous random variable.
For those $j\in M^{\ast c}$, if the event $\{j\in\widehat{M}\}$
occurs, then the KKT condition entails 
\begin{equation}
\frac{1}{n}x_{j}^{\prime}(y-X\hat{\beta})
= \mathrm{sgn}(\hat{\beta}_j) \frac{\lambda_{n}\hat{\tau}_{j}}{2n} .\label{eq:KKT}
\end{equation}
Notice that the order on the right-hand side is governed by 
\begin{equation}
\frac{\lambda_{n}\hat{\tau}_{j}}{n}=\frac{\lambda_{n}}{n|\widehat{\beta}_{j}^{\mathrm{ols}}|^{\gamma}}=\frac{\lambda_{n}n^{\gamma-1}}{|n\widehat{\beta}_{j}^{\mathrm{ols}}|^{\gamma}}=\frac{\lambda_{n}n^{\gamma-1}}{O_{p}\left(1\right)}\rightarrow\infty\label{eq:pure_ada4}
\end{equation}
given the rate of $\lambda_{n}$. However, using $y=X\beta_{n}^{\ast}+u$
and (\ref{eq:dev}), the left-hand side of (\ref{eq:KKT}) is 
\begin{align}
\frac{1}{n}x_{j}^{\prime}(y-X\hat{\beta}) & =\frac{1}{n}x_{j}^{\prime}(X\beta_{n}^{\ast}-X\hat{\beta}+u)=\frac{x_{j}^{\prime}X}{n^{2}}\cdot n(\beta_{n}^{\ast}-\hat{\beta})+\frac{x_{j}^{\prime}u}{n}\nonumber \\
 & =-\frac{x_{j}^{\prime}}{n^{2}}\left(X_{M^{*}}\hat{v}_{M^{\ast}}^{(n)}+X_{M^{*c}}\widehat{v}_{M^{\ast c}}^{\left(n\right)}\right)+\frac{x_{j}^{\prime}u}{n}\nonumber \\
 & \Longrightarrow-\Omega_{j,M^{*}}\Omega_{M^{\ast}}^{-1}\zeta_{M^{\ast}}-\Omega_{j,M^{*c}}o_{p}(1)+\zeta_{j}=O_{p}\left(1\right),\label{eq:pure_ada5}
\end{align}
where $\Omega_{j, M^*} = [\Omega_{j,l}]_{l\in M^*}$ and $\Omega_{j, M^{*c}}$ is defined similarly.
In other words, the left-hand side of (\ref{eq:KKT}) remains a non-degenerate
continuous random variable in the limit. For any $j\in M^{\ast c}$,
the disparity of the two sides of the KKT condition implies 
\[
P\left( j\in\widehat{M}_{n} \right)=P\left(\frac{1}{n}x_{j}^{\prime}(y-X\hat{\beta})
= \mathrm{sgn}(\hat{\beta}_j) \frac{\lambda_{n}\hat{\tau}_{j}}{2n} 
\right)\rightarrow0.
\]
That is, $P(M^{\ast c}\subseteq\widehat{M})\rightarrow0$ or equivalently
$P(\widehat{M}\subseteq M^{\ast})\rightarrow1$. We thus conclude
the variable selection consistency in Theorem \ref{thm:ada_unit}(a)
\end{proof}

\bigskip{}

\subsection{{\normalsize{}\label{subsec:Proofs-in-Section3}} Proofs in Section
\ref{sec:MRC}{\normalsize{} }}

To express the OLS asymptotic distribution, we need some extra notations.
Define $\Omega=\sum_{h=-\infty}^{\infty}\mathbb{E}\left(\xi_{i}\xi_{i-h}^{\prime}\right)$
as the long-run covariance matrix associated with the innovation vector.
It can be written as is 
\[
\underset{\left(p+1\right)\times\left(p+1\right)}{\Omega}=F(1)\Sigma_{\varepsilon}F(1)^{\prime}=\left(\begin{array}{cccc}
\Omega_{zz} & \Omega_{zv} & \Omega_{ze} & 0\\
\Omega_{zv}^{\prime} & \Omega_{vv} & \Omega_{ve} & 0\\
\Omega_{ze}^{\prime} & \Omega_{ve}^{\prime} & \Omega_{ee} & \Omega_{eu}\\
0 & 0 & \Omega_{eu}^{\prime} & \Omega_{uu}
\end{array}\right)
\]
according to the explicit form of $\Sigma_{\varepsilon}$, where $F(1)=\left(F_{z}^{\prime}\left(1\right),F_{v}^{\prime}\left(1\right),F_{e}^{\prime}\left(1\right),F_{u}\left(1\right)\right)^{\prime}$.
Moreover, define the sum of one-sided autocovariance as $\Lambda=\sum_{h=1}^{\infty}\mathbb{E}\left(\xi_{i}\xi_{i-h}^{\prime}\right),$
and $\text{ }\Delta=\Lambda+\mathbb{E}\left(\xi_{i}\xi_{i}^{\prime}\right)$.
We use the functional law \citep{phillips1992asymptotics} under Assumption
\ref{ass:INNOV coint} to derive 
\begin{align*}
\frac{1}{\sqrt{n}}\sum_{i=1}^{\left\lfloor nr\right\rfloor }\xi_{i} & =\left(B_{zn}'(r),B_{vn}'\left(r\right),B_{en}'\left(r\right),B_{un}(r)\right)'\\
 & \Longrightarrow\left(B_{z}'(r),B_{v}'\left(r\right),B_{e}'\left(r\right),B_{u}(r)\right)':=B_{\xi}\left(r\right) \sim  BM\left(\Omega\right),
\end{align*}
where $B_{\xi}\left(r\right)$ is a vector Brownian motion whose covariance
kernel is $\Omega$. Define the corresponding vector Ornstein Uhlenbeck
process $J_{c_{2}}\left(r\right)$ associated with $B_{v_{2}}\left(r\right)$,
which is a subvector of $B_{v}\left(r\right)=(B_{v_{1}}'\left(r\right),B_{v_{2}}'\left(r\right))^{\prime}$
in the standard way: 
\[
dJ_{c_{2}}\left(r\right)=c_{2}J_{c_{2}}\left(r\right)dr+dB_{v_{2}}\left(r\right),\text{ }J_{c_{2}}\left(0\right)=0\text{.}
\]
Similarly, define $J_{c_{x}}\left(r\right)$ as $dJ_{c_{x}}\left(r\right)=c_{x}J_{c_{x}}\left(r\right)dr+dB_{e}\left(r\right),\text{ }J_{c_{x}}\left(0\right)=0\text{.}$ 

\begin{lemma} \label{lem:OLS} If the linear model (\ref{DGP XYZ})
satisfies Assumption \ref{ass:INNOV coint}, then $\left(u'Z^{+}/\sqrt{n},u'X^{+}/n\right)$
converges in distribution to a non-degenerate stable law. \end{lemma}

\begin{proof} {[}Proof of Lemma \ref{lem:OLS}{]} The left-top $p\times p$
submatrix of $\Omega$ can be represented in a conformable manner
as \
\begin{equation}
\left(\begin{array}{ccc}
\Omega_{zz} & \Omega_{zv} & \Omega_{ze}\\
\Omega_{zv}^{\prime} & \Omega_{vv} & \Omega_{ve}\\
\Omega_{ze}^{\prime} & \Omega_{ve}^{\prime} & \Omega_{ee}
\end{array}\right)=\left(\begin{array}{cc}
\underset{\left(p_{z}+p_{1}\right)\times\left(p_{z}+p_{1}\right)}{\Omega_{zz}^{+}} & \Omega_{zx}^{+}\\
\Omega_{zx}^{+\prime} & \underset{\left(p_{2}+p_{x}\right)\times\left(p_{2}+p_{x}\right)}{\Omega_{xx}^{+}}
\end{array}\right).\label{eq:block_Omega}
\end{equation}
The Beveridge-Nelson (BN) decomposition and central limit theorem lead to 
\begin{eqnarray}
Z^{+\prime}u/\sqrt{n} & \Longrightarrow & \zeta_{z^{+}}\sim N\left(0,\Sigma_{uu}\Omega_{zz}^{+}\right).\label{eq:zeta_z}
\end{eqnarray}

The columns in $X^{+}$ are local unit root processes with no cointegration
relationship. Let the $i$-th row of $X^{+}$ be $X_{i}^{+}=\left(X_{2i\cdot}^{c}\text{ , }X_{i\cdot}\right)^{\prime}$, a $(p_2 + p_x)$-by-$1$ vector.
Using the component-wise BN decomposition, the scalar $u_{i}=  \varepsilon_{i}^{\prime} \times F_{u}(1)-\triangle\tilde{\varepsilon}_{ui}$,
where $\triangle\tilde{\varepsilon}_{ui}$ is the last $\left(p+1\right)$-th
scalar component of the vector $\triangle\tilde{\varepsilon}_{i}=\tilde{\varepsilon}_{i}-\tilde{\varepsilon}_{i-1}$
and $\tilde{\varepsilon}_{i}=\sum_{j=0}^{\infty}\tilde{F}_{j}\varepsilon_{i-j}$
with $\tilde{F}_{j}=\sum_{k=j+1}^{\infty}F_{k}$. Thus we have 
\[
\underset{(p_{2}+p_{x})\times1}{\frac{1}{n}X^{+\prime}u}=\frac{1}{n}\sum_{i=1}^{n}X_{i}^{+}u_{i}=\left(\frac{1}{n}\sum_{i=1}^{n}X_{i}^{+}\varepsilon_{i}^{\prime}\right)F_{u}(1)-\frac{1}{n}\sum_{i=1}^{n}X_{i}^{+}\triangle\tilde{\varepsilon}_{ui}.
\]
Functional CLT combined with the stationary $O_{p}(1)$ initial condition
provides $X_{i}^{+}/\sqrt{n} \Longrightarrow J_{x^{+}}(r)=\left(J_{c_{2}}^{\prime}\left(r\right),J_{c_{x}}^{\prime}\left(r\right)\right)^{\prime}$, and thus  $n^{-1}\sum_{i=1}^{n}X_{i}^{+}\varepsilon_{i}^{\prime}\Longrightarrow\int_{0}^{1}J_{x^{+}}(r)dB_{\varepsilon}(r)^{\prime}$.
The summation by parts now implies 
\[
\frac{1}{n}\sum_{i=1}^{n}X_{i}^{+}\triangle\tilde{\varepsilon}_{ui}=-\frac{1}{n}\sum_{i=1}^{n}u_{xi}^{+}\tilde{\varepsilon}_{ui-1}+o_{p}(1)\overset{p}{\rightarrow}\Delta_{+u}
\]
where $\Delta_{+u}=\sum_{h=0}^{\infty} E[ \tilde{u}_{i}u_{i-h}  ]$
is the corresponding submatrix of the one-sided long-run covariance
and $u_{xi}^{+}=X_{i}^{+}-X_{i-1}^{+}$. Combining these results, we have
\begin{equation}
\frac{X^{+\prime}u}{n} \Longrightarrow \zeta_{x^{+}}\sim\int_{0}^{1}J_{x^{+}}(r)dB_{\varepsilon}(r)^{\prime}F_{u}(1)+\Delta_{+u}.\label{eq:zeta_x}
\end{equation}
Since the marginal limit distributions are non-degenerate in (\ref{eq:zeta_z})
and (\ref{eq:zeta_x}), so does the joint limiting distribution. \end{proof}

\bigskip{}

\begin{proof} {[}Proof of Theorem \ref{thm:OLS}{]} We transform
and scale-normalize the OLS estimator as 
\begin{align}
R_{n}Q\left(\widehat{\theta}^{\mathrm{ols}}-\theta_{n}^{\ast}\right) & =R_{n}Q\left(W^{\prime}W\right)^{-1}W^{\prime}u\nonumber \\
 & =R_{n}Q\left(W^{\prime}W\right)^{-1}Q^{\prime}R_{n}\left(Q^{\prime}R_{n}\right)^{-1}W^{\prime}u\nonumber \\
 & =\left[R_{n}^{-1}Q^{\prime-1}W^{\prime}WQ^{-1}R_{n}^{-1}\right]^{-1}R_{n}^{-1}Q^{\prime-1}W^{\prime}u.\label{eq:ols1}
\end{align}
The first factor 
\begin{equation}
R_{n}^{-1}Q^{\prime-1}W^{\prime}WQ^{-1}R_{n}^{-1}=\left(\begin{array}{cc}
\frac{Z^{+\prime}Z^{+}}{n} & \frac{Z^{+\prime}X^{+}}{n^{3/2}}\\
\frac{Z^{+\prime}X^{+}}{n^{3/2}} & \frac{X^{+^{\prime}}X^{+}}{n^{2}}
\end{array}\right)\Longrightarrow\left(\begin{array}{cc}
\Omega_{zz}^{+} & 0\\
0 & \int_{0}^{1}J_{x^{+}}(r)J_{x^{+}}(r)^{\prime}dr
\end{array}\right)=:\Omega^{+}\label{eq:ols_term_quad}
\end{equation}
where $\Omega_{zz}^{+}$ is defined in (\ref{eq:block_Omega}), and $J_{x^{+}}(r)=\left(J_{c_{2}}^{\prime}\left(r\right),J_{c_{x}}^{\prime}\left(r\right)\right)^{\prime}$.
The second factor 
\begin{equation}
R_{n}^{-1}Q^{\prime-1}W^{\prime}u=\left(\begin{array}{c}
Z^{+\prime}u/\sqrt{n}\\
X^{+\prime}u/n
\end{array}\right)\Longrightarrow\zeta^{+},\label{eq:ols_term_cross}
\end{equation}
where the marginal distributions of $\zeta^{+}$ can be found in (\ref{eq:zeta_z})
and (\ref{eq:zeta_x}). Thus the stated conclusion follows by the
continuous mapping theorem. \end{proof}

\begin{remark} In the rotated coordinate system, (\ref{eq:OLS_Q_rate})
in Theorem \ref{thm:OLS} and the definition of $\zeta_{x^{+}}$ imply
that an asymptotic bias term $\Delta_{+u}$ appears in the limit distribution
of OLS with nonstationary predictors. This asymptotic bias arises
from the serial dependence in the innovations. However, the asymptotic
bias does not affect the rate of convergence as $Q(\hat{\theta}^{\mathrm{ols}}-\theta_{n}^{\ast})=O_{p}(\mathrm{diag}(R_{n}^{-1})).$
\end{remark}

\begin{remark} The OLS estimator will serve as the initial estimator
for Alasso. In principle either the fully-modified OLS \citep{phillips1990statistical}
or the canonical cointegrating regression estimator \citep{park1992canonical}
can be an initial estimator as well, because their rates of convergence
are the  same as those of OLS. \end{remark}

\bigskip{}

To simplify notation, define $R_{jn}$ as the $j$-th diagonal element of $R_{n}$,
$\widehat{C}=\widehat{M}^{\mathrm{A}}\cap\mathcal{C}$ and $C^{*}=M^{*}\cap\mathcal{C}$.
Recall $C^{*c}=M^{*c}\cap\mathcal{C}$.

\begin{proof} {[}Proof of Theorem \ref{thm:ada_3type}{]} Again,
we use $\widehat{\theta}=\widehat{\theta}^{\mathrm{A}}$ and $\widehat{M}=\widehat{M}^{\mathrm{A}}$
for conciseness in this proof. For a constant nonzero vector 
\begin{equation}
\tilde{v}=\left(\tilde{v}_{z}^{\prime},\tilde{v}_{1}^{\prime},\mathbf{0}_{p_{2}}^{\prime},\tilde{v}_{x}^{\prime}\right)^{\prime}\neq0\label{eq:v_tilde}
\end{equation}
where the elements associated with $\mathcal{C}_{2}$ are suppressed
as $\mathbf{0}$, we add a local perturbation 
\[
v_{n}=Q^{-1}R_{n}^{-1}\tilde{v}=\left(\frac{\tilde{v}_{z}^{\prime}}{\sqrt{n}},\frac{\tilde{v}_{1}^{\prime}}{\sqrt{n}},-\frac{\tilde{v}_{1}^{\prime}A_{1}}{\sqrt{n}},\frac{\tilde{v}_{x}^{\prime}}{n}\right)
\]
to $\theta_{n}^{*}$ so that the perturbed coefficient $\theta_{n}=\theta_{n}^{\ast}+v_{n}$.
Let 
\[
\Psi_{n}(\tilde{v})=\Vert Y-W\left(\theta_{n}^{*}+v_{n}\right)\Vert^{2}+\lambda_{n}\sum_{j=1}^{p}\hat{\tau}_{j}\left\vert \theta_{jn}^{\ast}+v_{jn}\right\vert ,
\]
and then define 
\begin{eqnarray*}
V_{n}(\tilde{v}) & = & \Psi_{n}(\tilde{v})-\Psi_{n}(0)=\left\Vert u-Wv_{n}\right\Vert ^{2}-\left\Vert u\right\Vert ^{2}+\lambda_{n}\sum_{j=1}^{p}\hat{\tau}_{j}\left(|\theta_{jn}^{\ast}+v_{jn}|-|\theta_{jn}^{\ast}|\right)\\
 & = & v_{n}^{\prime}W^{\prime}Wv_{n}-2v_{n}^{\prime}W^{\prime}u+\lambda_{n}\sum_{j=1}^{p}\hat{\tau}_{j}\left(|\theta_{jn}^{\ast}+v_{jn})|-|\theta_{jn}^{\ast}|\right).
\end{eqnarray*}
We have shown in the proof of Theorem \ref{thm:OLS} that the first
term 
\begin{align}
v_{n}^{\prime}W^{\prime}Wv_{n} & =\tilde{v}^{\prime}R_{n}^{-1}Q^{\prime-1}W^{\prime-1}R_{n}^{-1}\tilde{v}\Longrightarrow\tilde{v}^{\prime}\Omega^{+}\tilde{v}\label{eq:ala_quad}
\end{align}
by (\ref{eq:ols_term_quad}), and the second term 
\begin{equation}
2v_{n}^{\prime}W^{\prime}u=2\tilde{v}^{\prime}R_{n}^{-1}Q^{\prime-1}W^{\prime}u\Longrightarrow2\tilde{v}^{\prime}\zeta^{+}\label{eq:ala_cross}
\end{equation}
by (\ref{eq:ols_term_cross}).

We focus on the third term. Theorem \ref{thm:OLS} has shown that
the OLS estimator $\widehat{\theta}_{j}^{\mathrm{ols}}-\theta_{jn}^{\ast}=O_{p}\left(R_{jn}^{-1}\right)$
for each $j\in\mathcal{M}_{Q}$. Given any fixed $\tilde{v}_{j}\neq0$
and a sufficiently large $n$:
\begin{itemize}
\item For $j\in\mathcal{I}_{0}\cup\mathcal{C}_{1}$, by the definition of
$v$ each element $v_{jn}=\tilde{v}_{j}/\sqrt{n}$. If $\theta_{j}^{0*}\neq0$,
we have $|\theta_{j}^{*}+\tilde{v}_{j}/\sqrt{n}|-|\theta_{j}^{*}|=n^{-1/2}\tilde{v}_{j}\text{sgn}(\theta_{j}^{0*})$,
and thus $\lambda_{n}\hat{\tau}_{j} (|\theta_{j}^{0*}+ \tilde{v}_{j}/\sqrt{n}|-|\theta_{j}^{0*}|)=O_{p}\left(\lambda_{n}n^{-1/2}\right)=o_{p}\left(1\right).$
If $\theta_{j}^{0*}=0$, we have $\lambda_{n}\hat{\tau}_{j}(|\theta_{j}^{0\ast}+ \tilde{v}_{j} / \sqrt{n}|-|\theta_{j}^{0\ast}|)= \lambda_{n}n^{0.5(\gamma-1)} \left\vert \tilde{v}_{j}\right\vert /O_{p}(1)
 =O_{p}\left(\lambda_{n}n^{\left(\gamma-1\right)/2}\right)\rightarrow\infty$
when $\tilde{v}_{j}\neq0$.
\item For $j\in\mathcal{I}_{1}$, the true coefficient $\theta_{jn}^{*}=\theta_{j}^{0*}/n^{\delta_{j}}$
depends on $n$ while the perturbation $v_{jn}=\tilde{v}_{j}/n$.
If $\theta_{j}^{0*}\neq0$, then $\theta_{jn}^{\ast}$ dominates $\tilde{v}_{j}/n$
in the limit and $(|\theta_{jn}^{\ast}+\tilde{v}_{j}/n|-|\theta_{jn}^{\ast}|)=\text{sgn}(\theta_{j}^{0\ast})\tilde{v}_{j}/n$.
By the same derivation in (\ref{eq:pure_ala3}), $\lambda_{n}\hat{\tau}_{j}(|\theta_{jn}^{\ast}+ n^{-1} \tilde{v}_{j}|-|\theta_{jn}^{\ast}|)=O_{p}\left(\lambda_{n}n^{\left(\delta_{j}\gamma-1\right)}\right)=o_{p}\left(1\right)$
given the condition (\ref{eq:lambda_rate}). If $\theta_{j}^{0*}=0$,
according to the derivation in (\ref{eq:pure_ala4}) $\lambda_{n}\hat{\tau}_{j}(|\theta_{jn}^{\ast}+n^{-1}\tilde{v}_{j}|-|\theta_{jn}^{\ast}|)
=\lambda_{n}n^{\gamma-1} \left\vert \tilde{v}_{j}\right\vert / O_{p}(1) 
=O_{p}\left(\lambda_{n}n^{\gamma-1}\right)\rightarrow\infty$
when $\tilde{v}_{j}\neq0$. 
\end{itemize}
The above analysis indicates $V_{n}(\tilde{v})\Longrightarrow V(\tilde{v})$
for every fixed $\tilde{v}$ in (\ref{eq:v_tilde}), where 
\[
V(\tilde{v})=\begin{cases}
\tilde{v}^{\prime}\Omega^{+}\tilde{v}-2\tilde{v}^{\prime}\zeta^{+}, & \mbox{if }\tilde{v}_{M_{Q}^{*c}}=0\\
\infty, & \mbox{ otherwise}.
\end{cases}
\]
Let $\hat{v}^{(n)}=\hat{\theta}-\theta_{n}^{\ast}$. The same argument
about the strict convexity of $V_{n}\left(\tilde{v}\right)$ and $V\left(\tilde{v}\right)$
implies 
\begin{eqnarray}
(R_{n}Q\hat{v}^{(n)})_{M_{Q}^{*}} & \Longrightarrow & (\Omega_{M_{Q}^{*}}^{+})^{-1}\zeta_{M_{Q}^{*}}^{+}\label{eq:dev-1}\\
(R_{n}Q\hat{v}^{(n)})_{M_{Q}^{*c}} & \Longrightarrow & 0\label{eq:dev-2}
\end{eqnarray}
We have established Theorem \ref{thm:ada_3type}(a).

\bigskip{}

Next, we move on to discuss the effect of variable selection. Consider
any $j\in\widehat{M}$. The KKT condition with respect to $\theta_{j}$
entails 
\begin{equation}
W_{j}^{\prime}(y-W\hat{\theta})=\mathrm{sgn}(\hat{\theta}_j) \lambda_{n}\hat{\tau}_{j}/2.\label{eq:KKT-3types}
\end{equation}
We will invoke similar argument as in (\ref{eq:pure_ada4}) and (\ref{eq:pure_ada5})
to show the disparity of the two sides of the KKT condition. The left-hand
side of (\ref{eq:KKT-3types}) is the $j$-th element of the $p\times1$
vector $W^{\prime}(y-W\hat{\theta})$. Pre-multiply the diagonal matrix
$R_{n}^{-1}Q^{\prime-1}$ to the vector: 
\begin{align}
R_{n}^{-1}Q^{\prime-1}W^{\prime}(y-W\hat{\theta}) & =R_{n}^{-1}Q^{\prime-1}W^{\prime}\left(W(\theta_{n}^{\ast}-\hat{\theta})+u\right)\nonumber \\
 & =\left(R_{n}^{-1}Q^{\prime-1}W^{\prime}WQ^{-1}R_{n}^{-1}\right)R_{n}Q(\theta_{n}^{\ast}-\hat{\theta})+R_{n}^{-1}Q^{\prime-1}W^{\prime}u\nonumber \\
 & =R_{n}^{-1}Q^{\prime-1}W^{\prime}WQ^{-1}R_{n}^{-1}O_{p}\left(1\right)+R_{n}^{-1}Q^{\prime-1}W^{\prime}u\nonumber \\
 & =\left(\Omega^{+}+o_{p}\left(1\right)\right)O_{p}\left(1\right)+\left(\zeta^{+}+o_{p}\left(1\right)\right)=O_{p}\left(1\right)\label{eq:alasso_LHS}
\end{align}
where the third equality follows by (\ref{eq:dev-1}) and (\ref{eq:dev-2}),
and the fourth equality by (\ref{eq:ols_term_quad}) and (\ref{eq:ols_term_cross}).
Notice that (\ref{eq:alasso_LHS}) gives the biggest order of the left-hand side,
while some components of this $p\times 1 $vector may be degenerate (of order $o_p(1)$).

Suppose $j\in M^{*c}$. For $j\in\mathcal{\mathcal{I}}$, the rotation
$Q$ does not change these variables so the order of the left-hand
side of (\ref{eq:KKT-3types}) is the same as (\ref{eq:alasso_LHS}).
If $j\in\mathcal{I}_{0}$, multiply $n^{-1/2}$ to the right-hand
side of (\ref{eq:KKT-3types}):
\[
\frac{\mathrm{sgn}(\hat{\theta}_j) }{2\sqrt{n}}\lambda_{n}\hat{\tau}_{j}
=\mathrm{sgn}(\hat{\theta}_j) \frac{\lambda_{n}}{2\sqrt{n}|\widehat{\theta}_{j}^{\mathrm{ols}}|^{\gamma}}
=\mathrm{sgn}(\hat{\theta}_j) \frac{\lambda_{n}n^{0.5\left(\gamma-1\right)}}{2|\sqrt{n}\widehat{\theta}_{j}^{\mathrm{ols}}|^{\gamma}}
=\mathrm{sgn}(\hat{\theta}_j) O_{p}\left(\lambda_{n}n^{0.5\left(\gamma-1\right)}\right)\to\infty \mbox{ or } -\infty
\]
as $\gamma\geq1$ and $\lambda_{n}\to\infty$. Similarly, if $j\in\mathcal{I}_{1}$
we multiply $n^{-1}$ to the right-hand side of (\ref{eq:KKT-3types}):
\[
 \frac{\mathrm{sgn}(\hat{\theta}_j) }{2n}\lambda_{n}\hat{\tau}_{j}
=\mathrm{sgn}(\hat{\theta}_j) \frac{\lambda_{n}n^{\left(\gamma-1\right)}}{2|n\widehat{\theta}_{j}^{\mathrm{ols}}|^{\gamma}}
=\mathrm{sgn}(\hat{\theta}_j) O_{p}\left(\lambda_{n}n^{\left(\gamma-1\right)}\right)\to\infty \mbox{ or } -\infty.
\]
We have verified that for $j\in\mathcal{I}$ the right-hand side of
(\ref{eq:KKT-3types}) is of bigger order than its left-hand side.
It immediately follows that given the specified rate of $\lambda_{n}$,
for any $j\in M^{*c}\cap\mathcal{\mathcal{I}}$ we have (\ref{eq:sel_I01})
since $P(j\in\widehat{M}\cap M^{*c}\cap\mathcal{I})\to P\left(O_{p}\left(1\right)=\infty\ \mbox{ or} -\infty
\right)=0.$
We have established (\ref{eq:sel_I01}).

\medskip{}

Estimation consistency (\ref{eq:alasso_active_dist}) immediately
implies (\ref{eq:sel_C}). On the other hand, it is possible that
variables in $\mathcal{C}$ are wrongly selected into $\widehat{M}$.
If the event $\{j\in\widehat{M}\}$ occurs for some $j\in\mathcal{C}^{*c}$,
the KKT condition may still hold in the limit. To see this, pre-multiply
$n^{-1}$ to the the left-hand side of (\ref{eq:KKT-3types}) and
it is of order $O_{p}\left(1\right)$ according
to (\ref{eq:alasso_LHS}), and it can be degenerate. 
On the other hand, the right-hand side becomes
\[
 \frac{\mathrm{sgn}(\hat{\theta}_j) }{2n}\lambda_{n}\hat{\tau}_{j}
= \mathrm{sgn}(\hat{\theta}_j) \frac{\lambda_{n}n^{\left(0.5\gamma-1\right)}}{2|\sqrt{n}\widehat{\theta}_{j}^{\mathrm{ols}}|^{\gamma}}
 = \mathrm{sgn}(\hat{\theta}_j) \frac{\lambda_{n}n^{\left(0.5\gamma-1\right)}}{O_p(1)} \to 0
\]
and we cannot rule out the possibility that the two sides of (\ref{eq:KKT-3types})
being equal. 

\medskip{}

Finally, to show (\ref{eq:sel_CC}) we argue by contraposition. For
those $j\in\widehat{C}$, the counterpart of (\ref{eq:KKT-3types})
is the following KKT condition 
\begin{equation}
x_{j}^{c\prime}(y-W\hat{\theta})=\mathrm{sgn}(\hat{\theta}_j) \lambda_{n}\hat{\tau}_{j}/2,\ \ \forall\ j\in\widehat{C}.\label{eq:KKT_coint}
\end{equation}
(\ref{eq:sel_C}) already rules out $\mathrm{CoRk}(M^{*})>\mathrm{CoRk}(\widehat{M})$
in the limit. Now suppose $\mathrm{CoRk}(M^{*})<\mathrm{CoRk}(\widehat{M})$.
If so, for any $\widehat{M}$ there exists a constant cointegrating
vector $\psi\in\mathbb{R}^{p_{c}}$ such that $\psi^{\prime}\psi=1$
(scale normalization), $\psi_{C^{*c}\cap\widehat{C}}\neq0$ (must
involve wrongly selected inactive variables) and the linear combination
$x_{i\cdot}^{c}\psi$ is an I(0). Although $\psi$ is not unique in
general, we use $\psi$ to represent any one of them. Such a $\psi$
linearly combines the multiple equations in (\ref{eq:KKT_coint})
to generate a single equation 
\begin{equation}
\frac{1}{\sqrt{n}}\sum_{j\in\widehat{C}}\psi_{j}x_{j}^{c\prime}(y-W\hat{\theta})
=\frac{\lambda_{n}}{2\sqrt{n}}\sum_{j\in\widehat{C}} \mathrm{sgn}(\hat{\theta}_j)  \hat{\psi}_{j}\hat{\tau}_{j},\label{eq:KKT_comb}
\end{equation}
where a scaling factor $n^{-1/2}$ is multiplied on both sides to
facilitate the derivation of the rate, 
and we define $\hat{\psi}_j = \mathrm{sgn}(\hat{\theta}_j)  \psi_{j}$
to simplify notation.
 The left-hand side of (\ref{eq:KKT_comb})
is $O_{p}\left(1\right)$ since $\sum_{j\in\widehat{C}}\psi_{j}x_{ji}^{c\prime}=O_{p}\left(1\right)$
due to cointegration. The right-hand side of (\ref{eq:KKT_comb})
can be decomposed into two terms: 
\begin{align*}
\frac{\lambda_{n}}{\sqrt{n}}\sum_{j\in\widehat{C}}
\mathrm{sgn}(\hat{\theta}_j)  \hat{\psi}_{j}\hat{\tau}_{j} & =\frac{\lambda_{n}}{\sqrt{n}}\sum_{j\in\widehat{C}\cap C^{*}} \mathrm{sgn}(\hat{\theta}_j)  \hat{\psi}_{j}\hat{\tau}_{j}+\frac{\lambda_{n}}{\sqrt{n}}\sum_{j\in\widehat{C}\cap C^{*c}} \mathrm{sgn}(\hat{\theta}_j)  \hat{\psi}_{j}\hat{\tau}_{j}=:I_{1}+I_{2}.
\end{align*}
The first term 
\[
I_{1}\leq\frac{\lambda_{n}}{\sqrt{n}}\sum_{j\in\widehat{C}\cap C^{*}}\left|\hat{\psi}_{j}\right|\left|\hat{\tau}_{j}\right|\leq\frac{\lambda_{n}}{\sqrt{n}}\sum_{j\in\widehat{C}\cap C^{*}}\left|\hat{\tau}_{j}\right|\leq\frac{\lambda_{n}}{\sqrt{n}}\sum_{j\in C^{*}}\left|\hat{\tau}_{j}\right|=\frac{\lambda_{n}}{\sqrt{n}}O_{p}\left(1\right)=o_{p}\left(1\right)
\]
as $\widehat{\tau}_{j}=O_{p}\left(1\right)$ for $j\in\mathcal{C}^{*}$,
whereas the second term 
\[
I_{2}=\lambda_{n}n^{0.5\left(\gamma-1\right)}\sum_{j\in\widehat{C}\cap C^{*c}}\frac{ \mathrm{sgn}(\hat{\theta}_j)  \hat{\psi}_{j}}{|\sqrt{n}\widehat{\theta}_{j}^{\mathrm{ols}}|^{\gamma}}\to\infty\ \mbox{or }-\infty
\]
as $\text{\ensuremath{\sqrt{n}\widehat{\theta}_{j}^{\mathrm{ols}}}}=O_{p}\left(1\right)$
for $j\in C^{*c}$, $\psi_{j}$ is a constant with nonzero elements,
and $\lambda_{n}n^{0.5\left(\gamma-1\right)}\to\infty$ given the
specified rate for $\lambda_{n}\to\infty$ and $\gamma\geq1$. Whether
the right-hand side diverges to $+\infty$ or $-\infty$ depends on
the configuration of $\hat{\psi}_j$'s and the values of the sign functions. 
Eq.(\ref{eq:KKT_comb}) holds
with probability approaching 0 when $n$ is sufficiently large. That
is, no inactive cointegrating residuals can be formed within the selected
variables wpa1. Otherwise, a redundant cointegration group would induce
a cointegration vector $\psi$ that sends the right-hand side of (\ref{eq:KKT_comb})
to either $\infty$ or $-\infty$ in the limit. \end{proof}

\bigskip{}

\begin{proof} {[}Proof of Theorem \ref{thm:alasso_twin}{]} In Eq.(\ref{DGP XYZ})
we have separated the cointegrated variables into the active ones
and the inactive ones 
\begin{equation}
y=Z\alpha+X\beta+u=Z\alpha+\sum_{l\in C^{*}}X_{l}^{c}\phi_{l}+\sum_{l\in C^{*c}}X_{l}^{c}\phi_{l}+X\beta+u.\label{eq:full_eq_C1_1}
\end{equation}
We proceed our argument conditioning on the three events 
\[
S_{1}=\{M^{*}\cap\mathcal{\mathcal{I}}=\widehat{M}\cap\mathcal{\mathcal{I}}\},\ S_{2}=\{C^{*}\subseteq\widehat{C}\},\ S_{3}=\{\mathrm{CoRk}(M^{*})=\mathrm{CoRk}(\widehat{M})\}.
\]
According to Theorem \ref{thm:ada_3type}, these events occur wpa1,
given a sufficiently large sample size.

Under these three events, the regression equation (\ref{eq:full_eq_C1_1})
is reduced to 
\begin{equation}
y=Z_{M^{*}}\alpha_{M^{*}}+\sum_{l\in C^{*}}X_{l}^{c}\phi_{l}+(\sum_{l\in C^{*c}\cap\widehat{C}}X_{l}^{c}\phi_{l}+X_{M^{*}}\beta_{M^{*}})+u,\label{eq:full_eq_C1_2}
\end{equation}
where on the right-hand side of the above equation the first and the
fourth terms are present by the event $S_{1}$, the second and the
third terms by $S_{2}$. Due to event $S_{3}$, there is no cointegration
group in the third term, so we can re-write (\ref{eq:full_eq_C1_2})
as 
\begin{equation}
y=Z_{M^{*}}\alpha_{M^{*}}+\sum_{l\in\mathcal{C}^{*}}X_{l}^{c}\phi_{l}+\tilde{X}^{+}\tilde{\beta}^{+}+u,\label{eq:full_eq_C1_3}
\end{equation}
where $\tilde{X}^{+}=\left(\left(X_{l}^{c}\right)_{l\in C^{*c}\cap\widehat{C}},X_{M^{*}}\right)$
is the collection of \emph{augmented} non-cointegrating local unit
root processes in the post-selection regression equation (\ref{eq:full_eq_C1_2})
and $\tilde{\beta}^{+}=\left(\left(\phi_{j}\right)_{l\in C^{*c}\cap\widehat{C}},\beta_{M^{*}}\right)$
is the corresponding coefficient vector.

For $l\in C^{*c}\cap\widehat{C}$, the true coefficients are 0. Now
these variables appear as non-cointegrating local unit root processes
in (\ref{eq:full_eq_C1_3}), for which Theorem \ref{thm:ada_3type}
gives variable selection consistency. We implement the post-selection
Alasso in (\ref{eq:full_eq_C1_3}), or equivalently TAlasso as in
(\ref{eq:alasso_twin}). Among the variables in $\tilde{X}$ TAlasso
will eliminate $\left(X_{l}^{c}\right)_{l\in C^{*c}\cap\widehat{C}}$
wpa1 while the true active variables in $M^*$ will be maintained.
In the mean time, all variables in the first and second terms in (\ref{eq:full_eq_C1_3})
are active and Theorem \ref{thm:ada_3type}'s (\ref{eq:sel_C}) guarantees
their survival wpa1. The asymptotic distribution immediately follows
by applying Theorem \ref{thm:ada_3type}(a) to (\ref{eq:full_eq_C1_2}).
\end{proof}

\bigskip{}

\subsection{Proofs in Section \ref{sec:Conventional-LASSO-with}\label{subsec:Proofs-in-Section-4}}

In the following proofs concerning Plasso and Slasso, we use a compact
notation 
\[
D\left(s,v,\theta\right)=\sum_{j=1}^{\mathrm{dim}\left(\theta\right)}s_{j}\left[v_{j}\mathrm{sgn}(\theta_{j})I(\theta_{j}\neq0)+|v_{j}|I(\theta_{j}=0)\right]
\]
for three generic vectors $s$, $v$, and $\theta$ of the same dimension.
It takes the the scalar-based symbol $D\left(\cdot,\cdot,\cdot\right)$
in the main text as a special case.

\medskip{}

\begin{proof} {[}Proof of Corollary \ref{thm:lasso-mixture}{]} For
Part (a) and (b), we start with the local perturbation $\check{v}=\left(\check{v}_{z}^{\prime},\check{v}_{1}^{\prime},\check{v}_{2}^{\prime},\check{v}_{x}^{\prime}\right)^{\prime},$
$v_{n}=Q^{-1}R_{n}^{-1}\check{v}$ and $\theta_{n}=\theta_{n}^{\ast}+v_{n}$.
Notice that $\check{v}$ is different from $\tilde{v}$ in Theorem
\ref{thm:ada_3type} as we do not impose $\check{v}_{2}=0$. Define
\begin{equation}
V_{n}(\check{v})=v_{n}^{\prime}W^{\prime}Wv_{n}-2v_{n}^{\prime}W'u+\lambda_{n}\sum_{j=1}^{p}(|\theta_{jn}^{\ast}+v_{n}|-|\theta_{jn}^{\ast}|).\label{eq:VV_check}
\end{equation}
In view of (\ref{eq:ala_quad}) and (\ref{eq:ala_cross}), 
\[
V_{n}(\check{v})\Longrightarrow V(\check{v})=\check{v}'\Omega^{+}\check{v}-2\check{v}^{\prime}\zeta^{+}+\lim_{n\rightarrow\infty}\lambda_{n}\sum_{j=1}^{p}R_{jn}^{-1}D(1,\tilde{v}_{j},\theta_{j}^{0*}).
\]

Notice for $j\in\mathcal{I}_{1}$ with the local-to-zero true coefficient
$\theta_{jn}^{*}=\theta_{j}^{0*}/n^{\delta_{j}}$, if $\theta_{j}^{0*}\neq0$,
then $\theta_{jn}^{\ast}$ dominates $\check{v}_{j}/n$ in the limit
and $\lambda_{n}\left(|\theta_{jn}^{\ast}+n^{-1}\check{v}_{j}|-|\theta_{jn}^{\ast}|\right)= (\lambda_{n}/n) \text{sgn}(\theta_{j}^{0\ast})\cdot\check{v}_{j}\to0$
for any $\lambda_{n}/n\to0$. Similarly, if $\theta_{j}^{0*}=0$,
we have $\lambda_{n}(|\theta_{jn}^{\ast}+n^{-1}\check{v}_{j}|-|\theta_{jn}^{\ast}|)=  (\lambda_{n}/n)\left\vert \check{v}_{j}\right\vert \to0$
when $\check{v}_{j}\neq0$. Thus $\delta_{j}$ is irrelevant in the
limit for all parts (a), (b) and (c).

We invoke the Convexity Lemma. For Part (a), $\lambda_{n}/R_{jn}\to0$
for all $j$ so the penalty vanishes in the limit and the asymptotic
distribution is equivalent to that of OLS. For Part (b), the tuning
parameter's rate is $\lambda_{n}/\sqrt{n}\rightarrow c_{\lambda}\in(0,\infty)$
so that only the penalty term associated with $\mathcal{I}_{1}$ vanishes
in the limit. \medskip{}

Part (c) needs more elaboration. Let $v_{\lambda_{n}}=(\lambda_{n} / \sqrt{n})Q^{-1}R_{n}^{-1}\check{v}=(\lambda_{n} / \sqrt{n})v_{n}$
and $\theta_{\lambda_{n}}=\theta_{n}^{\ast}+v_{\lambda_{n}}$. Define
\[
V_{\lambda_{n}}(\check{v})=v_{\lambda_{n}}^{\prime}W^{\prime}Wv_{\lambda_{n}}^{\prime}-2v_{\lambda_{n}}^{\prime}W^{\prime}u+\lambda_{n}\sum_{j=1}^{p}(|\theta_{jn}^{\ast}+v_{\lambda_{n},j}|-|\theta_{jn}^{\ast}|),
\]
and multiply $n/\lambda_{n}^{2}$ on both sides, 
\[
\left(\frac{n}{\lambda_{n}^{2}}\right)V_{\lambda_{n}}(\check{v})=v_{n}^{\prime}W^{\prime}Wv_{n}-2\frac{\sqrt{n}}{\lambda_{n}}v_{n}^{\prime}W^{\prime}u+\frac{n}{\lambda_{n}}\sum_{j=1}^{p}(|\theta_{jn}^{\ast}+v_{\lambda_{n},j}|-|\theta_{jn}^{\ast}|).
\]
By the rate condition of $\lambda_{n}$, the second term $2(\sqrt{n}/\lambda_{n})v_{n}^{\prime}W^{\prime}u=o_{p}(1)$.
Given $\tilde{v}_{j}\neq0$ and $n$ large enough:
\begin{itemize}
\item If $j\in\mathcal{I}_{0}\cup\mathcal{C}$, the coefficients $\theta_{jn}^{*}=\theta_{j}^{0\ast}$
is invariant with $n$ so that 
\begin{equation}
\frac{n}{\lambda_{n}}(|\theta_{j}^{0\ast}+\frac{\lambda_{n}}{\sqrt{n}}\frac{\check{v}_{j}}{\sqrt{n}}|-|\theta_{j}^{0\ast}|)=\frac{n}{\lambda_{n}}D\left(1,\frac{\lambda_{n}}{n}\check{v}_{j},\theta_{j}^{0\ast}\right)=D\left(1,\check{v}_{j},\theta_{j}^{0\ast}\right).\label{eq:term_vz}
\end{equation}
\item If $j\in\mathcal{I}_{1}$, the reverse triangular inequality $\left\vert |a+b|-|a|\right\vert \leq\left\vert b\right\vert $
for any $a,b\in\mathbb{R}$ guarantees 
\begin{equation}
\frac{n}{\lambda_{n}}\bigg||\theta_{jn}^{\ast}+\frac{\lambda_{n}}{n^{3/2}}\check{v}_{j}|-|\theta_{jn}^{\ast}|\bigg|\leq\frac{n}{\lambda_{n}}\left\vert \frac{\lambda_{n}}{n^{3/2}}\check{v}_{j}\right\vert =O\left(n^{-1/2}\right),\label{eq:term_vx}
\end{equation}
which is dominated by $D\left(1,\check{v}_{j},\theta_{j}^{0\ast}\right)$
in the limit if $\check{v}_{j}\ne0$ and $\theta_{j}^{0*}\neq0$. 
\end{itemize}
Thus we conclude Part (c) \end{proof}

\bigskip{}

\bigskip{}

\begin{proof} {[}Proof of Corollary \ref{thm:std_lasso_3type}{]}
We first express the limit distributions of the sample standard
deviation.
\begin{itemize}
\item For a nonstationary variable $x_{l}$ associated with $\mathcal{I}_{1}$, we have its limit distribution $$d_{l}\sim\sqrt{\int_{0}^{1}J_{c_{x,l}}^{2}(r)dr-\left(\int_{0}^{1}J_{c_{x,l}}\left(r\right)dr\right)^{2}}.$$
\item For $x_{2,l}^{c}$ associated with $\mathcal{C}_{2}$, similarly $d_{l}\sim\sqrt{\int_{0}^{1}J_{c_{2,l}}^{2}(r)dr-\left(\int_{0}^{1}J_{c_{2,l}}\left(r\right)dr\right)^{2}}$.
\item For $x_{1l}^{c}=\sum_{s=1}^{p_{2}}A_{1,ls}x_{2s}^{c}+v_{1l}$ associated
with $\mathcal{C}_{1}$, for each $i$ we have $x_{1il}^{c}=A_{1,l}^{\prime}x_{2i}^{c}+v_{1il}$
where $A_{1,l}^{\prime}$ is the $l$-th row of $A_{1}$ matrix,
for $l\in\left(1,...,p_{1}\right)$. Note that, for each $i=\left\lfloor nr\right\rfloor $
with $r\in\left[0,1\right]$,
\[
\frac{x_{1il}^{c}}{\sqrt{n}}=A_{1,l}^{\prime}\frac{x_{2i}^{c}}{\sqrt{n}}+\frac{v_{1il}}{\sqrt{n}}\sim A_{1,l}^{\prime}J_{c_2}\left(r\right)+O_{p}(\frac{1}{\sqrt{n}})
\]
so that after some simple algebra,
\begin{eqnarray*}
\frac{\hat{\sigma}_{l}}{\sqrt{n}} & = & \sqrt{\frac{1}{n^{2}}\sum_{i=1}^{n}\left(x_{1il}^{c}\right)^{2}-\frac{1}{n}\left(\frac{1}{n}\sum_{i=1}^{n}x_{1il}^{c}\right)^{2}}\\
 & = & \sqrt{\frac{1}{n} \sum_{i=1}^{n}\left(A_{1,l}^{\prime}\frac{x_{2i}^{c}}{\sqrt{n}}\right)^{2}
 -\left(\frac{1}{n} \sum_{i=1}^{n}A_{1,l}^{\prime}\frac{x_{2i}^{c}}{\sqrt{n}}\right)^{2}}+o_{p}(1)\\
 & \Longrightarrow & d_{l} \sim \sqrt{A_{1,l}^{\prime}\left(\int J_{c2}\left(r\right)J_{c2}^{\prime}\left(r\right)dr\right)A_{1,l}-\left(A_{1,l}^{\prime}\int J_{c2}\left(r\right)dr\right)^{2}}.
\end{eqnarray*}
\end{itemize}
\medskip{}

Now we start the proof. Set the same local perturbation $\check{v}=\left(\check{v}_{z}^{\prime},\check{v}_{1}^{\prime},\check{v}_{2}^{\prime},\check{v}_{x}^{\prime}\right)^{\prime},$
$v_{n}=Q^{-1}R_{n}^{-1}\check{v}$ and $\theta_{n}=\theta_{n}^{\ast}+v_{n}$
as in the proof of Corollary \ref{thm:lasso-mixture}. We focus on
the counterpart of the terms in (\ref{eq:VV_check}).
\begin{itemize}
\item If $j\in\mathcal{I}_{0}$, we have $\widehat{\sigma}_{j}=O_{p}\left(1\right)$
and the coefficient $\theta_{j}^{0\ast}$ is independent of $n$ so
that 
\[
\widehat{\sigma}_{j}\left(|\theta_{j}^{0\ast}+\frac{v_{j}}{\sqrt{n}}|-|\theta_{j}^{0\ast}|\right)=D\left(\widehat{\sigma}_{j},\frac{v_{j}}{\sqrt{n}},\theta_{j}^{0*}\right)=D\left(O_{p}\left(1\right),O\left(\frac{1}{\sqrt{n}}\right),\theta_{j}^{0*}\right)\overset{p}{\to}0;
\]
\item If $j\in\mathcal{C}$, again $\theta_{j}^{0\ast}$ is independent
of $n$ and 
\[
\widehat{\sigma}_{j}\left(|\theta_{j}^{0\ast}+\frac{v_{j}}{\sqrt{n}}|-|\theta_{j}^{0*}|\right)=D\left(\widehat{\sigma}_{j},\frac{v_{j}}{\sqrt{n}},\theta_{j}^{0*}\right)=D\left(\frac{\widehat{\sigma}_{j}}{\sqrt{n}},v_{j},\theta_{j}^{0*}\right)\Longrightarrow D\left(d_{j},v_{j},\theta_{j}^{0*}\right)=O_{p}\left(1\right),
\]
since these indices are associated with a local unit root process
$x_{j}^{c}$ and therefore $ \widehat{\sigma}_{j} / \sqrt{n}\Longrightarrow d_{j}$
is non-degenerate;
\item If $j\in\mathcal{I}_{1}$, similarly we have 
\[
\widehat{\sigma}_{j}\left(|\theta_{jn}^{\ast}+\frac{v_{j}}{n}|-|\theta_{jn}^{\ast}|\right)=D\left(\widehat{\sigma}_{j},\frac{v_{j}}{n},\theta_{j}^{0*}\right)=D\left(\frac{\widehat{\sigma}_{j}}{\sqrt{n}},\frac{v_{j}}{\sqrt{n}},\theta_{j}^{0*}\right)=D\left(O_{p}\left(1\right),\frac{v_{j}}{\sqrt{n}},\theta_{j}^{0*}\right)\overset{p}{\to}0.
\]
\end{itemize}
The above analysis implies 
\begin{equation}
V_{n}(\check{v})\Longrightarrow V\left(\check{v}\right)=\check{v}^{\prime}\Omega^{+}\check{v}-2\check{v}'\zeta^{+}+c_{\lambda}\sum_{j\in\mathcal{C}}D\left(d_{j},v_{j},\theta_{j}^{0*}\right),\label{eq:Slasso_part_ab}
\end{equation}
and Part (a) and (b) follow. \medskip{}

For Part (c), let $\tilde{R}_{n}=R_{n}/\lambda_{n}$ and $\theta_{n}=\theta_{n}^{\ast}+\tilde{R}_{n}^{-1}\tilde{v}$.
Define 
\[
\tilde{V}_{n}(\check{v})=\check{v}^{\prime}\left(\tilde{R}_{n}^{-1}W^{\prime}W\tilde{R}_{n}^{-1}\right)\check{v}-2\check{v}^{\prime}\tilde{R}_{n}^{-1}W^{\prime}u+\lambda_{n}\sum_{j=1}^{p}\widehat{\sigma}_{j}(|\theta_{jn}^{\ast}+\tilde{R}_{jn}^{-1}\check{v}_{j}|-|\theta_{jn}^{\ast}|).
\]
Multiply $1/\lambda_{n}^{2}$ on both sides, 
\begin{eqnarray*}
\frac{\tilde{V}_{n}(\check{v})}{\lambda_{n}^{2}} & = & \check{v}^{\prime}\left(R_{n}^{-1}W^{\prime}WR_{n}^{-1}\right)\check{v}-\frac{2}{\lambda_{n}}\check{v}^{\prime}R_{n}W^{\prime}u+\frac{1}{\lambda_{n}}\sum_{j=1}^{p}\widehat{\sigma}_{j}(|\theta_{jn}^{\ast}+\tilde{R}_{jn}^{-1}\check{v}_{j}|-|\theta_{jn}^{\ast}|)\\
 & = & \check{v}^{\prime}\left(R_{n}^{-1}W^{\prime}WR_{n}^{-1}\right)\check{v}+o_{p}(1)+\frac{1}{\lambda_{n}}\sum_{j=1}^{p}\widehat{\sigma}_{j}(|\theta_{jn}^{\ast}+\tilde{R}_{jn}^{-1}\check{v}_{j}|-|\theta_{jn}^{\ast}|).
\end{eqnarray*}
by the rate condition of $\lambda_{n}$. Again we study the last term.
For $\check{v}_{j}\neq0$ and a sufficiently large $n$:
\begin{itemize}
\item For $j\in\mathcal{I}_{0}$, 
\[
\frac{1}{\lambda_{n}}\widehat{\sigma}_{j}\left(|\theta_{j}^{0*}+\frac{\lambda_{n}}{\sqrt{n}}\check{v}_{j}|-|\theta_{j}^{0*}|\right)=\frac{1}{\lambda_{n}}D\left(\widehat{\sigma}_{j},\frac{\lambda_{n}}{\sqrt{n}}\check{v}_{j},\theta_{j}^{0*}\right)=D\left(\widehat{\sigma}_{j},\frac{\check{v}_{j}}{\sqrt{n}},\theta_{j}^{0*}\right)\overset{p}{\to}0;
\]
\item For $j\in\mathcal{C}$, 
\begin{align*}
\frac{1}{\lambda_{n}}\widehat{\sigma}_{j}\left(|\theta_{j}^{0*}+\frac{\lambda_{n}}{\sqrt{n}}\check{v}_{j}|-|\theta_{j}^{0*}|\right) & =\frac{1}{\lambda_{n}}D\left(\widehat{\sigma}_{j},\frac{\lambda_{n}}{\sqrt{n}}\check{v}_{j},\theta_{j}^{0*}\right)\\
 & =D\left(\frac{\widehat{\sigma}_{j}}{\sqrt{n}},\check{v}_{j},\theta_{j}^{0*}\right)=D\left(d_{j},v_{j},\theta_{j}^{0*}\right)=O_{p}\left(1\right);
\end{align*}
\item For $j\in\mathcal{I}_{1}$, the reverse triangular inequality $\left\vert |a+b|-|a|\right\vert \leq\left\vert b\right\vert $
implies
\begin{align*}
\frac{1}{\lambda_{n}}\widehat{\sigma}_{j}\left||\theta_{jn}^{\ast}+\frac{\lambda_{n}}{n}\check{v}_{j}|-|\theta_{jn}^{\ast}|\right| & \leq\frac{1}{\lambda_{n}}\widehat{\sigma}_{j}|\frac{\lambda_{n}}{n}\check{v}_{j}|=\frac{\widehat{\sigma}_{j}}{\sqrt{n}}\frac{\check{v}_{j}}{\sqrt{n}}\overset{p}{\rightarrow}0.
\end{align*}
\end{itemize}
We obtain $  \lambda_{n}^{-2} \tilde{V}(\check{v}) \Longrightarrow\check{v}^{\prime}\Omega^{+}\check{v}+\sum_{j\in\mathcal{C}}D\left(d_{j},v_{j},\theta_{j}^{0\ast}\right)$
and the conclusion follows. \end{proof}


\bigskip \bigskip \bigskip

\singlespacing
\bibliographystyle{elsarticle-harv}
{
\bibliography{citation}
}

\end{document}